\documentclass[accepted=2026-08-11,a4paper]{quantumarticle}
\pdfoutput=1 

\usepackage{caption} 

\usepackage{braket}
\usepackage{graphicx} 
\usepackage{dcolumn} 
\usepackage{bm} 
\usepackage{siunitx}
\usepackage{xcolor}  
\usepackage{tabularx}
\usepackage{dsfont}					
\usepackage{mathrsfs}				
\usepackage{float}
\usepackage{hyperref} 
\usepackage{amsfonts} 
\usepackage{amsmath} 
\usepackage{amssymb}

\newcommand{\comment}[1]{}			
\newcommand{\tr}[0]{\mathbf{tr}}		
\newcommand{\Tr}[0]{\mathbf{Tr}}
\newcommand{\llangle}[0]{\langle\!\langle}
\newcommand{\rrangle}[0]{\rangle\!\rangle}

\usepackage[normalem]{ulem}

\begin{document}

\title{Non-Gaussian Noise Magnetometry Using Local Spin Qubits}
\author{Jonathan~B.~Curtis}
\email[]{jon.curtis.94@gmail.com}
\affiliation{Institute for Theoretical Physics, ETH Z{\"u}rich, Z{\"u}rich, 8093, CH}
\author{Amir~Yacoby}
\affiliation{Department of Physics, Harvard University, Cambridge, MA 02138, USA}
\author{Eugene~Demler}
\affiliation{Institute for Theoretical Physics, ETH Z{\"u}rich, Z{\"u}rich, 8093, CH}


\begin{abstract}
Atomic scale qubits, as may be realized in nitrogen vacancy (NV) centers in diamond, offer the opportunity to study magnetic field noise with nanometer scale spatial resolution. 
Using these spin qubits, one can learn a great deal about the magnetic-field noise correlations, and correspondingly the collective-mode spectra, in quantum materials and devices.
However, to date these tools have been essentially restricted to studying Gaussian noise processes\textemdash equivalent to linear-response. 
In this work we will show how to extend these techniques beyond the Gaussian regime and show how to unambiguously measure higher-order magnetic noise cumulants in a local, spatially resolved way. 
We unveil two protocols for doing this; the first uses a single spin-qubit and different dynamical decoupling sequences to extract non-Markovian and non-Gaussian spin-echo noise. 
The second protocol uses two-qubit coincidence measurements to study spatially non-local cumulants in the magnetic noise. 
We then demonstrate the utility of these protocols by considering a model of a bath of non-interacting two-level systems, as well as a model involving spatially correlated magnetic fluctuations near a second-order Ising phase transition.
In both cases, we highlight how this technique can be used to measure in a real many-body system how fluctuation dynamics converge towards the central limit theorem as a function of effective bath size.
We then conclude by discussing some promising applications and extensions of this method.
\end{abstract}

\maketitle

\section{Introduction}
\label{sec:intro}

Over the last few decades, a large and sustained effort has been made in developing qubits for the purposes of quantum information processing and computing. 
As a result, significant advances have been made not only towards isolating, but also controlling and manipulating qubits across a number of solid-state platforms ranging from superconducting qubits to color-center defect spins.
These capabilities also make solid-state defect spins, such as nitrogen-vacancy (NV) centers in diamond, particularly well-suited for advanced quantum sensing applications.
By utilizing the high sensitivity and small spatial extent of such solid-state spin qubits, small magnetic fields can be studied on nanometer length scales.

An emerging application of these nanoscale spin-qubits is their ability to also measure magnetic-field noise spectra~\cite{Rovny.2024}, allowing them to probe magnetization dynamics as well as statics. 
This modality of operation, known as noise magnetometry, essentially operates the spin-qubit as a spatially local probe of the magnetic noise bath through the induced effect of the magnetic noise on the qubit's energy- and phase-relaxation times ($T_1$ and $T_2$). 
Theoretically, noise magnetometry has been proposed to be a useful tool for sensing a wide range of interesting phenomena in condensed matter physics including superconductors~\cite{Curtis.2024,Dolgirev.2022,Chatterjee.2022,Kelly.2024,Konig.2020,Liu.2025}, spin-liquids and magnetic insulators~\cite{Khoo.2022}, Landau levels~\cite{De.2024}, Wigner crystals~\cite{Dolgirev.2023}, semimetals~\cite{Zhang.2022o8f}, dynamical critical phenomena~\cite{Quan.2005, Wei.2012,Chen.2013, Machado.2022}, as well as strongly-correlated electron systems~\cite{Agarwal.2017,Rodriguez-Nieva.2018,Zhang.2024}.
Experimentally, these techniques have recently been successfully applied to a number of interesting systems including metals~\cite{Kolkowitz.2015}, topological antiferromagnets~\cite{McLaughlin.2022}, ferromagnets with signatures of magnon hydrodynamics~\cite{Xue.2024}, critical points of magnetic systems~\cite{Ziffer.2024}, and strongly current biased graphene junctions~\cite{Andersen.2019}. 
Recently efforts have started to push towards using multiple spin-qubits in tandem in order to measure spatial correlations in magnetic noise across a wider range in distance~\cite{Rovny.2022,Huxter.2024}, or prepare correlated states of these spin-qubits via their shared magnetic bath~\cite{Ji.2024,Li.2025}.
In the context of bulk diamond, dense of ensembles of NV centers have also been used to probe quantum many-body dynamics~\cite{Choi.2017,Davis.2023,Hughes.2024,Zhou.2020lzi}.

While this has been a fairly successful program, it fundamentally explores only the second moment of the magnetic noise and therefore is essentially limited to probing linear response functions (via application of the fluctuation dissipation relation).
While the second moment of the magnetic field, which can essentially be captured by an effective Gaussian model for the field fluctuations, holds a large amount of information about the nature of collective excitations in a material, it is expected that the complete description of the magnetic field fluctuations should not be able to be captured by any effective Gaussian model.
A particularly striking example is that near a second-order phase transition large fluctuations in the order parameter lead to violations of the central limit theorem~\cite{Bramwell.2009}.
However, these violations of the central limit theorem are not directly visible unless the higher-order correlation functions are measured.  
This then raises the pressing questions of (i) whether we can access these higher moments using local noise magnetometry, and (ii) if so what information can we discern from them? 
We will answer both of the these questions in this work, and in doing so show that NV center spin-qubits are ideally suited for exploring many-body correlations in quantum materials. 

In principle, it is known that many systems exhibit nontrivial higher order counting statistics; an early example is the Hanbury Brown and Twiss (HBT) intensity interferometer, which measures correlations in the intensity of light and has been used to great success in measuring the angular size of distant stars~\cite{Brown.1954}.
Such intensity correlation functions, and in particular the second-order optical coherence function $g^{(2)}(\tau) = \langle:\hat{I}(\tau)\hat{I}(0):\rangle/( \langle\hat{I}(0)\rangle\langle\hat{I}(\tau)\rangle)$, where $\hat{I}(\tau)$ is the quantized intensity operator for the light field and is itself second order in the electric field operator $\hat{E}(\tau)$, have since played a central role in the development of quantum optics.
An important example was the observation of {\it photon antibunching}~\cite{Kimble.1977}, which roughly corresponds to measuring $g^{(2)}(0)<1$, and implies that measuring a photon at time $t=0$ precludes the observation of a second photon at short times $\tau\to 0$.
This is known to only be possible if the incident light arrives in discrete units and therefore confirms the quantum nature of light; functionally, measuring $g^{(2)}(0) < 1$ is important in confirming that the emitting light source can operate in the ``single-photon" regime. 

In systems of ultracold atomic gases HBT and higher-order correlations have been measured in both bosonic~\cite{Folling.2005} and fermionic~\cite{Rom.2006} gases, wherein noise correlations have been used to reveal signatures of many-body correlations~\cite{Altman.2004} such as Cooper pairing~\cite{Greiner.2005} or to directly compare particle statistics~\cite{Jeltes.2007}.
The full distribution function of interference fringes in a one-dimensional Bose-Einstein condensate has also been measured~\cite{Hofferberth.2008}, with the dynamics of non-Gaussian correlations having even been extracted~\cite{Schweigler.2021}.
In solid-state physics however, the situation is much less developed. 
While theoretically it has been proposed that measuring the full counting statistics of charge conductance in a one-dimensional channel can be measured~\cite{Levitov.1996,Ubbelohde.2012,Bednorz.2010} and used to reveal the physics of e.g. Luttinger liquids~\cite{Kuhne.2015}, measuring this in a solid-state system has been challenging.
More recently, it has been proposed that HBT-type optical coherence functions could be used to probe higher-order correlations in quantum materials which interact with detected photons~\cite{Joubaud.2008,Nambiar.2024,Cheung.2024,Randi.2017,Kass.2024}, however these techniques are expected to be diffraction limited to essentially zero momentum-space resolution, and operate at relative high frequencies.

If in a similar manner, the higher cumulants of the magnetic noise could be extracted using a nanoscale local spin-qubit such as an NV center, it could potentially unearth a great deal of information about the nature of magnetic excitations and their interactions in quantum materials where it is otherwise inconceivable to obtain the level of detail that can be seen in quantum simulators and ultracold atom systems. 
For instance, were it possible to obtain such a degree of spatial resolution in a condensed matter material system, it might be possible to study braiding and fractionalization in spin-liquid systems~\cite{Feng.2023,McGinley.2024,Potts.2024}, a task which has been notoriously difficult through existing probes.

In this work we will propose a number of protocols which can be implemented in local spin-qubit systems such as NV centers in order to isolate higher-order (i.e. non-Gaussian) noise cumulants, very much in the spirit of the original proposal by Levitov, Lee, and Lesovik~\cite{Levitov.1996}.
It has long been known that non-Gaussian noise processes can have a particularly detrimental effect on the coherence of qubits, leading to the vexing ``two-level system noise" which can limit the coherence times of superconducting qubits~\cite{Galperin.2006,Paladino.2002,Cai.2020,Cywinski.2008,Bergli.2007,Klimov.2018}. 
Concomitantly, it has been proposed that qubits may be useful for measuring non-Gaussian noise correlations~\cite{Wang.2019,Norris.2016,Dong.2023nel,Bergli.2007,Szankowski.2016,Kuffer.2024,Cywinski.2014}, as recently demonstrated in the case of single superconducting qubits~\cite{Sung.2019}, single NV-centers~\cite{Meinel.2022,Laraoui.2011,Xia.2025,Zhang.2025}, and trapped ions~\cite{Kotler.2013}.
However, the utility and application of these protocols for sensing non-Gaussian correlations in quantum material settings has been limited and relatively unexplored.
Similarly, a few protocols for using multiple qubits in order to extract spatial correlations in noise~\cite{Szankowski.2016} have been demonstrated in quantum-dot spin qubits~\cite{Boter.2020}, superconducting qubits~\cite{Lupke.2020}, and recently NV centers as well~\cite{Rovny.2022,Rovny.2025}; however, these protocols have been so far restricted to Gaussian noise correlations. 
For instance, Ref.~\cite{Hosseinabadi.2025} comprehensively studies the second-order cross-correlations between two NV centers in great detail, but does not explore the higher order cumulants. 
Ultimately, despite these significant advances, a comprehensive set of protocols which can be used to systematically isolate and probe non-Gaussian noise correlations in quantum materials has remained elusive. 

In this work we present a simple, comprehensive, and extensible set of protocols which can be applied to NV center spin-qubits in order to extract non-Gaussian noise cumulants from magnetic field noise across a variety of length scales.
Building on the spin-echo protocols of Refs.~\cite{Wang.2019,Norris.2016,Dong.2023nel,Bergli.2007,Kuffer.2024}, our protocols fundamentally hinge on the novel capabilities offered by the local, scanning nature of atomic spin-defects like NV centers which makes them ideally suited for studying non-Gaussian noise correlations in quantum materials and devices. 
In the case of a single spin-qubit, we will show how dynamical decoupling spin-echo sequences can be combined with local noise magnetometry in order to study {\bf scale-dependent} non-Gaussian correlations in a material system. 
We also address multi-qubit systems~\cite{Hosseinabadi.2025,Le.2025}, such as the one recently implemented in Refs.~\cite{Rovny.2022,Rovny.2025,Le.2025,Zhou.2025}, and show that either coincidence measurements or entangled ``Bell-state echo" techniques (similar to those employed in Refs.~\cite{Boter.2020,Lupke.2020,Rovny.2025,Zhou.2025}) can be used to isolate spatially non-local multipoint correlations in non-Gaussian magnetic noise. 
This is particularly exciting given the demonstration of two-photon interference effects~\cite{Bernien.2012}, and more recently of entanglement between two NV centers and the usage of this entanglement as a resource for measuring noise~\cite{Rovny.2025,Zhou.2025}.
Finally, we consider applying these techniques to specific examples of interesting magnetic systems which may exhibit non-Gaussian magnetic noise.
First, we consider an ensemble of independent bath spins exhibiting a random telegraph noise process and show how by varying both the spin echo-time and qubit-sample distance, non-Gaussian correlations can be extracted from the magnetic noise in a paramagnetic system and how, in the thermodynamic and long-time limit Gaussianity is recovered, as anticipated from the central-limit theorem. 
We then show how this technique can be used to study the non-Gaussian nature of fluctuations at a critical point, where spatial correlations become important; specifically, here we will consider the case of a second-order Ising-type magnetic phase transition and show how this technique is able to measure the scale-dependent correlated magnetic noise.
{While critical noise has been a subject of intense study~\cite{Quan.2005,Wei.2012,Chen.2013,Joubaud.2008,Cheung.2024,Bramwell.2009}, here we will argue that local noise probes offer new and valuable insights into the growth of spatial correlations.}
Finally, we conclude by offering a number of fascinating perspectives on how these new techniques may be implemented in solid-state systems in order to probe new aspects of quantum many-body physics in real material systems.

The remainder of this paper is organized as follows.
In Sec.~\ref{sec:gaussian} we provide a brief review of how noise spectroscopy works in the case of Gaussian noise processes, and why it is difficult to extract higher cumulants. 
Then in Sec.~\ref{sec:higher-cum} we show how spin-echo sequences can be used to isolate non-Markovian non-Gaussian higher noise cumulants.
In Sec.~\ref{sec:spatial} we provide two protocols useful for extracting spatially non-local non-Gaussian noise cumulants which are amenable to multi-qubit setups. 
Then, in Sec.~\ref{sec:app} we provide examples of how higher-order cumulants can be extracted from model systems and highlight the novel capabilities offered by local noise spectroscopy when studying these nonlinear noise correlations. 
Finally, in Sec.~\ref{sec:conc} we conclude and offer a perspective on further uses and applications of this technique for studying quantum materials.

\section{Gaussian Noise Spectroscopy}
\label{sec:gaussian}

{While there are already comprehensive formal theories of noise spectra in open quantum systems~\cite{Gasbarri.2018,Wang.2019,Norris.2016}, these theories remain very complex and difficult to implement experimentally.
Here we recast these analysis in order to present simple, intuitive, and extensible protocols for extracting specific higher noise cumulants.}
It will be helpful to first briefly review how typical Ramsey spectroscopy works.
We first consider the case of a single qubit with Hamiltonian 
\begin{equation}
\label{eqn:hamiltonian}
    \hat{H}(t) =  \frac12\Delta(t)\hat{\sigma}_z,
\end{equation}
where $\Delta(t)$ is a fluctuating time-dependent splitting which, in the case of the spin-qubit is given in terms of the magnetic field at the qubit location $\mathbf{B}(\mathbf{r}_{\rm NV},t)$ by $\Delta(t) = g \mu_B \mathbf{n}\cdot\mathbf{B}(\mathbf{r}_{\rm NV},t)$, where $g\mu_B$ is the spin magnetic moment and $\mathbf{n}$ is the spin quantization axis of the NV spin. 
We will consider the case where $\Delta(t)$ is fluctuating in time with zero mean, which can also be accomplished by passing to a rotating frame.
Furthermore, we will assume the fluctuations of $\Delta(t)$ is a classical time-translation invariant stochastic process; 
{the general theory of quantum dephasing processes and their unravelling and characterization using multispectra has been worked out~\cite{Gasbarri.2018,Wang.2019}. 
Details of how our specific protocols can be generalized to quantum processes can also be found in Appendix~\ref{app:quantum}.}

\begin{figure}
    \centering
    \includegraphics[width=\linewidth]{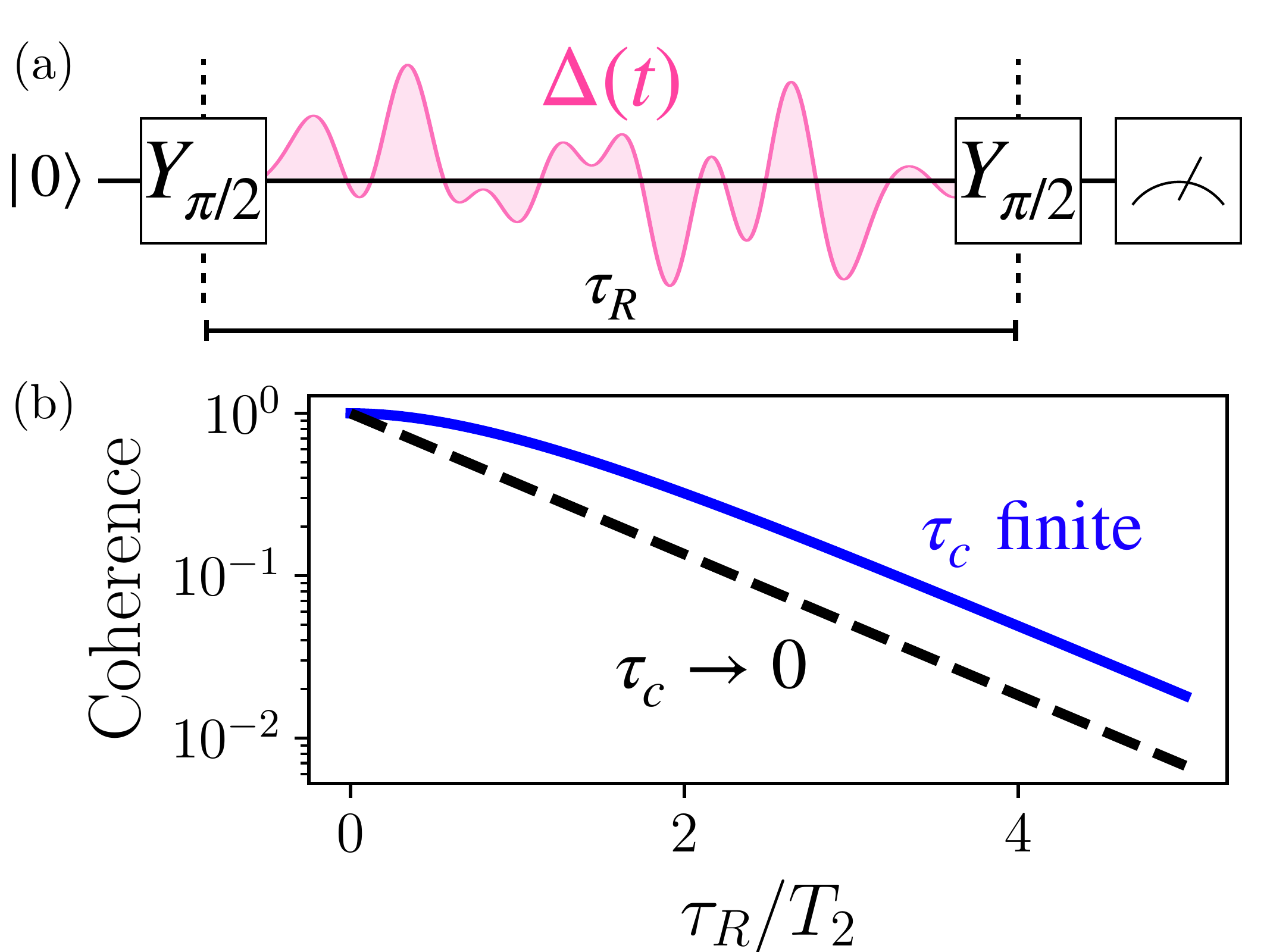}
    \caption{(a) Basic Ramsey sequence for duration $t_R$ in the presence of fluctuating level splitting $\Delta(t)$.
    (b) Resulting density-matrix coherence plotted as a function of Ramsey time $\tau_R$ for both white noise (dashed black) and correlated noise with correlation time $\tau_c$ (solid blue).
    }
    \label{fig:Ramsey}
\end{figure}

In order to probe the noise spectrum we will perform a Ramsey spectroscopy sequence, illustrated in Fig.~\ref{fig:Ramsey}(a).
Here and throughout we will assume the qubit is decoupled from the system for $t<0$; in reality this is not true, but rather it is expected that qubit will be in an unpolarized mixed state of $\hat{\rho}(-\infty) = \frac12\mathds{1}$.
Then, at time $t= 0$ the qubit is initialized in the $|+\rangle = \frac{1}{\sqrt{2}}(|0\rangle+|1\rangle)$ state using a $\pi/2$ rotation around the $y$-axis, enacted by the unitary operator $\hat{Y}_{\pi/2} = (\mathds{1} + i \hat{\sigma}_y)/\sqrt{2}$.
The qubit is then allowed to evolve under the Hamiltonian in Eq.~\eqref{eqn:hamiltonian} for a time $\tau_R$ (the Ramsey time), at which point a second $\pi/2$ rotation is applied and the qubit population is measured.
At time $\tau_R$ just before measuring, for a particular realization (unravelling) of $\Delta(t)$, the state of the qubit is 
\begin{equation}
    \hat{\rho}(\tau_R^-) = \frac12 \left[ \mathds{\hat{1}} + \cos X \hat{\sigma}_z + \sin X \hat{\sigma}_y  \right],
\end{equation}
where the phase acquired is 
\begin{equation}
    X = \int_0^{\tau_R}\Delta(t) dt.
\end{equation}
As this is a fluctuating over realizations of the noise, the density matrix must be averaged over the bath in order to predict the behavior of a measurement.
We denote here the bath averaging by the symbols $\llangle \cdot \rrangle$, which averages the enclosed functional over the distribution of $\Delta(t)$.

Averaging and measuring the qubit population gives access to 
\begin{equation}
    \langle \hat{\sigma}_z \rangle = \llangle \cos X\rrangle = \textrm{Re}\llangle  e^{-iX } \rrangle . 
\end{equation}
Note that, in particular this measurement protocol described gives access to $\llangle \cos X\rrangle $.
If the noise processes has reflection symmetry such that odd cumulants vanish, this will be equivalent to the full function $\llangle e^{-iX}\rrangle $.
If this is not the case, then one must be careful of the fact that in this case the quantities measured are all missing information about the odd cumulants.
Here we will not dwell on this distinction and assume we can deal with the full cumulant function $\llangle e^{-iX}\rrangle$ unless otherwise specified. 

We then find the standard result that the coherence of the density matrix dephases over time, and is characterized by the decoherence function
\begin{equation}
    \mathcal{C}_{X} = \log \llangle e^{-iX}\rrangle.
\end{equation}
Within a Gaussian, but potentially non-Markovian, approximation for the correlations of $\Delta(t)$, we find that the coherence function generically decays as 
\begin{multline}
    \mathcal{C}_X = -\frac12 \llangle X^2 \rrangle_c\\
    = -\frac12 \int_0^{\tau_R} dt_1 \int_0^{\tau_R} dt_2 \llangle \Delta(t_1) \Delta(t_2) \rrangle_c. 
\end{multline}
Here we introduce the notation that $\llangle X^n \rrangle_c$ is the connected correlation function for $X^n$, obtained by subtracting lower order cumulants of $X$.
This is illustrated schematically in Fig.~\ref{fig:Ramsey}(b), which shows typical decay profiles for the Ramsey coherence both in the case of Markovian white noise (dashed black curve), which exhibits purely exponential decay, as well as exponentially correlated noise (solid blue curve), with correlation function 
\begin{equation}
    \llangle \Delta(t_1) \Delta(t_2) \rrangle = \frac{1}{T_2\tau_c} e^{-|t_1-t_2|/\tau_c},
\end{equation}
which leads to the well-known Kubo lineshape decoherence function~\cite{Mukamel.1995,Hamm.2011}
\begin{equation}
    \mathcal{C}_X = -\frac{\tau_R}{T_2} + \frac{\tau_c}{T_2} \left[ 1 - e^{-\tau_R/\tau_c}\right].
\end{equation}
Here $\tau_c$ is the correlation (memory) time of the bath, whereas $T_2$ is the resulting dephasing time (valid for Ramsey times longer than the correlation time). 
More generally, we find the well-known result~\cite{Mukamel.1995,Hamm.2011} that in terms of the bath spectral function, defined by  
\begin{equation}
    \llangle \Delta(t_1) \Delta(t_2) \rrangle = \int \frac{d\omega}{2\pi} J(\omega) e^{-i\omega(t_1-t_2)},
\end{equation}
the decoherence function is given in the Gaussian approximation as  
\begin{equation}
    \mathcal{C}_X =  \int \frac{d\omega}{\pi} \frac{J(\omega)}{\omega^2} \sin^2\left(\frac{\omega \tau_R}{2}\right).
\end{equation}
Therefore, by studying the time-dependence of the dephasing it is possible to infer properties about the noise correlation function, such as the amplitude of the noise as well as the correlation time.

In the case of the spin-qubit this allows us to interrogate the properties of the magnetic field noise (and therefore spectrum) at the qubit location, which can be controlled and manipulated with nanometer spatial resolution.
For instance, the spectrum of magnetic field noise (projected onto the $z$ axis) at a distance $z$ above a two-dimensional sample can be related to the near-field $s$-polarized reflection coefficient $r_s(q)$~\footnote{In the non-retarded limit of $\omega \ll c/z$. In general, this will also involve $p$-polarized reflection coefficients which become appreciable once retardation and electrodynamic effects become relevant.} of the sample via (see Ref.~\cite{Agarwal.2017}) 
\begin{multline}
    \mathcal{N}_{zz}(\omega,z) \\
    = \frac{\mu_0}{2}\coth\left(\frac{\beta \omega}{2}\right) \textrm{Im} \int_0^{\infty} \frac{dq}{2\pi} q^2  r_s(q) e^{-2zq} .
\end{multline}
We can then therefore learn about the dispersion of collective excitations in the material by studying how this noise depends on the distance $z$ and Ramsey time $\tau_R$, which in turn control the momenta and frequencies sampled via $q\sim 1/z$ and $\omega \lesssim 1/\tau_R$.

\subsection{Difficulty in Extracting Higher Moments}
It is important here to remark that ultimately, the insights gained through this protocol are based on the Gaussian approximation that 
\begin{equation}
\label{eqn:gaussian-approx}
    \log \llangle e^{-iX}\rrangle \approx -\frac12 \llangle X^2 \rrangle_c.
\end{equation}
It is the main goal of this work to show how to systematically go beyond this approximation, so that in addition to the second moments of the noise distribution, further higher-order moments can also be characterized.
This would potentially be of great interest, as these higher moments may uncover a wealth of information not only about the collective excitations of a material, but also their interactions, potentially their statistics, and finite-size and mesoscopic fluctuations.  
In the spirit of the fluctuation-dissipation relation, this is also equivalent to probing the {\bf nonlinear} response functions of the system by virtue of their imprint on the non-Gaussian fluctuations. 

In principle, it is clear why it is possible to obtain these moments; this relies on the simple observation that the decoherence function $\mathcal{C}_X$ is in fact nothing by the cumulant function of the phase, which in general admits an expansion beyond Gaussian order in terms of higher-order cumulants (up to a phase) as 
\begin{equation}
    \mathcal{C}_X = \sum_{n=0}^{\infty} C_n \frac{1}{n!}.
\end{equation}
For a Gaussian process with zero mean, this expansion truncates at second order, justifying the approximation in Eq.~\eqref{eqn:gaussian-approx}.
However, for a general non-Gaussian process this expansion will not terminate, and higher-order cumulants are known to characterize deviations of the distribution from Gaussianity (such as the so-called kurtosis which is related to the fourth moment of the distribution).

The main difficult thus far has been in isolating these higher order cumulants unambiguously. 
This is ultimately due to the fact that, for a particular Ramsey time $\tau_R$, one can only measure the decoherence function $\mathcal{C}_X$ at one argument.
It is in principle always possible to use this to reconstruct the lowest second-order moment (which may still be non-Markovian), but without more knowledge about the distribution of the phase it is not possible to reconstruct the higher moments unambiguously.  
Put differently, in order to extract the higher-moments one must in fact know the cumulant {\bf generating function}, defined in terms of a parameter $\lambda$
\begin{equation}
    \mathcal{C}_X(\lambda) = \log \llangle \exp\left(-i\lambda X\right)\rrangle = \sum_{n=0}^{\infty} {C}_n \frac{\lambda^{n}}{n!}.
\end{equation}
The $n^{\rm th}$ cumulant is then obtained by differentiating with respect to $\lambda$ (not $\tau_R$) $n$ times; physically however, we only ever have access to $\lambda = 1$ and therefore it is not possible to obtain the necessary derivatives to extract more than one cumulant (which is typically taken to be the lowest non-trivial cumulant arising from Gaussian noise).
In the next section, we will show how by using a set of {\bf multiple different} pulse sequences, it is possible to circumvent this issue and systematically probe higher-cumulants of the magnetic noise. 

\section{Single-Qubit Non-Gaussian Echo}
\label{sec:higher-cum}

\subsection{Simple Model}
In order to solve this problem, it is first instructive to consider a simpler example.
Suppose we have random variables $X_1$ and $X_2$, and we wish to characterize their higher moments. 
For instance, these may be thought of as corresponding to the phase accumulated by the qubit during two successive time intervals $[-\tau_1,0]$ and $[0,\tau_2]$.
Then, the cumulant functions may be used to characterize $X_1$ and $X_2$ via 
\begin{subequations}
    \begin{align}
    & \mathcal{C}_{X_1} = \log \llangle e^{-iX_1}\rrangle \\ 
    & \mathcal{C}_{X_2} = \log \llangle e^{-iX_2}\rrangle .
    \end{align}
\end{subequations}
If the variables are drawn from the same distribution (as we have assumed here), then these two cumulants will be the same. 
Motivated by the observation that if $X_1$ and $X_2$ are both Gaussian variables, then $X_1\pm X_2$ are also Gaussian, we then can also consider the cumulant functions for these variables, which are 
\begin{subequations}
    \begin{align}
    & \mathcal{C}_{X_2+X_1} = \log \llangle e^{-i(X_2+X_2)}\rrangle \\ 
    & \mathcal{C}_{X_2-X_1} = \log \llangle e^{-i(X_2-X_2)}\rrangle .
    \end{align}
\end{subequations}
Now, let us consider first consider the case that the variables {\bf are} Gaussian distributed.
In this case the cumulant functions are known in closed form as they truncate at quadratic order, and read
\begin{subequations}
    \begin{align}
    & \mathcal{C}_{X_1} \equiv \mathcal{C}^{(2)}_{X_1} = -\frac12 \llangle X_1^2\rrangle_c \\ 
    & \mathcal{C}_{X_2}\equiv \mathcal{C}^{(2)}_{X_2} = -\frac12 \llangle X_2^2\rrangle_c \\
    & \mathcal{C}_{X_2+X_1} \equiv \mathcal{C}^{(2)}_{X_2+X_1}= -\frac12 \llangle (X_2+X_1)^2\rrangle_c \\ 
    & \mathcal{C}_{X_2-X_1}\equiv \mathcal{C}^{(2)}_{X_2-X_1} = -\frac12 \llangle (X_2-X_1)^2\rrangle_c .
    \end{align}
\end{subequations}
Here we have indicated the truncation of $\mathcal{C}_X$ to the second order cumulant by $\mathcal{C}_X^{(2)}$ which we again point out is exact in the case where the variable $X$ is Gaussian distributed.  

In this case, we note that these quantities are not independent, but instead satisfy the relation
\begin{equation}
    0 = \mathcal{C}_{X_2+X_1} + \mathcal{C}_{X_2-X_1} - 2\mathcal{C}_{X_1} - 2\mathcal{C}_{X_2}.
\end{equation}
This is simply seen by expanding out the cross-terms in the covariances. 
We then see this immediately gives us a diagnostic for determining if $X_1$ and $X_2$ are {\bf not Gaussian} since we can simply check whether the quantity 
\begin{multline}
    \Gamma_{X_1,X_2} \equiv \mathcal{C}_{X_2+X_1}+ \mathcal{C}_{X_2-X_1} - 2\mathcal{C}_{X_2} - 2\mathcal{C}_{X_1} \\
    = \log \left[ \llangle e^{-i(X_2+X_1)}\rrangle \llangle e^{-i(X_2-X_1)} \rrangle / \left( \llangle e^{-iX_1}\rrangle \llangle e^{-iX_2}\rrangle \right)^2 \right] 
\end{multline}
deviates from zero. 
For instance, in the case where the distribution of the variables $X_1$ and $X_2$ can be characterized by both quadratic and quartic moments~\footnote{We will reserve the notation of $\Gamma^{(4)}$ for the case where we have expanded $\Gamma_{X_1,X_2}$ up to fourth order.} (i.e. the cumulant function truncates at fourth order without skewness), we can directly compute 
\begin{equation}
    \Gamma^{(4)} = \frac12 \llangle X_1^2 X_2^2 \rrangle_{c}.
\end{equation}
This therefore gives us direct access to at least some of the higher order correlation functions.
In fact, as will be important later, the echo spectroscopies described here naturally give access to the real-part of the coherence, $\llangle \cos X \rrangle$.
As a result, we find that even if there is a nonvanishing third cumulant, it is not extracted using this protocol, and the leading contribution to $\Gamma_{X_1,X_2}$ will still be due to the fourth cumulant. 

We also remark here that in fact, this diagnostic is much stronger than simply probing the non-Gaussianity.
Indeed, while it is evident that if $X_1$ and $X_2$ are Gaussian distributed, then $\Gamma = 0$, the converse is not true.
This is easily seen by noting that if $X$ and $Y$ are arbitrarily distributed but {\bf statistically independent} then 
\begin{equation}
    \llangle e^{-i(X_2\pm X_1)} \rrangle = \llangle e^{-iX_2}\rrangle \llangle e^{\mp iX_1}\rrangle
\end{equation}
and as a result, the cumulant functions will remain additive even if they are not quadratic. 
As such $\Gamma_{X_1,X_2}$ will also be identically zero in this case.
Therefore this diagnostic is in fact a very powerful probe for the distributions as it is non-zero {\bf only if} $X_1$ and $X_2$ are {\bf correlated and non-Gaussian} variables.
{Finally, we will mention that this protocol can in principle be extended to extract higher order cumulants as well.
As an example which is illustrative of the more general framework, see Appendix~\ref{app:6order}, which shows how to extract sixth-order cumulants. }

\subsection{Application to Qubit Dephasing}
By making use of our original analogy, we should then be able to probe the non-Gaussian correlations in the magnetic noise provided we can measure the four different cumulants corresponding to $X_1,X_2,X_2+X_1,X_2-X_1$.
We will here consider the case where the two periods of the echo sequence may have unequal durations, of $\tau_1$ and $\tau_2$ respectively, with the phases acquired over these echo times indicated by $X_1$ and $X_2$. 

\begin{figure}
    \centering
    \includegraphics[width=\linewidth]{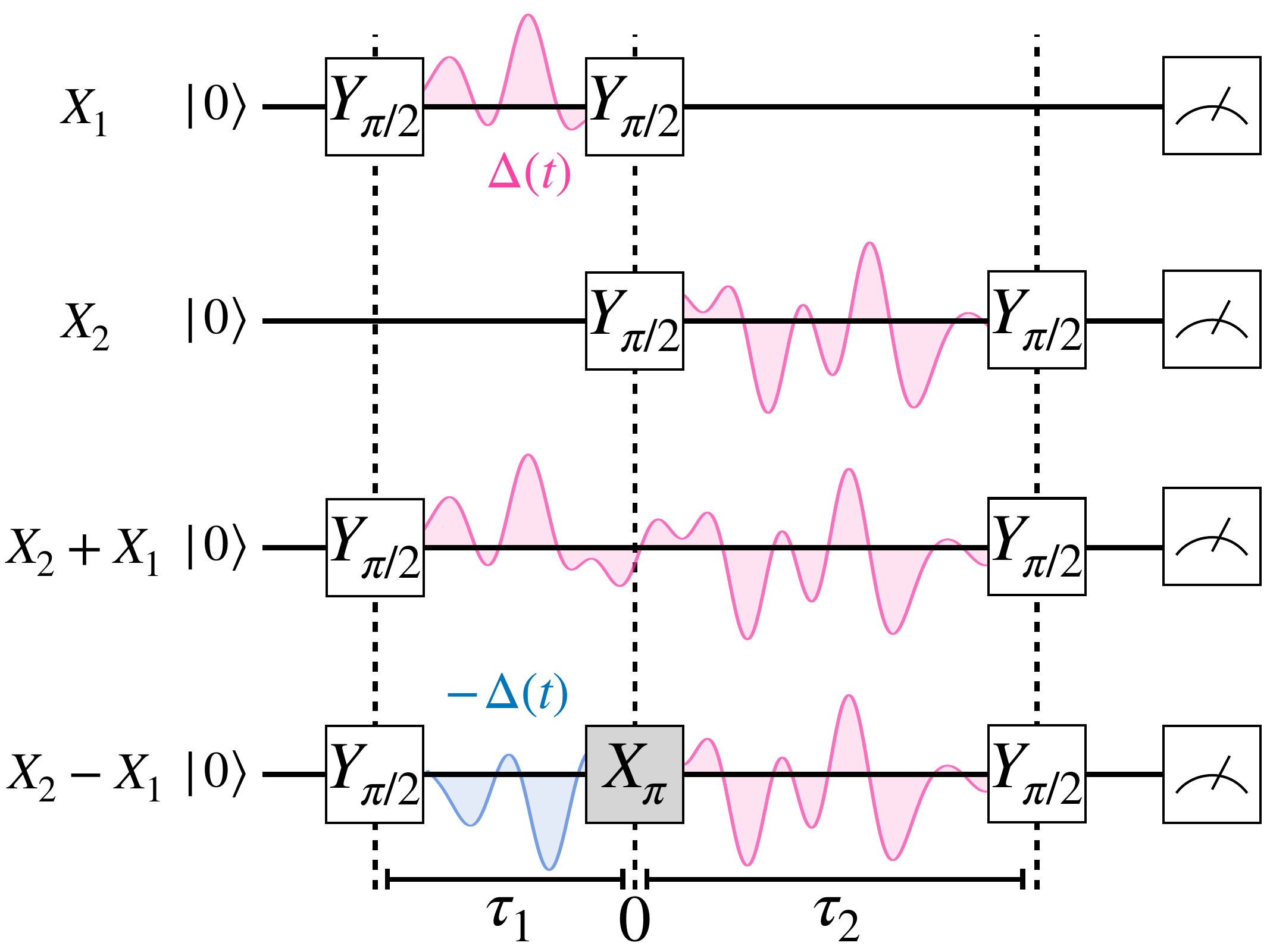}
    \caption{Pulses used to isolate non-Gaussian correlations of signal.
    Accumulated phases $X_1,X_2$ and $X_2+X_1$ are measured by three Ramsey sequences of duration $\tau_1,\tau_2$ and $\tau_1+\tau_2$.
    The difference phase $X_2-X_1$ is acquired using a Hahn echo sequence with echo at time $t = 0$, which essentially reverses the phase acquired before the $\pi$-pulse, obtaining $X_2-X_1$. 
    }
    \label{fig:NGpulses}
\end{figure}

If we consider 
\begin{subequations}
    \begin{align}
        & X_1 = \int_{-\tau_1}^{0} \Delta(t) dt \\ 
        & X_2 = \int_{0}^{\tau_2} \Delta(t) dt ,    
    \end{align}
\end{subequations}
we then see that $\mathcal{C}_{X_1}$ and $\mathcal{C}_{X_2}$ simply correspond to regular Ramsey sequences of duration $\tau_1,\tau_2$ respectively. 
Likewise, the additive cumulant $\mathcal{C}_{X_2+X_1}$ corresponds simply to the Ramsey sequence of duration $\tau_1 + \tau_2$ by the additivity of integrals. 
The only new ingredient needed is therefore the cumulant $\mathcal{C}_{X_2-X_1}$. 
In order to obtain this, we need obtain the phase evolution with an effective time-reversal applied at $t=0$. 
In fact, this is also not difficult to obtain, and corresponds to a simple Hahn spin-echo.
These four combinations are illustrated in Fig.~\ref{fig:NGpulses}.

The Hahn echo sequence we will employ is very similar to the Ramsey sequence.
After preparing the initial state and evolving for time $\tau_1$ we have again the density matrix 
\begin{equation}
     \hat{\rho}(0^-)  = \frac12 \left[ \mathds{\hat{1}} + e^{-iX_1} |1\rangle \langle 0| + e^{iX_1} |0\rangle\langle 1| \right] .    
\end{equation}
Now we perform a $\pi$ rotation around $x$, which implements the Hahn echo protocol.
This acts by $\hat{X}_\pi = i\hat{\sigma}_x$ on the left-and right of the density matrix and therefore swaps the coherences so that immediately afterwards we have
\begin{equation}
     \hat{\rho}(0^+)  = \hat{\sigma}_x \hat{\rho}(0^-) \hat{\sigma}_x = 
     \frac12 \left[ \mathds{\hat{1}} + e^{-iX_1} |0\rangle \langle 1| + e^{iX_1} |1\rangle\langle 0| \right].    
\end{equation}
Now, we proceed to further evolve for a time $\tau_2$ before applying a second $\pi/2$ rotation to arrive at the final density-matrix 
\begin{equation}
     \hat{\rho}(\tau_2^-)  =  \frac12 \left[ \mathds{\hat{1}} + e^{-i(X_2-X_1)} |-\rangle \langle +| + e^{i(X_2-X_1)} |+\rangle\langle -| \right].    
\end{equation}
We see that the phases before and after the echo acquire opposite signs.
If we now perform the final step of the Hahn echo protocol and measure the population of the density matrix (which again must be bath averaged), we will obtain the required object of  
\begin{equation}
    \langle \hat{\sigma}_z\rangle = \llangle \cos (X_2 - X_1) \rrangle \Leftrightarrow \mathcal{C}_{X_2-X_1} = \log \llangle e^{-i(X_2-X_1)}\rrangle.
\end{equation}

From the Hahn echo and the three different Ramsey echoes we can obtain the four cumulant functions 
\begin{widetext}
\begin{subequations}
    \begin{align}
        & \mathcal{C}_{X_1} \equiv \log \llangle \exp\left(-i\int_{-\tau_1}^{0} dt \Delta(t)\right) \rrangle \\
        & \mathcal{C}_{X_2} \equiv \log \llangle \exp\left(-i\int_{0}^{\tau_2} dt \Delta(t)\right) \rrangle \\
        & \mathcal{C}_{X_2+X_1} \equiv \log \llangle \exp\left(-i\int_{-\tau_1}^{\tau_2} dt \Delta(t)\right) \rrangle \\
        & \mathcal{C}_{X_2-X_1} \equiv \log \llangle \exp\left(-i\left[\int_{0}^{\tau_2} dt \Delta(t)-\int_{-\tau_1}^{0} dt \Delta(t) \right]\right) \rrangle .
    \end{align}
\end{subequations}
\end{widetext}
We can then use this to compute the diagnostic $\Gamma_{X_2,X_1}$.
If we again assume the distributions have only second and fourth moments we can relate this to the fourth moment of the magnetic field noise as  
\begin{multline}
    \Gamma^{(4)} \equiv \frac12 \int_{-\tau_1}^{0} dt_1\int_{-\tau_1}^{0} dt_2 \int_{0}^{\tau_2} dt_3\int_{0}^{\tau_2} dt_4 \\
    \llangle \Delta(t_1)\Delta(t_2)\Delta(t_3)\Delta(t_4) \rrangle_c. 
\end{multline}
This then allows us direct access to the fourth order moments of the magnetic field noise, and in turn the higher moments of the magnetization or currents that generate this correlated noise.  

It is worth commenting that here we have shown how, by using different Ramsey and Hahn echo sequences, it is possible to isolate a particular non-Gaussian noise cumulant. 
However, in a realistic setting, the efficacy of this procedure may be diluted by typical static inhomogeneous broadening mechanisms which are expected to also affect such NV center spin qubits. 
These mechanisms can be viewed as sources of magnetic noise that is extremely low frequency and therefore essentially static but varies shot-to-shot.
These lead to a broadening of the linewidth however can be rephased using the echo sequence, since over the duration of the echo experiment the noise is essentially static. 
As a result, this can lead to a highly non-Gaussian (if the distribution of the level broadening is also non-Gaussian) noise cumulant which is however fairly uninteresting. 
This has been remedied in the context of Gaussian (linear) noise spectroscopies by replacing Ramsey sequences with Hahn echo, or higher dynamical decoupling sequences~\cite{Mukamel.1995,Hamm.2011}, such as the Carr-Purcell-Meiboom-Gill (CPMG) sequence, which filters out low-frequency noise.

\begin{figure}
    \centering
    \includegraphics[width=\linewidth]{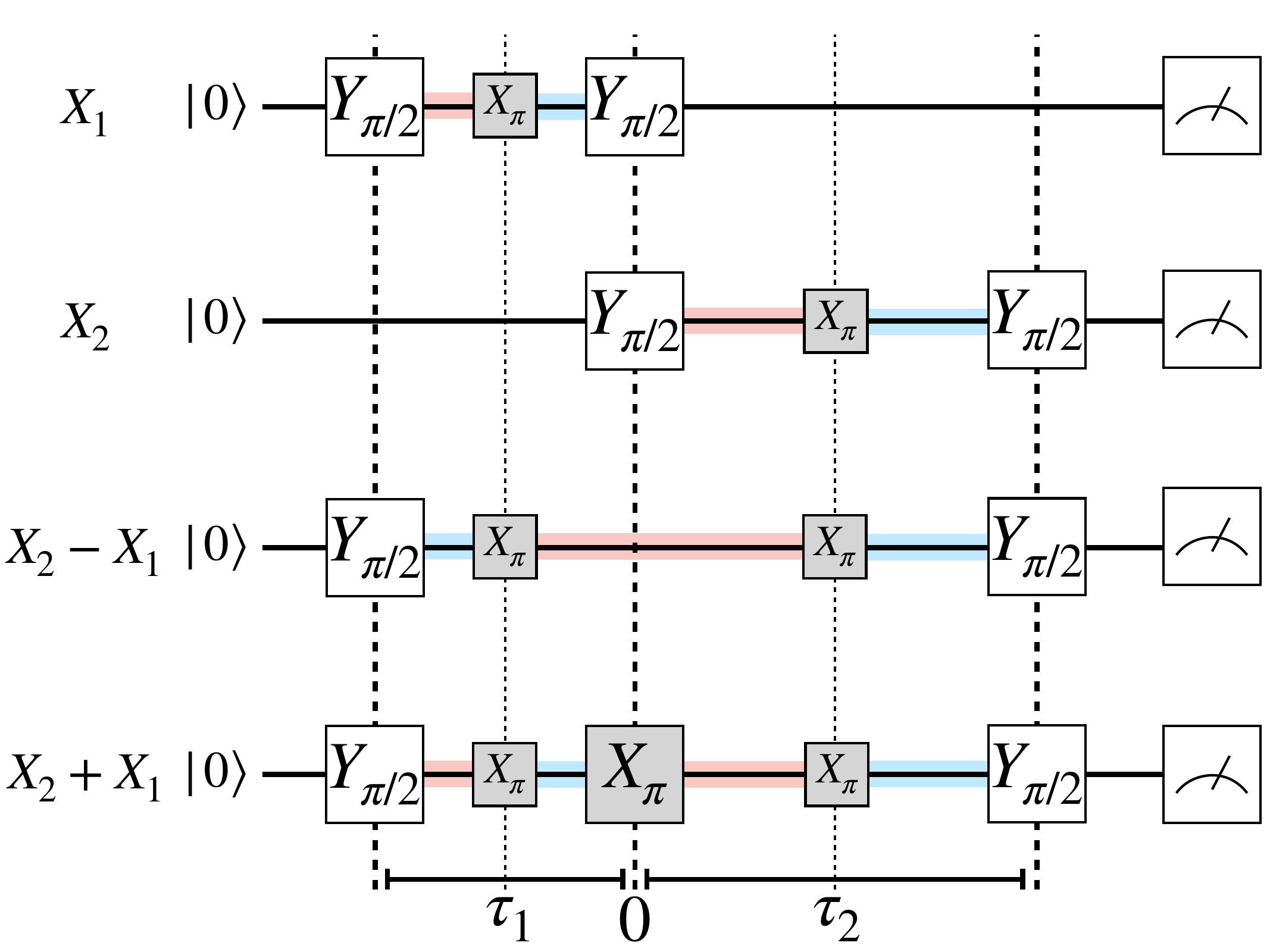}
    \caption{Pulses used to isolate non-Gaussian correlations and filter out static noise, with additional $X_\pi$ gates (shown smaller) used to cancel out static (non-Gaussian) noise.
    Evolution durations are colored according to the sign of the phase evolution on that interval (with red for $-1$ and blue for $+1$ evolution polarity).
    Every sequence is seen to be overall compensated.
    }
    \label{fig:NGpulses-Hahn}
\end{figure}

Though we will not examine this in depth here, a similar approach should be possible to isolate the non-Gaussian noise cumulants by utilizing appropriate higher dynamical decoupling echoes, as shown in Ref.~\cite{Norris.2016}.
In particular, by replacing the simple Ramsey sequences with Hahn echoes we can make the identifications
\begin{subequations}
    \begin{align}
        & X_1 \equiv -\int_{-\tau_1}^{-\tau_1/2} dt \Delta(t)  +\int_{-\tau_1/2}^{0}\Delta(t) dt  \\
        & X_2 \equiv -\int_{0}^{\tau_2/2} dt \Delta(t)  + \int_{\tau_2/2}^{\tau_2}\Delta(t) dt  .
    \end{align}
\end{subequations}
This is equivalent to the pulse sequences shown in Fig.~\ref{fig:NGpulses-Hahn}.
The non-Gaussian noise can then be computed explicitly in the frequency domain up to fourth order.
This is expressed in terms of a non-Gaussian noise polyspectrum~\cite{Norris.2016} 
\begin{equation}
    \Gamma^{(4)}(\omega_1,\omega_2,\omega_3,\omega_4) = \llangle \Delta(\omega_1) \Delta(\omega_2) \Delta(\omega_3) \Delta(\omega_4) \rrangle_c,
\end{equation}
which is expected in the presence of time-translational symmetry to be proportional to $\delta(\sum_j\omega_j)$ and fully symmetric under interchange of frequencies, and a non-Gaussian filter function. 
For simplicity, we will only present this for the case where $\tau_1 = \tau_2 = \tau_R$. 
In the simple lowest-order {\bf Ramsey} protocol illustrated in Fig.~\ref{fig:NGpulses} this reads 
\begin{multline}
\label{eqn:filter-Ramsey}
    W_{\rm Ramsey}^{(4)}(\{\omega\};\tau_R) = \\
    \frac16 \sum_{j<k=1}^4 \cos(\omega_j \tau_R + \omega_k \tau_R) \prod_{j=1}^4 \frac{2}{\omega_j} \sin(\omega_j \tau_R/2) .
\end{multline}
This then gives the non-Gaussian contribution to the Ramsey echo of 
\begin{equation}
    \Gamma_{\rm Ramsey}^{(4)}(\tau_R) = \frac12 \int_{\omega_1,...\omega_4} \Lambda^{(4)}(\{\omega\})W^{(4)}_{\rm Ramsey}(\{\omega\};\tau_R) .
\end{equation}
This is sensitive to the noise at zero frequency and thus is often undesirable. 
By using the compensated Hahn-echo sequences illustrated in Fig.~\ref{fig:NGpulses-Hahn} we instead find the filter function of 
\begin{multline}
\label{eqn:filter-Hahn}
    W_{\rm Hahn}^{(4)}(\{\omega\};\tau_R) = \\
    \frac16 \sum_{j<k=1}^4 \cos(\omega_j \tau_R + \omega_k \tau_R) \prod_{j=1}^4 \frac{4}{\omega_j} \sin^2(\omega_j \tau_R/4),
\end{multline}
and the non-Gaussian noise in the Hahn echo of 
\begin{equation}
    \Gamma_{\rm Hahn}^{(4)}(\tau_R) = \frac12 \int_{\omega_1,...\omega_4} \Lambda^{(4)}(\{\omega\})W^{(4)}_{\rm Hahn}(\{\omega\};\tau_R) .
\end{equation}
Unlike the Ramsey protocol, this is then seen to be insensitive to static noise, and therefore is likely more useful in a realistic experiment.  

{{
Now we briefly outline routes towards experimental implementation of these protocols and various difficulties which may be encountered. 
Since this work is focused on magnetic noise spectroscopy, we will focus our attention to various spin-qubit and magnetometry platforms. 
One of the most promising such platforms are NV centers due to their high spatial resolution and dynamic range~\cite{Rovny.2024,Casola.2018}.

In principle, the single-qubit protocols we outline here are no more difficult to implement than standard spin-echo and dynamical decoupling sequences.
In practice however, due to the nonlinear postprocessing protocols, in order to accurately evaluate the non-Gaussian echo, it is important to minimize the uncertainty on the bare spin-echo contrast as much as possible.
For a particular pulse sequence if we assume the NV center contrast $N \sim e^{-C}$ has an additive uncertainty of $\sigma_N$, then using simple error propagation this will lead to an uncertainty of $\sigma_C = \frac{1}{N}\times \sigma_N$ on the cumulant $C \sim -\log N$. 
At long times $N$ typically vanishes exponentially, making even small errors perilous for extracting the value of the cumulants once they become of order one.
This points to the benefit of using an ensemble of NV centers, which can be used to reduce the uncertainty on the echo signal and boost sensitivity, at the expense of spatial resolution and addressability.
Over- or under-rotation errors or systematic pulse timing errors can also play a highly detrimental role, though this has always been the case and has lead to the development of protocols such as XY8 which attempt to mitigate these problems.
Fortunately, many of the results of this analysis carry through unchanged in the case of these more complicated dynamical decoupling sequences as well.
}}

Finally, we comment that while the treatment thus far has assumed a classical noise source, in Appendix~\ref{app:quantum} we show that this analysis essentially {\bf carries through unchanged} even if the noise is of quantum mechanical origin~\footnote{Though in practice it may become substantially more difficult to compute theoretically.}, at least provided the following {\bf criteria are satisfied} during the evolution: (i) the state of the qubit is completely decoupled before initialization (it should suffice for it to also be a completely mixed state), (ii) the transverse coupling to the bath is negligible at the working frequency scale, (iii) only Clifford-type $\pi$-rotations are used in the dynamical decoupling sequence.
It would be interesting in future works to consider relaxing this last assumption, as allowing for dynamical decoupling sequences which also feature, e.g. $\pi/2$ rotations, could lead to coupling between the ``quantum" and ``classical" spins configurations in the language of the Keldysh formalism.
This could then allow for probing the response functions of the bath as well as the noise correlation functions, similar to what was shown in Ref.~\cite{Wang.2021fej}.
We now move on to consider a complementary case; that of spatially correlated non-Gaussian noise, and how this may be extracted using multiple qubits.

\section{Multi-Qubit Sensing of Spatial Non-Gaussian Correlations}
\label{sec:spatial}
In the previous section, we showed how by using a combination of spin echo sequences it was possible to isolate the non-Markovian and non-Gaussian cumulants of the noise spectrum.
It is also interesting to consider whether a similar procedure can be implemented but access non-local spatial correlations instead of non-local temporal (i.e. non-Markovian) correlations; in short, we will show that this is possible provided one can perform coincidence measurements between the two qubits. 
This is difficult but has been experimentally implemented in the case of NV centers~\cite{Rovny.2022}, including a recent demonstration that entangled states can be prepared and used for noise spectroscopy~\cite{Rovny.2025} and sensing~\cite{Zhou.2025}.
It is also possible to remotely-entangle distinct qubits using two-photon interference effects~\cite{Bernien.2012}.

\begin{figure}
    \centering
    \includegraphics[width=\linewidth]{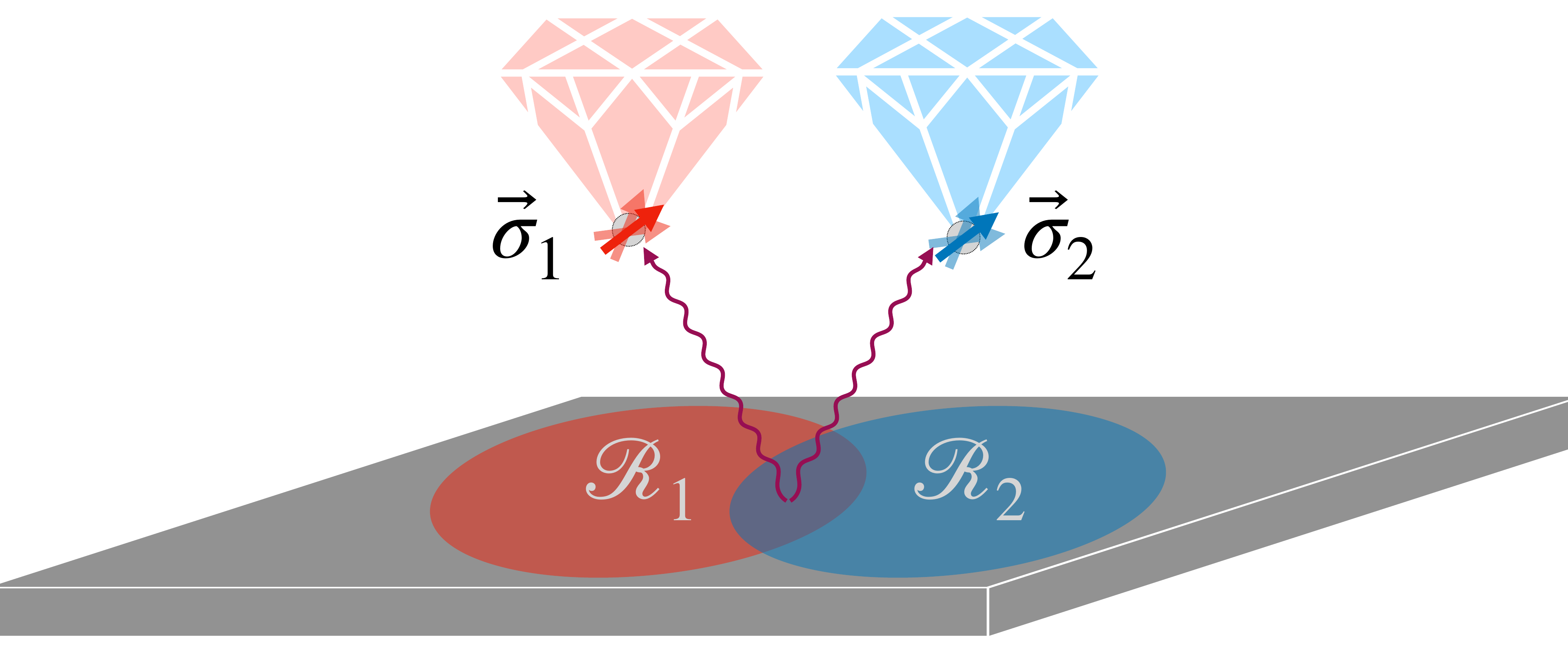}
    \caption{Schematic depiction of spectroscopy for spatially correlated noise. 
    Each qubit (color coded to red for 1 or blue for 2) is sensitive to the field fluctuations in a quasi-local area $\mathcal{R}_1,\mathcal{R}_2$ respectively.
    If these qubits are entangled they can be used to probe the mutually correlated magnetic field fluctuations of an underlying material (purple wavy lines emanating from the overlap).}
    \label{fig:2qubits-schematic}
\end{figure}

To do this, we consider a situation with two identical qubits placed at different locations in the sample.
This is shown schematically in Fig.~\ref{fig:2qubits-schematic}, which shows each qubit along with the region of fluctuations each one is sensitive to (in reality this is not a sharply defined area). 
Rather than utilizing a Hahn echo sequence, we will show that by performing Ramsey spectroscopy using appropriately prepared or measured qubits, one can access the non-Gaussian {\bf spatial} correlations.

We will present two pathways towards realizing this.
The first method requires only single-qubit control gates as well as the ability to perform two-qubit coincidence measurements and is likely easier to implement in NV center platforms, as it has already been demonstrated in Ref.~\cite{Rovny.2022}.
The second protocol instead can be performed using only local population measurements, but it requires the ability to implement entangling gates between the qubits, which may be difficult for NV centers but possible in, e.g. quantum dots~\cite{Boter.2020}\textemdash though recent experiments are very encouraging in this direction~\cite{Bernien.2012,Rovny.2025,Zhou.2025}.

Let us indicate the qubits by $1,2$ with system-bath coupling Hamiltonian 
\begin{equation}
\label{eqn:hamiltonian-2qu}
    \hat{H}(t) = \frac12 \Delta_1(t) \hat{\sigma}_1^z + \frac12 \Delta_2(t) \hat{\sigma}_2^z.
\end{equation}
The two splittings $\Delta_{1,2}(t)$ are in principle correlated and related to the magnetic field noise at the location of each qubit via 
\begin{equation}
    \Delta_j(t) = g_j \mu_B \mathbf{n}_j\cdot\mathbf{B}_j(\mathbf{r}_j,t),
\end{equation}
where $g_j,\mathbf{n}_j$ are the magnetic moment and quantization axis of the $j^{\rm th}$ spin, located at $\mathbf{r}_j$. 
For the moment we leave these general however.

\subsection{Unentangled Echo Spectroscopy}
\label{sub:unentangled}
We now show how to isolate spatially correlated non-Gaussian noise cumulants using only single-qubit control and coincidence measurements. 
First, let us consider applying the unitary which is a $\pi/2$ rotation separately on each qubit, with gate 
\begin{equation}
    \hat{U} = \hat{Y}_{\pi/2}\otimes\hat{Y}_{\pi/2}.
\end{equation}
This prepares the qubit in to the product state $|++\rangle$.

Let us again, for the moment assume a classical magnetic field bath, although as shown in Sec.~\ref{app:quantum} this should also hold for quantum noise sources.
Then we introduce the random variables 
\begin{equation}
    X_j = \int_0^{\tau_R} \Delta_j(t) dt ,
\end{equation}
which are the total phases acquired by qubit $j$ during the Ramsey sequence. 
After evolution for time $\tau_R$ the qubits will then be in the state 
\begin{multline}
    |\psi_{12}(\tau_R^-)\rangle = \\
    \frac12\left[ e^{-iX_1/2} |1\rangle + e^{iX_1/2}|0\rangle\right]\left[ e^{-iX_2/2} |1\rangle + e^{iX_2/2}|0\rangle\right].
\end{multline}
Applying $\hat{U}$ again, as in a Ramsey sequence, gives 
\begin{multline}
    |\psi_{12}(\tau_R^+)\rangle =  \\
    \frac12\left[ e^{-iX_1/2} |-\rangle + e^{iX_1/2}|+\rangle\right]\left[ e^{-iX_2/2} |-\rangle + e^{iX_2/2}|+\rangle\right].
\end{multline}

We now compute the density matrix of this system after bath averaging.
For simplicity, we will assume the odd cumulants vanish. 
We find the full result
\begin{multline}
    \llangle \hat{\rho}_{12}(\tau_R)\rrangle = \frac14 \mathds{\hat{1}}\otimes\mathds{\hat{1}} \\
     +\frac14 e^{\mathcal{C}_{X_1}} \hat{\sigma}_z\otimes\mathds{\hat{1}} + \frac14 e^{\mathcal{C}_{X_2}} \mathds{\hat{1}}\otimes \hat{\sigma}_z \\
     + \frac14 e^{\mathcal{C}_{X_2+X_1}}\left[ |++\rangle\langle -- | +|--\rangle\langle ++ | \right]\\
     +\frac14 e^{\mathcal{C}_{X_2-X_1}}\left[ |+-\rangle\langle -+ |  +|-+\rangle\langle +- | \right] .
\end{multline}

We see that in principle the relevant cumulants can all be extracted from this single density matrix.
It is easy to see that the individual cumulants $\mathcal{C}_{X_1},\mathcal{C}_{X_2}$ can be extracted by measuring the local populations via 
\begin{equation}
    \mathcal{C}_{X_j} = \log\left( \tr [\hat{\sigma}_z^j\llangle \hat{\rho}_{12}(\tau_R)\rrangle]\right),
\end{equation}
where $\sigma_z^j$ is the population operator on site $j$ tensored with the identity on the other site. 
The joint cumulants are however more difficult to extract as these are encoded in the non-local coherences. 
Therefore, any single-site measurement will not be sensitive to these terms.
We therefore consider measuring the joint-observables $\hat{\sigma}_a\otimes \hat{\sigma}_b$ where $a,b = x,y,z$.
We can furthermore see that the relevant coherences are between states of opposite parity, requiring overlaps of the form $\langle \pm | \hat{\sigma}_a |\mp \rangle$. 
Therefore, these will vanish if any of the $\hat{\sigma}_a$ are $\hat{\sigma}_x$. 
We can therefore restrict to measurements of $\hat{\sigma}_{z}\otimes \hat{\sigma}_{z},\hat{\sigma}_{z}\otimes \hat{\sigma}_{y},\hat{\sigma}_{y}\otimes \hat{\sigma}_{z},\hat{\sigma}_{y}\otimes \hat{\sigma}_{y}$.
Furthermore, for the particular combination above it is seen that the expectation values $\langle \hat{\sigma}_{y}\otimes \hat{\sigma}_{z}\rangle = \langle \hat{\sigma}_{z}\otimes \hat{\sigma}_{y}\rangle = 0$.
We therefore reduce the problem to the two coincidence measurements of $\hat{\sigma}_{z}\otimes \hat{\sigma}_{z}$ and $\hat{\sigma}_{y}\otimes \hat{\sigma}_{y}$. 

We then find  
\begin{subequations}
\begin{align}
& \tr [\hat{\sigma}_{z}\otimes \hat{\sigma}_{z}\llangle \hat{\rho}_{12}(\tau_R)\rrangle] = \frac12 e^{\mathcal{C}_{X_2+X_1}}+\frac12 e^{\mathcal{C}_{X_2-X_1}} \\ 
& \tr [\hat{\sigma}_{y}\otimes \hat{\sigma}_{y}\llangle \hat{\rho}_{12}(\tau_R)\rrangle] = -\frac12 e^{\mathcal{C}_{X_2+X_1}}+\frac12 e^{\mathcal{C}_{X_2-X_1}},
\end{align}
\end{subequations}
such that all of the needed cumulants can be obtained from the simple two-qubit Ramsey scheme by reconstructing the following measurements 
\begin{subequations}
\begin{align}
& \mathcal{C}_{X_1} =  \log \left( \tr [\hat{\sigma}_{z}\otimes \mathds{1} \llangle \hat{\rho}_{12}(\tau_R)\rrangle]  \right) \\
& \mathcal{C}_{X_2} =  \log \left( \tr [ \mathds{1}\otimes \hat{\sigma}_{z}\llangle \hat{\rho}_{12}(\tau_R)\rrangle]  \right) \\
& \mathcal{C}_{X_2+X_1} =  \log \left( \tr [ (\hat{\sigma}_z\otimes \hat{\sigma}_{z} -  \hat{\sigma}_y\otimes \hat{\sigma}_{y} ) \llangle \hat{\rho}_{12}(\tau_R)\rrangle]  \right) \\
& \mathcal{C}_{X_2-X_1} =  \log \left( \tr [ (\hat{\sigma}_z\otimes \hat{\sigma}_{z} +  \hat{\sigma}_y\otimes \hat{\sigma}_{y} )\llangle \hat{\rho}_{12}(\tau_R)\rrangle] \right) .
\end{align}
\end{subequations}
As before, we have assumed there are only even cumulants; in the case of odd cumulants, one must be careful in distinguishing between $\llangle \cos X\rrangle $ and $\llangle e^{-iX}\rrangle $.
From these we can once again compute the non-Gaussian cross-correlations as 
\begin{equation}
\label{eqn:NG-2qu}
    \Gamma_{12} = \mathcal{C}_{X_2 +X_1} + \mathcal{C}_{X_2 - X_1} - 2\mathcal{C}_{X_2} - 2\mathcal{C}_{X_1}.
\end{equation}

If we again assume fourth order cumulant model this gives the formula 
\begin{multline}
\Gamma_{12} = \frac12 \int_{0}^{\tau_R}dt_1 \int_0^{\tau_R} dt_2 \int_{0}^{\tau_R}dt_3 \int_{0}^{\tau_R}dt_4 \\
    \llangle \Delta_1(t_1) \Delta_1(t_2) \Delta_2(t_3) \Delta_2(t_4)\rrangle_{c}.
\end{multline}
This is very similar to our previous result but rather than sensitive to the temporally non-local correlations which exist across the time-intervals $[-\tau_1,0]$ and $[0,\tau_2]$, it is sensitive to spatially non-local correlations between the regions $\mathcal{R}_1$ and $\mathcal{R}_2$.
If we assume the qubits are spin-qubits at lateral locations $\mathbf{r}_1$ and $\mathbf{r}_2$ and heights $z_1,z_2$ from the sample, then this heuristically will probe the correlations across a distance $|\mathbf{r}_1 - \mathbf{r}_2|$ which are relevant at length scales $z_1,z_2$. 
For completeness, we will now show how these same measurements can be performed using entangled Ramsey spectroscopy.

\subsection{Entangled Echo Spectroscopy}
\label{sub:entangled}

\begin{figure}
    \centering
    \includegraphics[width=\linewidth]{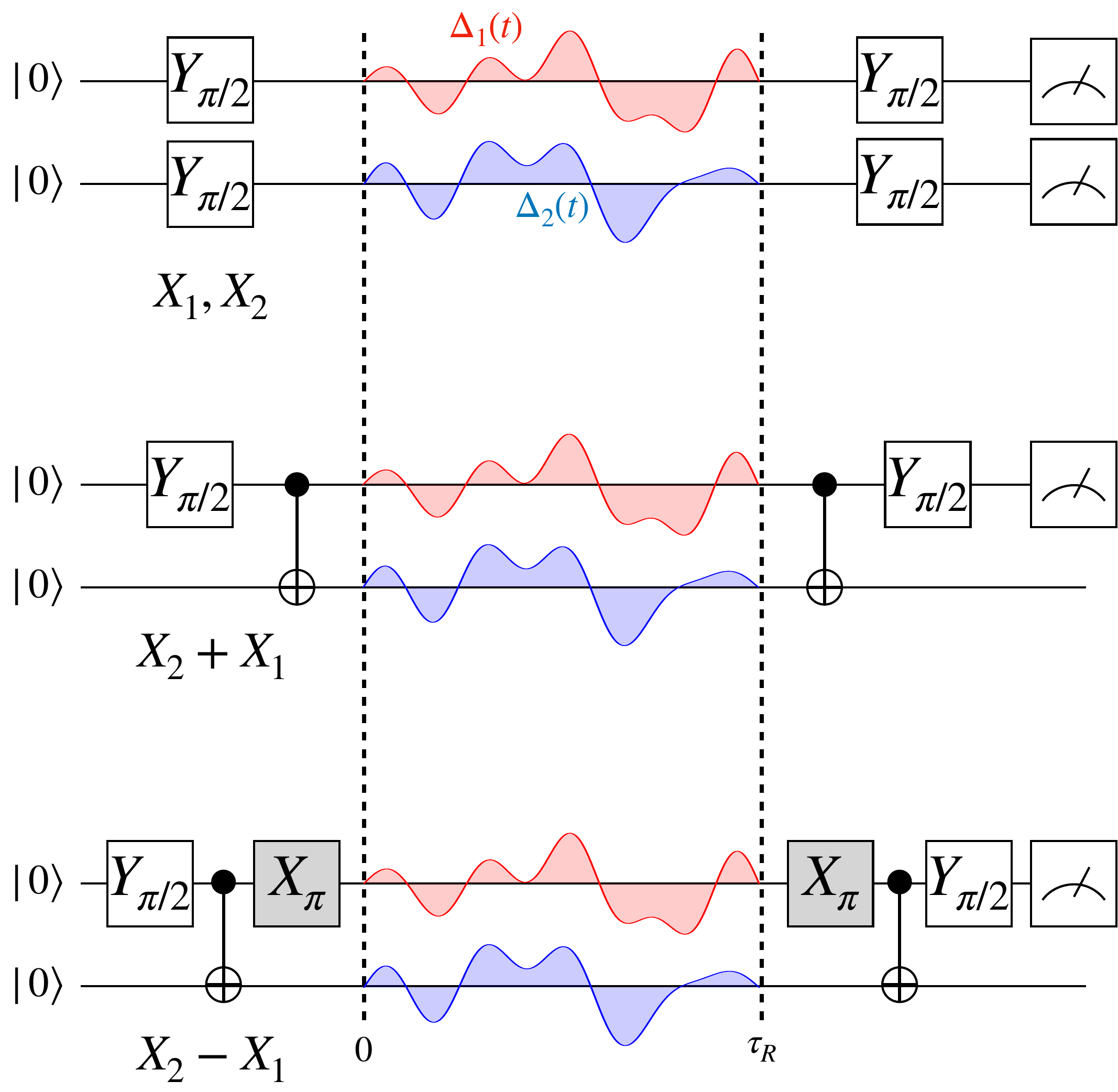}
    \caption{Protocol for measuring non-Gaussian noise using entangled qubits. 
    The first two measurements are independent local Ramsey measurements of duration $\tau_R$.
    To extract the cross cumulant $X_2+X_1$ the qubit pair is prepared in the Bell state $|\Phi^+\rangle$ by first applying a local $\pi/2$ rotation gate on one qubit followed by a controlled-NOT gate, which is then undone after the Ramsey echo, with the collective phase then encoded in the local population measurement of a single qubit.
    The cross cumulant $X_2 - X_1$ is obtained similarly, but with an additional $\pi$ rotation before and after the echo.
    }
    \label{fig:2qubits-gates}
\end{figure}

In certain cases, it may actually be easier to perform certain entangling gates rather than perform qubit coincidence measurements.
It was also recently shown in the context of NV centers that entanglement may provide a useful advantage in terms of sensitivity~\cite{Rovny.2025}.
In this case, we will now show how, using entangling gates, one can extract the same non-Gaussian correlations with only local measurements.
To accomplish this, we will use four sets of gates; these are 
\begin{subequations}
\label{eqn:unitaries}
    \begin{align}
        & \hat{U}_1 = \hat{Y}_{\pi/2} \otimes \mathds{\hat{1}} \\
        & \hat{U}_2 = \mathds{\hat{1}}\otimes \hat{Y}_{\pi/2}  \\
        & \hat{U}_\Phi =  \hat{\textrm{CNOT}}_1 \hat{U}_1 \\
        & \hat{U}_\Psi = (\hat{X}_{\pi} \otimes \mathds{\hat{1}})\hat{U}_\Phi.
    \end{align}
\end{subequations}
Here $\hat{\textrm{CNOT}}_1$ is the controlled-NOT gate acting on qubit 2 with control qubit 1.
These correspond to the echo circuits depicted in Fig.~\ref{fig:2qubits-gates}.

These will each be applied onto the fiducial state $|00\rangle$ in order to obtain four initial states
\begin{subequations}
    \begin{align}
        & |\psi_1(0)\rangle = \hat{U}_1 |00\rangle =  |+0\rangle \\
        & |\psi_2(0)\rangle = \hat{U}_2 |00\rangle = |0+\rangle \\
        & |\Phi(0)\rangle = \hat{U}_\Phi |00\rangle =  \frac{1}{\sqrt{2}}\left[ |00\rangle  + |11\rangle\right]\\
        & |\Psi(0)\rangle = \hat{U}_{\Psi}|00\rangle = \frac{i}{\sqrt{2}}\left[ |01\rangle + |10\rangle\right].
    \end{align}
\end{subequations}

Given these four states we then evolve under the Hamiltonian in Eq.~\eqref{eqn:hamiltonian-2qu} for a time $\tau_R$.
After the evolution, and immediately before the second gate application, the qubits are in the states 
\begin{subequations}
    \begin{align}
        & |\psi_1(\tau_R^-)\rangle = \frac{1}{\sqrt{2}}\left[ e^{-iX_1/2}|1\rangle + e^{+iX_1/2}|0\rangle \right] e^{iX_2/2}|0 \rangle \\
        & |\psi_2(\tau_R^-)\rangle = e^{iX_1/2}|0\rangle \frac{1}{\sqrt{2}}\left[e^{-iX_2/2}|1\rangle  + e^{iX_2/2} |0\rangle \right]\\
        & |\Phi(\tau_R^-)\rangle = \frac{1}{\sqrt{2}}\left[ e^{-i(X_1 +X_2)/2}|11\rangle  + e^{i(X_1 +X_2)/2}|00\rangle\right]\\
        & |\Psi(\tau_R^-)\rangle = \frac{i}{\sqrt{2}}\left[ e^{-i(X_1-X_2)/2}|10\rangle + e^{i(X_1 - X_2)/2}|01\rangle\right].
    \end{align}
\end{subequations}

In the case of the first two measurements, which are unentangled and therefore independent, we can simply average these over the bath as the measurement will commute with the averaging.
The result will factorize and we can simple measure each qubit independently. 
We then find the coherences of the density matrices are characterized by the local noise cumulants  
\begin{equation}
    \mathcal{C}_j = \log \llangle e^{iX_j}\rrangle .
\end{equation}

The two entangled states are now handled more carefully. 
After the Ramsey time $\tau_R$, we first undo the $\hat{X}_\pi$ gate (if it was applied), and then apply another CNOT gate in order to disentangle the qubits, followed by the second Ramsey $\pi/2$ rotation as shown in Fig.~\ref{fig:2qubits-gates}. 
In each of the two cases, we then have the wavefunctions 
\begin{subequations}
    \begin{align}
         & |\Phi(\tau_R^+)\rangle =\left[ \cos\left((X_1 +X_2)/2\right)|10 \rangle + i\sin\left((X_1 +X_2)/2\right)  |00\rangle \right] \\
        & |\Psi(\tau_R^+)\rangle = -\left[ \cos\left((X_1 -X_2)/2\right)|1 0\rangle - i\sin\left((X_1 -X_2)/2\right)  |00\rangle \right].
    \end{align}
\end{subequations}
Now, we can average the density matrix to obtain the result of a local measurement. 
By measuring the population of the control qubit we find the remaining cumulants  
\begin{subequations}
    \begin{align}
        &\langle \hat{\sigma}_z \rangle_\Phi = \llangle \cos (X_1 + X_2) \rrangle \Leftrightarrow \mathcal{C}_\Phi(\tau_R) = \log \llangle e^{i(X_1+X_2)}\rrangle\\
        &\langle \hat{\sigma}_z \rangle_\Psi = \llangle \cos (X_1 - X_2) \rrangle \Leftrightarrow \mathcal{C}_\Psi(\tau_R) = \log \llangle e^{i(X_1-X_2)}\rrangle.
    \end{align}
\end{subequations}
In the end, this yields essentially the same information as that obtained from the coincidence measurements.
By combining the various cumulants, we can measure the non-local correlations as 
\begin{equation}
    \Gamma_{1,2} = \mathcal{C}_{\Psi} + \mathcal{C}_{\Phi} - 2\mathcal{C}_1 - 2\mathcal{C}_2.
\end{equation}

While this protocol requires only local measurements in the computational basis, it requires the ability to perform appropriate entangling gates at least twice.
{Though for NV center spin-qubits, this is difficult, it is certainly within the realm of feasibility. 
A recent experiment~\cite{Rovny.2025} was able to entangle two nearby ($\sim 10$nm) NV centers using magnetic dipole-dipole interactions between the two qubits. 
Given that the protocol we outline here is sensitive only to noise which is correlated between the two qubits, this experiment~\cite{Rovny.2025} is a crucial benchmark, but in order to probe the emergence of interesting spatial correlations, being able to entangle qubits which are further away from each other will be important\textemdash ideally with separations ranging up to 100's of nanometers. 
While this is a significant challenge, it may be possible to implement this by using dynamical-decoupling sequences carefully designed to generate entanglement in dipolar-interacting NV center ensembles~\cite{Zheng.2022}, or possibly by using magnon mediated interactions~\cite{Fukami.2021}.
For larger separations it becomes possible to optically resolve the NV centers, which may enable the protocol discussed in Sec.~\ref{sub:unentangled}, or potentially other photonics-based entangling protocols~\cite{Bernien.2012,Humphreys.2018,Pompili.2021}.
Furthermore, looking more broadly to other platforms such as other color centers~\cite{Knaut.2024}, quantum dots~\cite{Delteil.2016,Boter.2020}, or Josephson junctions, multiqubit sensing protocols which use entangling operations should already be feasible.
}
We will now proceed on to consider how these techniques may be used to study novel physics by considering a simple example of independent two-level systems comprising a bath. 

In the future it would be interesting to extend this protocol to include more qubits, or allow for combinations of coincidence measurements and echo sequences. 
In particular, it is probably useful to replace the Ramsey sequences used in these two-qubit protocols with Hahn echo sequences in order to also filter out low-frequency noise.
Additionally, using ideas from classical shadow tomography~\cite{Huang.2020h0e}, it may be possible to implement these protocols more efficiently by utilizing randomized single-qubit rotations or echos.
Finally, whether these techniques can be extended to more qubits or qubit ensembles, and whether entanglement can be used to increase sensitivity~\cite{Wang.2024}, are interesting directions to move towards.
Having now established how these protocols may be implemented in NV spin-qubits, we will now explore how these tools can be used to study nontrivial noise correlations in magnetic materials. 

\section{Applications}
\label{sec:app}

To see how these techniques are useful, we will consider two specific cases of magnetic noise which exhibits interesting correlations. 
The first is a model of independent classical telegraph noise processes which add together to generate correlated noise\textemdash a standard model of dynamic non-Gaussian noise.
We will then move on to consider correlated magnetic noise which might arise near a second-order phase transition.
In this case, we will show how this technique can be used to sense the strong non-Gaussian noise correlations which are present at length scales within the correlation length.

\subsection{Two-Level System Noise}
\label{sub:TLS}

We will now be interested in studying the noise that comes from a simple independent fluctuator model, where each fluctuator has only two possible states.
In this case, the origin of the non-Gaussian correlations is the intrinsic nonlinearity of the two-level systems which generate the noise.
Such a telegraph process was recently studied experimentally in an ensemble of NV centers~\cite{Davis.2023}, and has long been an issue which affects superconducting qubits~\cite{Galperin.2006,Paladino.2002,Cai.2020,Cywinski.2008,Bergli.2007,Klimov.2018}.
While simple, such models are quite natural for describing magnetic noise in diamond NV systems in particular, where such noise processes can arise from ``dark" $P_1$ spins, which are optically inactive spin defects which also reside in the diamond lattice, or other impurity spin baths~\cite{Lange.2012,Meinel.2022,Xia.2025,Zhang.2025}.

\begin{figure}
    \centering
    \includegraphics[width=\linewidth]{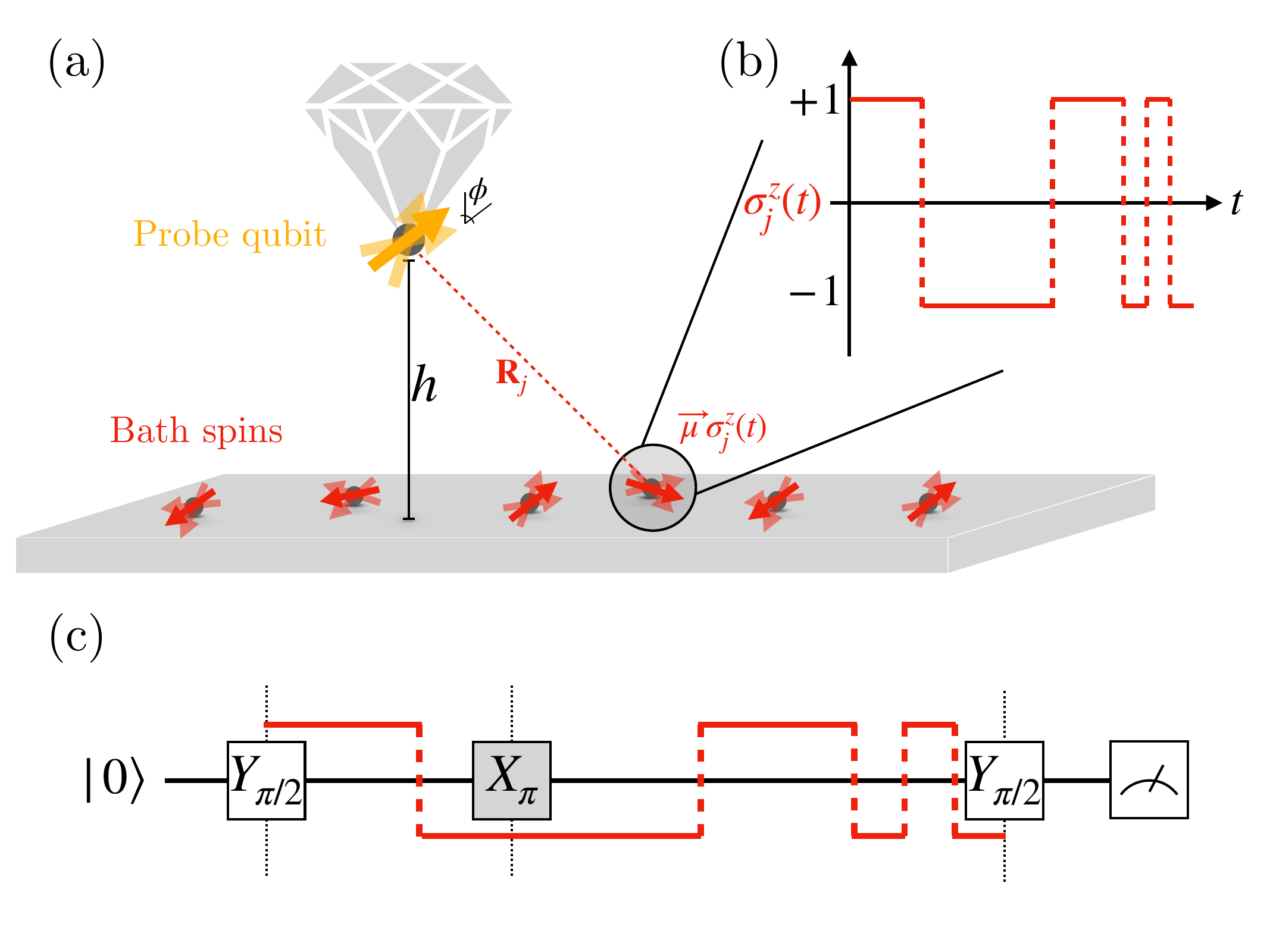}
    \caption{(a) Schematic depiction of system. 
    The large (yellow) spin is the spin qubit we use to probe the bath. 
    It has quantization axis at an angle $\phi$ from vertical, and is located a distance $h$ above the plane.
    It is coupled to a large ensemble of small bath spins (red) which lie in the plane with uniform average density $n_{\rm 2D}$ and random orientation of magnetic moment $\bm{\mu}_j$.
    The magnetic moment is modeled to fluctuate along this axis via a classical telegraph noise processes.
    (b) Telegraph noise example processes, highlighted for a particular bath spin located at $\mathbf{R}_j$ with moment $\bm{\mu}_j \sigma_j(t)$. 
    Each bath spin independently flips between two angular momentum states with $\sigma_j(t) = \pm 1$, with a characteristic switching rate $\gamma$. 
    This generates a non-Markovian, and non-Gaussian dynamical noise processes with memory time of order $1/\gamma$.
    (c) Connected correlations the probe qubit detects during the echo sequence.
    The fourth cumulant involves connected correlations at four times which are such that two time-points lie in each echo interval; a typical sequence of times is shown with the blue arrows. 
    The cumulant will only be non-vanishing if the noise process remains strongly correlated across the echo, such that $\tau_1\gamma,\tau_2\gamma \ll 1$. 
    In this case, only a few switches are likely to happen before the echo reverses the dynamics. }
    \label{fig:TLS-noise}
\end{figure}

In our case we will consider an example where the bath spins are randomly distributed across the surface of a two-dimensional sample with average density $n_{\rm 2D}$\textemdash shown schematically in Fig.~\ref{fig:TLS-noise}(a). 
Each spin is modeled as undergoing a classical independent random telegraph noise process with switching rate $\gamma$ (which sets the memory time) between two magnetic states $\pm 1$. 
This dynamics is schematically depicted in Fig.~\ref{fig:TLS-noise}(b). 

Each of these is taken to be located at position $\mathbf{R}_j$ relative to the qubit, with dipole moment $\bm{\mu}_j \sigma_j(t)$, where $\bm{\mu}_j$ characterizes the size and orientation of the magnetic moment, and $\sigma_j(t) = \pm 1$ is a classical Ising-like variable which encodes the magnetic state of the impurity as a function of time.

The magnetic field the NV qubit experiences is then given by
\begin{equation}
    \hat{B}_z(t) = \sum_{j=1}^N V_j {\sigma}_j^z(t)
\end{equation}
where 
\begin{equation}
    V_j = \frac{\mu_0}{4\pi}\frac{3 \mathbf{R}_j\cdot\mathbf{e}_z \mathbf{R}_j \cdot\bm{\mu}_j - \mathbf{R}_j^2 \bm{\mu}_j\cdot\mathbf{e}_z}{R_j^5}
\end{equation}
is the field projection along the qubit quantization axis (taken to be $\mathbf{e}_z$) due to the $j^{\rm th}$ fluctuator.

We then know that our non-Gaussian diagnostic will measure the connected correlation function 
\begin{multline}
    \Gamma^{(4)} = \frac12 \llangle \left(\int_{-\tau_1}^0 dt \sum_j V_j {\sigma}_j(t) \right)^2 \left(\int_0^{\tau_2} dt \sum_j V_j {\sigma}_j(t) \right)^2 \rrangle_c \\
    = \frac12\sum_{jklm}V_jV_kV_lV_m\int_{-\tau_1}^0 dt_1 \int_{-\tau_1}^0 dt_2 \int_{0}^{\tau_2} dt_3\int_{0}^{\tau_2} dt_4  \\
    \llangle {\sigma}_j(t_1) {\sigma}_k(t_2){\sigma}_l(t_3){\sigma}_m(t_4)\rrangle_c. 
\end{multline}
This involves the connected fourth-order correlation function of the classical spin states $\sigma_j(t)$.

We note that unless all four indices for the fluctuator are the same, the connected part of this correlation function will vanish. 
Therefore we can reduce this to 
\begin{multline}
    \Gamma = \frac12 \sum_{j}V_j^4 \int_{-\tau_1}^0 dt_1 \int_{-\tau_1}^0dt_2 \int_{0}^{\tau_2} dt_3 \int_{0}^{\tau_2}dt_4  \\
    \llangle {\sigma}_j(t_1) {\sigma}_j(t_2) {\sigma}_j(t_3){\sigma}_j(t_4)\rrangle_c. 
\end{multline}
Therefore, this is sensitive only to processes which couple $t_1,t_2$ to $t_3,t_4$ which reside on opposite sides of the Hahn echo sequence, shown in Fig.~\ref{fig:TLS-noise}(c).

To evaluate this we need the disconnected correlations of second and fourth order. 
We obtain for a single telegraph noise channel
\begin{multline}
    g^{(2)}(t,t') = \llangle \sigma^z(t) \sigma^z(t') \rrangle \\
    = \theta(t-t')\sum_{s,s'} ss' P(s,t | s',t') P(s',t')  + t\leftrightarrow t'.
\end{multline}
The conditional probability is obtained from a master equation to be 
\begin{equation}
P(s,t|s',t') = \frac12 +\frac12 ss' e^{-2\gamma (t-t')}.
\end{equation}
The absolute probability is simply $P(s,t) = \frac12$.
This then corresponds to the infinite temperature ensemble for the bath spins. 
We then find, taking care to sum over both $t>t'$ and $t<t'$ contributions 
\begin{equation}
g^{(2)}(t,t') = e^{-2\gamma |t-t'|}.
\end{equation}
We introduce the random variables $X_1 = \int_{-\tau_1}^0 dt \sigma^z(t) $ and $X_2 = \int_{0}^{\tau_2} dt \sigma^z(t) $.
From this, we obtain
\begin{subequations}
\begin{align}
    & \llangle X_1^2 \rrangle = \frac{\tau_1}{\gamma} + \frac{1}{2\gamma^2} \left(1 - e^{-2 \gamma \tau_1} \right)  \\
    & \llangle X_2^2 \rrangle = \frac{\tau_2}{\gamma} + \frac{1}{2\gamma^2} \left(1 - e^{-2 \gamma \tau_2} \right) \\
    & \llangle X_1 X_2 \rrangle  = \frac{1}{4\gamma^2}\left(1 - e^{-2 \gamma \tau_1} \right)\left(1 - e^{-2 \gamma \tau_2} \right) .
    \end{align}     
\end{subequations}

The fourth order correlation function is 
\begin{multline}
    g^{(4)}(t_1,t_2,t_3,t_4) = \llangle \sigma(t_1) \sigma(t_2) \sigma(t_3) \sigma(t_4) \rrangle \\
    = \sum_{s_1s_2s_3s_4} s_1s_2s_3s_4 \\
    P(s_4,t_4|s_3,t_3)P(s_3,t_3|s_2,t_2)P(s_2,t_2|s_1,t_1) P(s_1,t_1) .
\end{multline}
We consider this for one particular ordering of $t_4 > t_3 > t_2 > t_1$, and then we will sum over all possible orderings, which there are four of such that $t_1,t_2 < 0 < t_3, t_4$. 

We get 
\begin{multline}
    g^{(4)}(t_1,t_2,t_3,t_4) =\frac{1}{16} \sum_{s_1s_2s_3s_4} s_1s_2s_3s_4 \\
    (1 + s_4s_3 e^{-2\gamma(t_4-t_3)})\times \\
    (1+ s_3 s_2 e^{-2\gamma(t_3-t_2)})(1 + s_1 s_2 e^{-2\gamma(t_2-t_1)}) .
\end{multline}
The only term which survives the sum is the one where all spin labels appear an even number of times.
There are 16 such combinations which cancels the denominator to give 
\begin{equation}
    g^{(4)}(t_1,t_2,t_3,t_4) = e^{-2\gamma(t_4-t_3)}e^{-2\gamma(t_2-t_1)}.
\end{equation}
Taking into account all orderings of the time allowed we find 
\begin{multline}
   g^{(4)}(t_1,t_2,t_3,t_4) = e^{-2\gamma|t_4-t_3|}e^{-2\gamma|t_2-t_1|} \\
    = g^{(2)}(t_4,t_3)g^{(2)}(t_2,t_1).
\end{multline}
This then integrates and cancels one of the disconnected contributions from $\llangle X^2_1\rrangle \llangle X^2_2\rrangle$ to yield 
\begin{multline}
\Gamma^{(4)} = -\sum_j V_j^4 \llangle X_{1j} X_{2j} \rrangle^2_c \\
= -\sum_j \left(\frac{V_j}{2\gamma}\right)^4 \left( 1 - e^{-2\gamma \tau_1} \right)^2\left( 1 - e^{-2\gamma \tau_2} \right)^2. 
\end{multline}
This is computed under a model of randomly oriented magnetic moments distributed uniformly in a plane with density $n_{\rm 2D}$ in Appendix~\ref{app:angular}.
We will focus on the case where the NV quantization axis is along $z$ for simplicity.
The angular dependence of this is not important, leading to small numerical variations in the noise as function of the quantization axis angle. 
The result is calculated in Appendix~\ref{app:angular}; we find the fourth-order non-Gaussian noise contribution is (restoring units of magnetic moment)
\begin{multline}
\label{eqn:TLS-kurtosis-raw}
\Gamma^{(4)} = -\left(\frac{\mu_0 g\mu_B \mu}{4\pi}\right)^4 \frac{87\pi n_{\rm 2D} }{175 h^{10}} \\
\times \left(\frac{1 - e^{-2\gamma \tau_1} }{2\gamma}\right)^2\left(\frac{1 - e^{-2\gamma \tau_2} }{2\gamma}\right)^2. 
\end{multline}

It is more convenient to study the normalized signal, which we will compare against the leading (Gaussian) contribution to the dephasing, which is expressed in terms of a $T_2$ dephasing time which depends on distance $h$, bath spin density $n_{\rm 2D}$, and the memory time $\tau_c = 1/(2\gamma)$.
The Gaussian contribution to dephasing for time $\tau_R$ is 
\begin{multline}
    \mathcal{C}_X= -\frac12 \llangle X^2 \rrangle = \\
    - \frac{\pi n_{\rm 2D}}{2h^4}\left(\frac{ g\mu_0\mu_B \mu}{4\pi} \right)^2 \left[ \frac{\tau_R}{\gamma} + \frac{1 - e^{-2\gamma \tau_R}}{2\gamma^2} \right]\\
    = -\frac{\tau_R}{T_2(h)} + \frac{1}{T_2(h)2\gamma}(1-e^{-2\gamma \tau_R}).
\end{multline}
Here, we find the $T_2$ time of 
\begin{equation}
     \frac{1}{T_2(h)}= \frac{\pi n_{\rm 2D}}{2h^4}\left(\frac{ g\mu_0\mu_B \mu}{4\pi} \right)^2 \frac{1}{2\gamma}.
\end{equation}
Expressed in terms of this, we find the non-Gaussian noise is 
\begin{multline}
\label{eqn:TLS-kurtosis-normed}
\Gamma^{(4)}(\tau_1,\tau_2) =\\
- \frac{174}{175}\frac{2}{\pi n_{\rm 2D}h^2} \left(\frac{\tau_c}{T_2(h)} \right)^2  \left( 1 - e^{-\tau_1/\tau_c} \right)^2\left( 1 - e^{-\tau_2/\tau_c} \right)^2. 
\end{multline}

We then see two notable facts. 
First, the prefactor of the signal can be seen to be scale as 
\begin{equation}
\frac{1}{\pi n_{\rm 2D} h^2} = (\xi/h)^2,
\end{equation}
where $\xi^{-1} \equiv \sqrt{\pi n_{\rm 2D}}$ is the typical distance between impurities in the plane. 
This is also related to the so-called ``inverse participation ratio" of the distribution of bath spin magnetic dipole couplings, and is essentially a measure of the effective size of the bath probed at distance $h$. 

This essentially reflects the fact that the non-Gaussian aspect of the noise is dominated by the contributions of the closest bath spins and indicates that all-else-equal a lower density of bath spins leads to a larger non-Gaussian contribution.
This is expected based on the central limit theorem, which would imply that many independent non-Gaussian noise processes should converge towards an effective Gaussian process. 
Therefore, the lower the density of bath spins, the larger the outlier effects are and the non-Gaussian noise becomes more visible. 
Experimentally, this indicates this would be best seen if the probe spin were placed very close to a small handful of bath spins.
{This closely parallels the results from several recent studies on the interactions between isolated NV centers and nuclear spin baths~\cite{Meinel.2022,Xia.2025,Zhang.2025}, which have demonstrated that non-Gaussian noise correlations are particularly effective for characterizing finite-size baths and their deviations from the central limit theorem.}

It is important to note that for extremely small $h/\xi \ll 1$ it is expected that the self-averaging assumption of the impurity distribution will break down.
If the probe qubit is sufficiently close to a single impurity, which is likely to happen when $h \ll \xi$, then the average over angle and position is no longer appropriate.
Rather, in this case, the noise should not be averaged over the distribution of the magnetic impurity locations, since these will likely not vary shot-to-shot.
Exploring this effect would be interesting, but is beyond the scope of this work. 

\begin{figure}
    \centering
    \includegraphics[width=\linewidth]{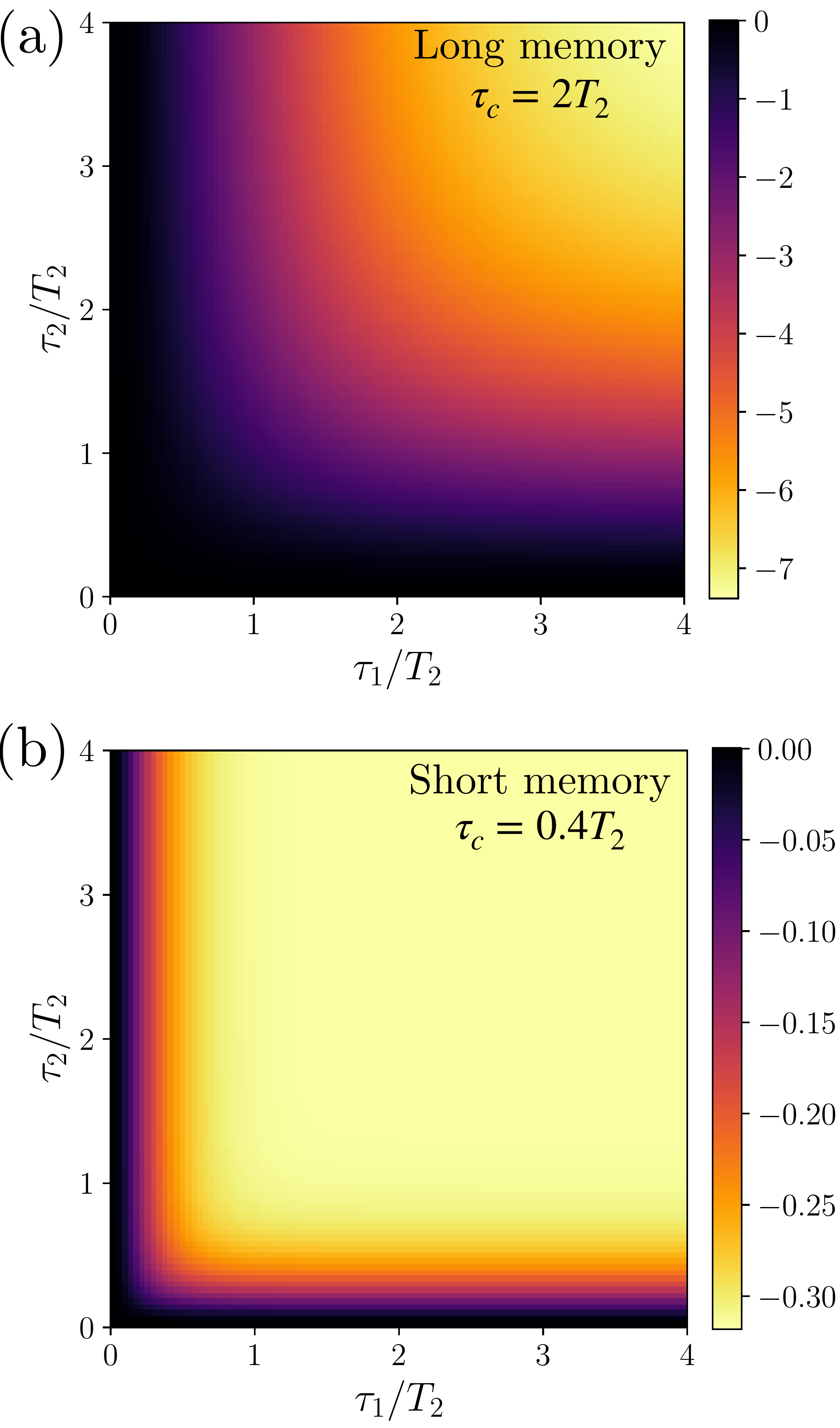}
    \caption{Fourth-order non-Gaussian noise cumulant $\Gamma^{(4)}(\tau_1,\tau_2)$ as a function of echo-times $\tau_1,\tau_2$ for fixed $\pi n_{\rm 2D} h^2 = 1$ measured relative to Ramsey $T_2(h)$ time.  
    (a) Long bath-memory case of $\tau_c = 2T_2(h)$. 
    At long echo times this result saturates.
    (b) Short bath-memory case of $\tau_c = 0.4T_2(h)$. 
    We see that the long bath-memory time leads to a larger result, as well as a slower and more pronounced approach to the saturation.}
    \label{fig:cumulants-compare}
\end{figure}

The second important aspect of this result is the dependence on the echo times $\tau_1,\tau_2$. 
This is illustrated in Fig.~\ref{fig:cumulants-compare}, which presents the two-echo-time dependent cumulant $\Gamma^{(4)}$ for the case of long and short bath memory times for a fixed value of $n_{\rm 2D} h^2$, and normalized according to $T_2(h)$. 
We see that the non-Gaussian noise cumulant essentially grows (negatively) from zero over a time of order $\tau_c$ before it saturates at long echo times.
The over size of the cumulant (at fixed distance $h$) is controlled by the memory of the bath, codified in $\tau_c/T_2(h)$; the longer the memory, the larger the saturation of the cumulant as well as the slower the approach to the maximum. 
Whereas the non-Gaussian noise cumulant ultimately saturates at long echo times, the Gaussian cumulants grow with echo time linearly at long-times. 
For instance, the Gaussian contribution to the Ramsey sequence for duration $\tau_R$ will ultimately grow as 
$\tau_R/T_2(h)$, which can be viewed as a consequence of the central limit theorem.
If we break up the phase evolution in to time chunks of order the memory time $\tau_c$, then these roughly form an independent random walk process, and this will ultimately grow like a Gaussian diffusion process.
At long times, the dephasing therefore has a tendency to self-average and recover Gaussian dynamics, with higher-order non-Gaussian cumulants ultimately saturating while the Gaussian cumulant does not. 
Therefore, by studying the dependence of the non-Gaussian noise cumulants on the echo times, it is possible to experimentally measure the convergence of the noise process towards the central limit theorem.  

Finally, it is interesting to observe that this non-Gaussian noise signal, which is a distinct signature of correlated bath dynamics, is observed even in the effectively infinite-temperature ensemble with any particular spin in the bath being in a completely unpolarized $50:50$ mixed state.
Nevertheless, by studying dynamical correlations one can gain a great deal of insight into the dynamics which underlie even the infinite-temperature ensemble.
Even though this was a model of independent two-level systems, we still saw that the non-Gaussian echos can encode rich and important physics pertaining to the dynamics and correlations of the bath degrees of freedom. 
{We now briefly discuss in more detail the validity of a perturbative treatment of the higher cumulants, and when this may fail.

\subsection{Justification of Cumulant Expansion}
\label{sub:perturbative}
In the previous example, we saw that the distance to the sample $h$ played an important role in the overall scaling of the higher cumulants relative to the Gaussian contributions. 
While the protocols we have introduced here actually measure the full nonperturbative echo $\Gamma_{1,2}$, it is often simpler to model and interpret the results using a perturbative cumulant expansion to the lowest non-vanishing order, which in most cases is the fourth cumulant $\Gamma^{(4)}$. 
One may wonder whether this perturbative procedure is justified, or conversely when it might be expected to fail.
Within the scope of independent two-level systems, we will consider the effect of higher order cumulants and show that, under certain circumstances the perturbative treatment can be formally justified. 
In particular, we will consider again a bath of independent two-level systems with magnetic moments $\mu$ and density $n_{\rm 2D} \equiv \frac{1}{\pi \xi_c^2}$ undergoing telegraph noise with correlation time $\tau_c$.
In order to simplify the treatment slightly in comparison to the previous section, we will consider their moments fixed out-of-plane. 

In this case, we can express the magnetic field at the distance $h$ away from the plane by the sum 
\begin{equation}
    B_{z}(t) =  \sum_j  \frac{\mu_0 \mu}{4\pi (\mathbf{R}_j^2 + h^2 )^\frac32} \left[\frac{3 h^2}{ h^2 + \mathbf{R}_j^2 } -1 \right] \sigma_j^z(t) .
\end{equation}
The relevant echo sequences have generating functionals for a particular pulse $f(t) $ of 
\begin{multline}
    \mathcal{C}[f] = \log \llangle \exp\left( i \int dt f(t) g\mu_B B_z(t) \right) \rrangle \\
    = \sum_j \log \llangle \exp\left( i \lambda_j \int dt f(t)\sigma_j^z(t)    \right) \rrangle,
\end{multline}
where the coupling strength for spin $j$ is given by 
\begin{equation}
    \lambda_j =    \frac{g\mu_B \mu_0 \mu}{4\pi (\mathbf{R}_j^2 + h^2 )^\frac32} \left[\frac{3 h^2}{ h^2 + \mathbf{R}_j^2} -1 \right]  .
\end{equation}
Here we have used the fact that the fluctuators are independent.
Ultimately, this stems from the presence of a finite magnetic correlation length, such that one can coarse grain to an effective description in terms of a density $n_{\rm 2D} \sim 1/\pi\xi_c^2$ of independent two-level systems. 

This now depends on the full nonperturbative form of the telegraph-noise cumulant generating functional
\begin{equation}
    \log \llangle e^{i\lambda \int dt f(t) \sigma(t) } \rrangle \equiv \mathcal{C}_\lambda[f].
\end{equation}
To evaluate this, we will refer to Appendix~\ref{app:TLS_cumulant}.
After evaluating the three Ramsey sequences over $[-\tau_R,0],[0,\tau_R],[-\tau_R,\tau_R]$, and the Hahn echo sequence over $[-\tau_R,0,\tau_R]$ we obtain the fully non-perturbative expression for the non-Gaussian echo of 
\begin{multline}
\label{eqn:exact_tls_ng}
\Gamma_{1,2} = n_{\rm 2D}\int d^2 r\\
\log\left[ 1 - \left( \frac{\lambda(r) \tanh \sqrt{\gamma^2 - \lambda^2(r)}\tau_R}{\sqrt{\gamma^2 - \lambda^2(r)} + \gamma \tanh \sqrt{\gamma^2 - \lambda^2(r)}\tau_R}\right)^4\right], 
\end{multline}
where here we have converted the average over positions $\sum_j \to n_{\rm 2D} \int d^2 r $ into an integral over space, assuming a self-averaging result, and the coupling strength is now a function of position as 
\begin{equation}
    \lambda(r) =  \frac{g\mu_B \mu_0 \mu}{4\pi (r^2 + h^2 )^\frac32} \left[\frac{3 h^2}{ h^2 + r^2} -1 \right]  .
\end{equation}

This can be better understood by scaling $r = h u$ and expressing in terms of the unitless coupling strength 
\begin{equation}
    \alpha = \frac{g\mu_B \mu_0 \mu}{2\pi  h^3 \gamma },
\end{equation}
bath correlation time 
\begin{equation}
    s = \tau_R \gamma,
\end{equation}
and coupling kernel 
\begin{equation}
    f(u) = \frac{3(u^2 + 1)^{-1}-1}{2(u^2+1)^{3/2}}.
\end{equation}
Performing a substitution to $v = u^2 + 1$ the integral becomes 
\begin{multline}
\Gamma_{1,2} = \pi n_{\rm 2D}h^2 \int_1^\infty dv \\
\log\left[ 1 - \left( \frac{\alpha f(v) \tanh \sqrt{1 - \alpha^2 f^2(v)}s}{\sqrt{1 - \alpha^2f^2(v)} +  \tanh \sqrt{1 - \alpha^2f^2(v)}s}\right)^4\right], 
\end{multline}
When $\alpha$ is small, this can be seen to go as $\alpha^4$ to leading order, which is the perturbative result; namely,
\begin{multline}
\Gamma^{(4)} = -\pi n_{\rm 2D}h^2\alpha^4 \left( \frac{ \tanh s}{1 +  \tanh s}\right)^4  \int_1^\infty dv f^4(v) \\
\sim - \frac{1}{\xi_c^2 h^{10} \gamma^4}\frac{ \tanh^4(\gamma\tau_R)}{(1+\tanh\gamma \tau_R)^4}. 
\end{multline}
However, we can now check the validity of the perturbative expansion. 
The difference between the exact and first-order expressions are shown in Fig.~\ref{fig:pert_compare} for different values of $\alpha$.

\begin{figure}
    \centering
    \includegraphics[width=\linewidth]{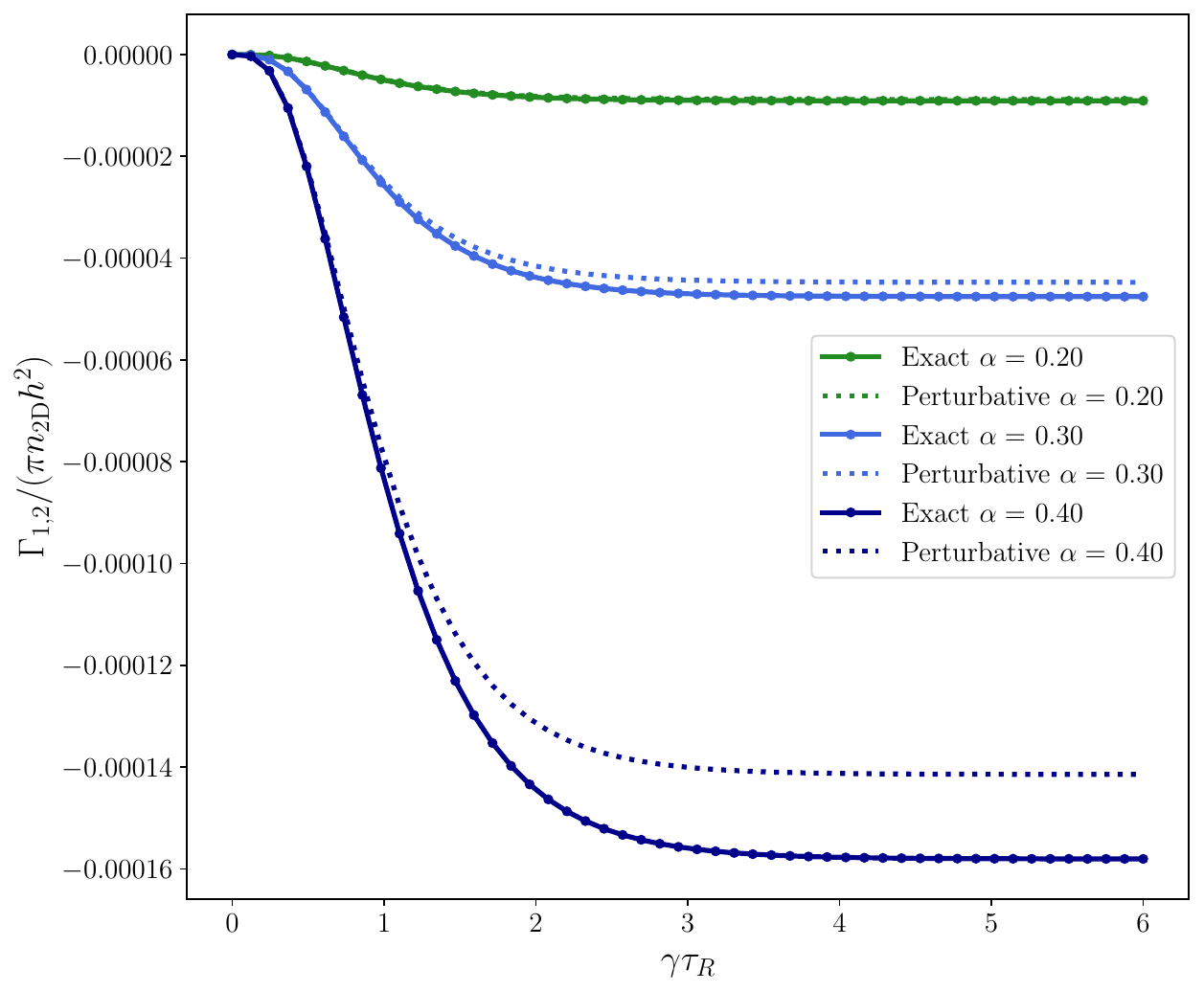}
    \caption{Comparison of numerically exact non-Gaussian echo against lowest-order cumulant expansion result. 
    For small $\alpha$ the agreement is excellent.
    For larger $\alpha$ a small discrepancy appears at long times $\tau_R \gamma$.
    As $\alpha \to 1$ the discrepancy becomes worse.}
    \label{fig:pert_compare}
\end{figure}

Evidently, the agreement is excellent for $\alpha\ll 1$. 
At long-times the difference between the full and perturbative results becomes more noticeable with increasing $\alpha < 1$.
As $\alpha \to 1$ perturbation theory is ultimately expected to fail completely, as non-perturbative contributions to the echo become significant. 

Therefore, we see that in the end, what ultimately controls the validity of a perturbative treatment is that the coherent phase accumulated is small\textemdash i.e. this is a weak-coupling approximation.
More specifically, since the qubit phase is a compact measurable, it is expected that when the coherent phase acquired approaches $\Delta \theta \sim 2\pi$, nonperturbative effects will onset. 
The rate of phase acquisition is set by the magnetostatic dipole interaction between the sensor and sample, and is of order 
\begin{equation}
    \dot \theta \sim \frac{\mu_0 \mu g\mu_B}{4\pi h^3}. 
\end{equation}
On the other hand, the magnetic field fluctuates on a time scale set by $\tau_c \sim 1/\gamma$.
If we therefore estimate the amount of phase acquired before the field fluctuates, we find the condition  
\begin{equation}
    \tau_c\times \dot \theta \sim \frac{\mu_0 \mu g\mu_B}{2 h^3\gamma} \ll 2\pi , 
\end{equation}
which is essentially what we find from the exact calculation.

Experimentally, this is best controlled by tuning the distance of the sensor to the sample, $h$, as $\alpha \sim 1/h^3$.
This yet again demonstrates a benefit of using local spin-qubits over, e.g. optical methods~\cite{Cheung.2024,Nambiar.2024,Randi.2017,Joubaud.2008} to probe magnetic noise, since this local spin-qubits offer a route towards tuning into the strong-coupling regime {\it in situ}. 

This also points to the potentially special importance of a second-order phase transition~\cite{Quan.2005,Wei.2012,Chen.2013,Bramwell.2009,Hohenberg.1977,Joubaud.2008}; near the critical point $\tau_c\to \infty$ due to critical slowing down.
In this regime, a perturbative treatment is not necessarily justified and corrections to $\Gamma_{1,2}$ from cumulants beyond $\Gamma^{(4)}$ might need to be included.
Understanding the behavior in this regime is much more complicated however, and a full non-perturbative treatment is beyond the scope of this work; in the following section we will simply consider what happens as we start to approach criticality and perturbatively explore a system which exhibits strongly-interacting bath degrees of freedom, as is expected near a second-order phase transition.
}

\subsection{Critical Magnetic Fluctuations}
\label{sub:modelA}

The example from Sec.~\ref{sub:TLS} showed that even local (i.e. uncorrelated in space) noise sources could present interesting deviations from Gaussianity due to intrinsic nonlinearities.
Here we will show that this technique is also well suited for studying systems which feature spatially correlated noise.
An interesting example of such a case appears near a second-order phase transition, whereupon the correlation length diverges.
While a full treatment of such a complicated case deserves its own dedicated investigation in future works, we will show here that this technique is well-suited to studying these dynamic critical magnetic fluctuations~\cite{Hohenberg.1977}, which are expected to be relevant to a number of two-dimensional Van der Waals magnetic materials~\cite{Thiel.2019,Ghiasi.2023,Xue.2024,Ziffer.2024,Hohenberg.1967,Mermin.1966}.
{Theoretical studies have also extensively explored (largely within the Gaussian regime) how critical noise manifests in the Loschmidt and spin echoes~\cite{Quan.2005,Wei.2012,Chen.2013}, which are particularly interesting as they may be used to study signatures of quantum criticality at finite temperature.}

We start by consider a two-dimensional magnetic system which has an out-of-plane magnetization density $m(\mathbf{r},t)$ that undergoes an Ising $\mathbb{Z}_2$ symmetry breaking transition.
Near the transition, this can be described by a free-energy Ginzburg-Landau functional 
\begin{equation}
    \mathcal{F}[m] = \int d^2 r \left[ \frac12 K(\nabla m)^2  + \frac12 r m^2 + \frac14 u m^4 \right],
\end{equation}
where $K,u > 0 $ and at the critical point $r$ changes from positive for $T>T_C$ to negative for $T < T_C$. 
These coefficients are derived from a microscopic Ising model in Appendix~\ref{app:GL}; in terms of the lattice constant $a$ and critical temperature $T_C = 4J$ (with superexchange interaction $J$), we find $u = T_C/(3a^2)$, $K = T_C/8$, and $r = (T-T_C)/a^2$.
The full equilibrium critical theory for this model is well-known, but complicated; we will consider here a simple treatment of the dynamic critical theory which is valid not too close to the critical point. 
In particular, we will assume the order parameter exhibits overdamped dynamics of the type ``Model-A" in the classification of Hohenberg and Halperin~\cite{Hohenberg.1977}.
This amounts to a Langevin type of time-dependent Ginzburg-Landau equation of motion for the order parameter $m(\mathbf{r},t)$ of the form 
\begin{equation}
    \Gamma \partial_t m(\mathbf{r},t) = -\frac{\delta \mathcal{F}}{\delta m(\mathbf{r},t)} + \eta(\mathbf{r},t),
\end{equation}
where the field $\eta(\mathbf{r},t)$ is a fluctuating force that exhibits Gaussian fluctuations in accordance with the fluctuation dissipation relation, such that $\llangle \eta(\mathbf{r},t) \eta(\mathbf{r}',t')\rrangle = 2\Gamma T \delta^2(\mathbf{r} - \mathbf{r}') \delta(t-t')$. 
Note that, while the field $\eta$ is Gaussian distributed, the magnetization $m(\mathbf{r},t)$ will not be due to the nonlinearity. 

Given a planar (out-of-plane polarized) magnetic order parameter $m(\mathbf{r},t)$, one can solve the magnetostatic equations to obtain the magnetic field at the qubit location a distance $z$ above the sample to be
\begin{equation}
    B_z(z,t) = \frac{\mu_0\mu_B S}{2a^2} \int_{\bf q} |\mathbf{q}| e^{-z|\mathbf{q}|} m_{\mathbf{q}}(t).
\end{equation}
Here we have ascribed a magnetic moment of $\mu_BS$ to each Ising spin, with $S$ an effective spin length, such that $\mu_B S/a^2$ is the areal density of magnetic moments. 
We have also assumed the qubit to lie at the origin in the $x-y$ plane, which is valid if we have translational invariance, and is applicable therefore only to the single-qubit non-Gaussian echo protocol. 
We will leave the two-qubit echo protocol to future works. 
Under these assumptions, if we expand to quartic order the non-Gaussian echo will measure the correlation function 
\begin{multline}
    \Gamma^{(4)} = \frac12 \left( \frac{gS \mu_B^2\mu_0}{2a^2} \right)^4\int_{-\tau_R}^0 dt_1 \int_{-\tau_R}^0 dt_2\int_0^{\tau_R} dt_3 \int_0^{\tau_R}dt_4 \\
     \int_{ {\bf q}_1,\mathbf{q}_2,\mathbf{q}_3,\mathbf{q}_4}  \prod_{j=1}^{4} |\mathbf{q}_j| e^{-z|\mathbf{q}_j|}\\
     \llangle m_{\mathbf{q}_1}(t_1) m_{\mathbf{q}_2}(t_2) m_{\mathbf{q}_3}(t_3) m_{\mathbf{q}_4}(t_4)\rrangle_c.
\end{multline}
Here we have, for simplicity assumed a compensated echo with $\tau_1 = \tau_2 = \tau_R$. 
We now go to the frequency domain, and write 
\begin{multline}
    \Gamma^{(4)} = \frac12 \left( \frac{gS \mu_B^2\mu_0}{2a^2} \right)^4 \int_{q_1,q_2,q_3,q_4} \\
    {W}^{(4)}_{\rm Ramsey}(\{q\};\tau_R,z) \times \underbrace{\llangle m(q_1) m(q_2) m(q_3)m(q_4)\rrangle_c}_{\Lambda^{(4)}(\{q\})},
\end{multline}
where we recall the appropriate non-Gaussian fourth-order echo ``filter-function" from Eq.~\ref{eqn:filter-Ramsey} (which now has also been expressed in terms of the spatial filter functions) of  
\begin{multline}
    {W}^{(4)}(\{q\};\tau_R,z)  = \prod_{j=1}^{4} |\mathbf{q}_j| e^{-z|\mathbf{q}_j|} \\
    \times  \int_{-\tau_R}^0 dt_1 \int_{-\tau_R}^0 dt_2\int_0^{\tau_R} dt_3 \int_0^{\tau_R}dt_4  e^{-i\sum_{j}\omega_j t_j }. 
\end{multline}
We have also defined the fourth-order vertex function 
\begin{equation}
\Lambda^{(4)}(\{q\}) \equiv \llangle m(q_1) m(q_2) m(q_3)m(q_4)\rrangle_c .
\end{equation}
It should be expected that, while this nominally has a dependence on four momenta and frequencies, these will be constrained by spatial and translational invariance to be overall conserved. 
We also expect that this function is invariant under permutation of the external momenta and frequencies $\{q\}$. 
Under these assumptions, we can simplify the filter function to the form 
\begin{multline}
    {W}^{(4)}(\{q\};\tau_R,z)  = \tau_R^4 \prod_{j=1}^{4} |\mathbf{q}_j| e^{-z|\mathbf{q}_j|} \textrm{sinc}(\omega_j \tau_R/2) \\
    \times  \frac16 \sum_{j < k} \cos[(\omega_j + \omega_k)\tau_R ]. 
\end{multline}

\begin{figure}
    \centering
    \includegraphics[width=\linewidth]{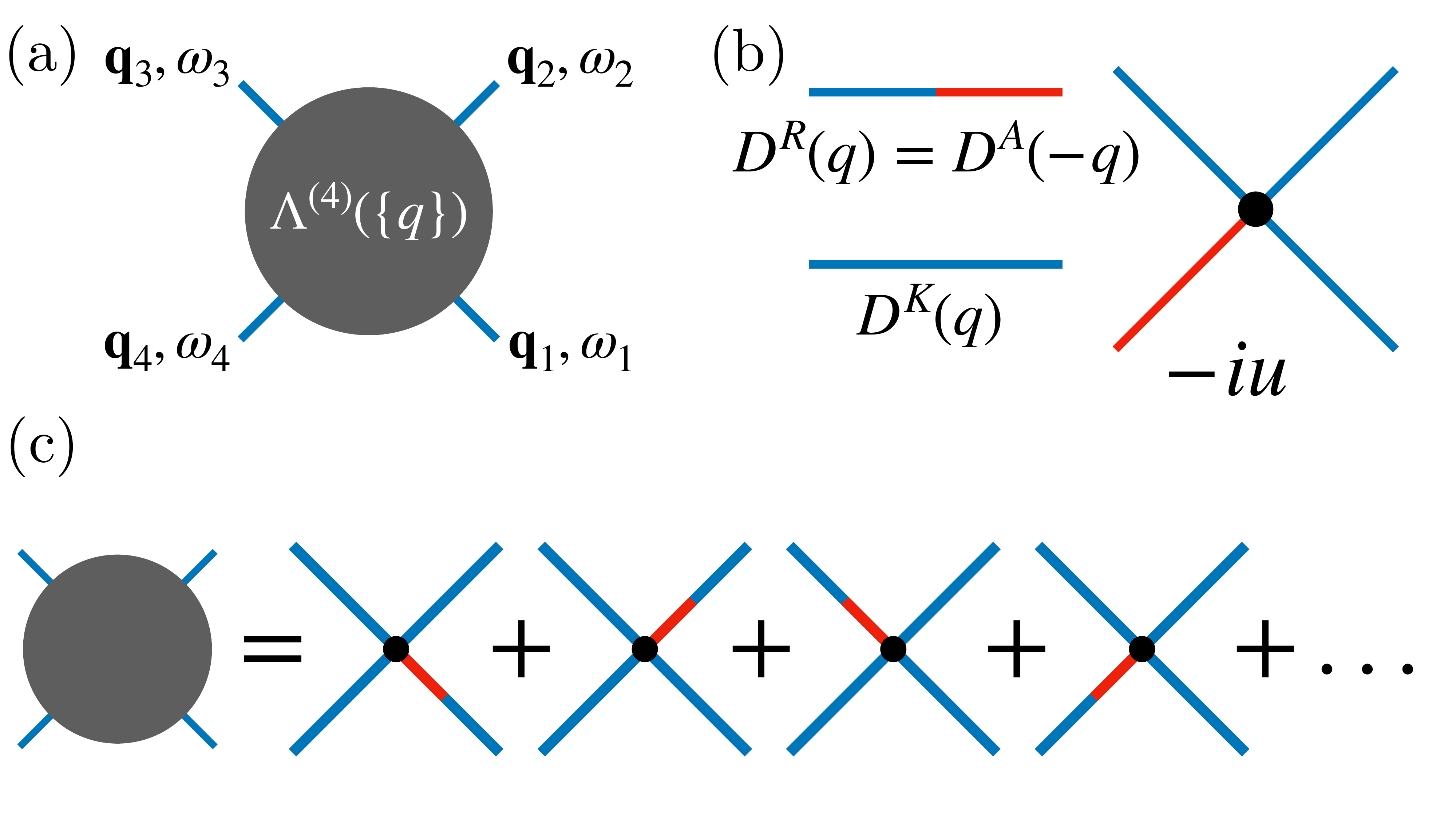}
    \caption{(a) Feynman diagram representation of connected four-point scattering amplitude measured in noise spectroscopy. 
    (b) Basic Feynman rules for the model. 
    Mixed propagators between the $m$ and $\zeta$ fields (shown by blue and red lines respectively) generate causal response functions, whereas $\llangle mm\rrangle$ generates the Keldysh correlation function. 
    Interactions due to the nonlinearity generate the vertex shown, which involves three classical $m$ fields and one quantum $\zeta$ field.
    (c) Perturbative expression used for the vertex function in this work.
    Higher diagrams, such as vertex loop corrections, are neglected here. }
    \label{fig:feynman}
\end{figure}

In order to compute the correlation function $\Lambda^{(4)}$, we will use the procedure of Martin, Siggia, Rose, Jannsen, and de Dominicis (MSRJD) (see Ref.~\cite{Kamenev.2011}), which allows us to map a classical stochastic process onto a path integral on the Schwinger-Keldysh contour, thereby connecting with the previously employed formalism.
The relevant action describing the dynamical magnetic fluctuations is then given by
\begin{equation}
    \mathcal{S} = \int d^3 x \zeta \left[ -\Gamma \partial_t + K \nabla^2 - r - um^2 \right]m + i T \Gamma \zeta^2 
\end{equation}
where $\zeta$ plays the role of the ``quantum" field and $m$ is the ``classical" magnetization. 
We are now tasked with calculating the connected four-point function $\Lambda^{(4)}(\{q\})$.
This is shown in terms of Feynman diagrams in Fig.~\ref{fig:feynman}(a), with the basic Feynman rules shown in Fig.~\ref{fig:feynman}(b). 

In this work we will calculate $\Lambda^{(4)}$ only to lowest order in the nonlinearity $u$, which is just the tree-level vertex. 
This neglects the effects of vertex corrections, which are expected to become important very close to the critical point, but provided fluctuations are not too strong we may expect this approach to yield reasonable results. 
The perturbative expansion is summarized by the Feynman diagrams presented in Fig.~\ref{fig:feynman}(c), with neglected higher order diagrams left in the ellipsis. 
We will also restrict ourselves to the high-temperature disordered phase, such that $T > T_C$.

We then find the correlation function shown in Fig.~\ref{fig:feynman}(c) of 
\begin{multline}
\Lambda^{(4)}(q_1,q_2,q_3,q_4) \\
=  \llangle \left( -i u \int d^3 x \zeta(x) m^3(x)  \right)m(q_1) m(q_2)m(q_3)m(q_4)\rrangle_c .
\end{multline}
We evaluate this using Wick's theorem.
From the action we find the relevant correlation and response functions to appropriate order in $u$ are 
\begin{subequations}
    \begin{align}
        & D^K(q) \equiv \langle m(q) m(-q)\rangle = -2i\frac{\Gamma T}{\omega^2 \Gamma^2 + \gamma_{\bf q}^2} \\
        & D^{R}(q)\equiv \langle m(q) \zeta(-q)\rangle = \frac{1}{i\omega \Gamma - \gamma_{\bf q}}.
    \end{align}
\end{subequations}
Here we have introduced
\begin{equation}
    \gamma_{\bf q} = K (\mathbf{q}^2 + r ) = K (\mathbf{q}^2 + \xi_{c}^{-2} ),
\end{equation}
with correlation length $\xi_{c} = \sqrt{K/r}$ and correlation time $\tau_c = \Gamma/ K \xi_{c}^2 $.
Note the definition of $D^R$ here differs slightly from convention in that for $t < 0$ it coincides instead with the advanced function. 

By conservation of energy and momentum, the four-point function will overall be proportional to a $\delta$-function.
By Wick's theorem we obtain the form 
\begin{multline}
\Lambda^{(4)}(\{q\})= -6iu (2\pi )^3\delta^{(3)}\left(\sum_j q_j\right) \times \\
 \left[D^R(q_1) D^K(q_2) D^K(q_3) D^K(q_4) + \textrm{permutations of $\{q\}$} \right] .
\end{multline}
This corresponds a sum over the four diagrams in the first line shown in Fig.~\ref{fig:feynman}(c).
Simplifying this, we find
\begin{multline}
\Lambda^{(4)}(\{q\})= (2\pi )^3\delta^{(3)}\left(\sum_j q_j\right) \times \\
-6u (2\Gamma T)^3\left[ \sum_j \gamma_{{\bf q}_j}\right] \prod_j \frac{1}{\omega_j^2\Gamma^2 + \gamma_{{\bf q}_j}^2 }.
\end{multline}
Note this sign is also negative, which is consistent with the simpler model of classical telegraph noise.
Heuristically, this signifies that the noise is ``bunched" and has a distribution that is narrower than a Gaussian with the same variance.

By resolving the $\delta$-function and utilizing the permutation symmetry of the integrand, we can express the noise as 
\begin{multline}
    \Gamma^{(4)} = -12 u (2\Gamma T)^3 \left( \frac{g S\mu_B^2\mu_0}{2a^2} \right)^4 \tau_R^4 \int_{q_1,q_2,q_3} \\
    \frac{ \gamma_{\bf Q} }{\Omega^2 \Gamma^2 + \gamma_{\bf Q}^2} |\mathbf{Q}| e^{-z |\mathbf{Q}|}\textrm{sinc}(\Omega \tau_R /2) \\
    \prod_{j = 1}^{3} \frac{1}{\omega_j^2 \Gamma^2 + \gamma_{{\bf q}_j}^2} |\mathbf{q}_j| e^{-z |\mathbf{q}_j|}\textrm{sinc}(\omega_j \tau_R/2) \\
    \frac13 \sum_{j<k=1}^{3} \cos[\omega_j + \omega_k]\tau_R,
\end{multline}
where we have introduced the notation 
\begin{equation}
\Omega = \sum_{j=1}^3 \omega_j,\quad \mathbf{Q} = \sum_{j=1}^3 \mathbf{q}_j.     
\end{equation}
We will express the overall non-Gaussian noise in terms of a unitless function which depends only on $\tau_R/\tau_c$ and $z/\xi_c$, up to prefactors. 

The result is given as 
\begin{equation}
    \Gamma^{(4)} = -12 u \left(\frac{2\Gamma T}{K}\right)^3 \left( \frac{gS \mu_B^2\mu_0}{2Ka^2} \right)^4 \tau_c \xi_c^{4}F(z/\xi_c,\tau_R/\tau_c),
\end{equation}
where the unitless function is expressed as an integral 
\begin{multline}
F(x,s) \equiv \frac{1}{(2\pi)^9} \frac{s}{x^{10} }\int d^2 \tilde{q}_1 d^2 \tilde{q}_2 d^2 \tilde{q}_3\int d\tilde{\omega}_1 d\tilde{\omega}_2d \tilde{\omega}_3 \\
\tilde{q}_1  e^{-\tilde{q}_1}\tilde{q}_2  e^{-\tilde{q}_2}\tilde{q}_3  e^{-\tilde{q}_3} \tilde{Q} e^{- \tilde{Q}}\\
\textrm{sinc}(\tilde{\omega}_1/2)\textrm{sinc}(\tilde{\omega}_2/2)\textrm{sinc}(\tilde{\omega}_3/2)\textrm{sinc}(\tilde{\Omega}/2) \\
\frac13 \sum_{j<k=1}^{3} \cos(\tilde{\omega}_j + \tilde{\omega}_j) \\
 \frac{1 + \tilde{Q}^2/x^2}{ ( \tilde{\Omega}/s)^2 + ( 1 + \tilde{Q}^2/x^2 )^2}\prod_{j=1}^{3} \frac{1}{ ( \tilde{\omega}_j/s)^2 + ( 1 + \tilde{q}_j^2/x^2 )^2}.
\end{multline}

\begin{figure}
    \centering
    \includegraphics[width=\linewidth]{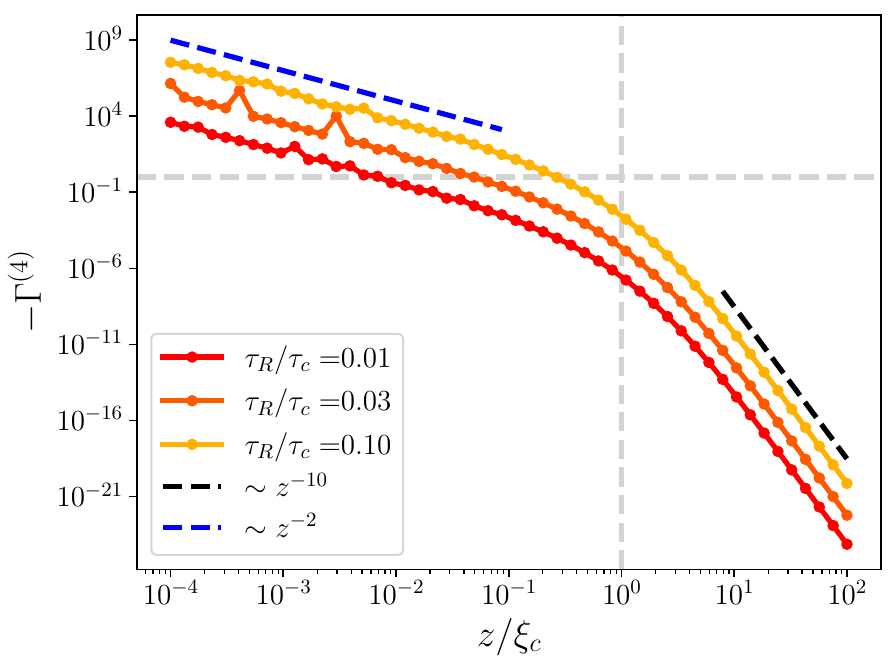}
    \caption{Short-time non-Gaussian noise due to critical magnetic fluctuations as a function of distance $z/\xi_c$ and echo time $\tau_R/\tau_c$.
    We use model parameters of $\xi_c/a = 10^2$ and $\tau_{\rm dp}/\tau_c = 10^{-2}$, with prefactors coming from the Ginzburg-Landau derivation.
    In this regime the scaling with echo time is $(\tau_R/\tau_c)^4$.
    At short distances, this noise decays slowly with liftoff distance as $(z/\xi_c)^{-2}$, signifying the strong spatial correlations in the bath below the correlation length.
    At long distaces, the noise rapidly drops as $(z/\xi_c)^{-10}$, which is what is expected also based on the local telegraph noise model. 
    Gray dashed lines show guides for the eye at unity.
    }
    \label{fig:model-A-static}
\end{figure}

Nominally this is a difficult integral to evaluate even numerically as it is high-dimensional and highly oscillatory. 
Here we will simplify this significantly by focusing on the ``static" noise regime with $\tau_R \ll \tau_c$ such that the noise is essentially of a $T_2^*$ type. 
This is equivalent to a short-time expansion in the echo time $\tau_R$, and is evaluated in Appendix~\ref{app:modelA}.
We find the leading behavior of $\tau_R^4$, as expected based on a static magnetic field measured using Ramsey spectroscopy with $\int_0^{\tau_R} B(t) dt \sim \tau_R B(0)$.
We present the dependence on the non-Gaussian fourth cumulant as a function of distance and time in Fig.~\ref{fig:model-A-static}.

In the short-time limit we can write the non-Gaussian cumulant as 
\begin{equation}
    \Gamma^{(4)} = - C \left(\frac{\tau_{c}}{\tau_{\rm dp}}\right)^4 \frac{1}{(\xi_c/a)^2} F(z/\xi_c,\tau_R/\tau_c)
\end{equation}
where we have introduced the ``dipole" time 
\begin{equation}
    \tau_{\rm dp} = a^3/(g S \mu_B^2 \mu_0)
\end{equation}
and the numerical prefactor $C$ is estimated based on the Ginzburg-Landau expansion to be of order $C = 2^{13}$.
We estimate for reasonable parameters of $g = S = 1$ and lattice constant of $a = $1nm that $\tau_{\rm dp} \sim $1ns.
Near a critical point we may also reasonably expect $\xi_c/a\sim 100$.
For the correlation time, we have less reliable estimates, but we will consider $\tau_c = 100 \tau_{\rm dp}\sim$100ns although this should be more carefully computed for a particular model of magnetization relaxation dynamics.

We can understand the distance dependence by recalling the results from Sec.~\ref{sub:perturbative} on independent fluctuators.
For long distances $z \gg \xi_c$ the bath degrees of freedom are essentially independent and therefore we expect the noise to scale as if the fluctuators were independent, which yields the scaling of $1/z^{10}$.
In this case, each independent region has a size which is roughly scaling as $(\xi_c/a)^2$ that ultimately determines the effective size of the magnetic bath. 

In contrast, at short distances compared to the correlation length, the bath degrees of freedom are highly correlated and fluctuate in unison. 
As a result, the noise is dramatically enhanced within the correlation length volume which is now determined not by the correlation length $\xi_c$, but rather by the NV qubit distance itself, such that the effective bath size is $(z/a)^2$.
This crossover effect becomes particularly interesting upon further approach to the critical point, where it is expected that $\xi_c$ and $\tau_c$ will both diverge.
At the critical point we expect the statistics of the fluctuations to be described by genuine non-Gaussian fixed point of the renormalization group flow (with some caveats~\cite{Bramwell.2009}).

{In contrast to probes based on, {\it e.g.} fluctuations of optical properties~\cite{Joubaud.2008,Cheung.2024}, the protocols we outline here operator with spatial resolution that can be both smaller or larger than the correlation length. 
While bulk (optical) probes are powerful, and have a deep connection to the renormalization group flow~\cite{Bramwell.2009}, it is also of great interest to be able to resolve how these correlations develop with a finer degree of spatial resolution; this is one of the main benefits of local spin-qubit noise spectroscopy.  
}

Using the non-Gaussian spin-echo techniques outlined in this work, it would then be possible to experimentally measure the departure from the central limit theorem which is theoretically predicted by the renormalization group analysis.
One may hope that given the divergent nature of noise at the critical point, it may be possible to apply these techniques to study real two-dimensional magnetic materials such as CrCl$_3$~\cite{Xue.2024}, CrI$_3$~\cite{Thiel.2019}, or CrSBr~\cite{Ghiasi.2023,Ziffer.2024}, where non-Gaussian noise signatures are expected to be most prominent~\cite{Bramwell.2009}.
It is even more interesting to consider extending these techniques to systems which exhibit strong quantum fluctuations~\cite{Quan.2005,Wei.2012,Chen.2013} which cannot be modeled by the classical Langevin-type dynamics employed here.

\section{Conclusion}
\label{sec:conc}

To conclude, we have seen how local noise-magnetometry as realized by, for instance NV center spin-qubits, can be used to study higher order cumulants of magnetic noise in a frequency- and momentum-resolved way\textemdash {simplifying and building on the general characterizations of Refs.~\cite{Gasbarri.2018,Norris.2016,Wang.2019}}.
We have presented a number of protocols, which can be implemented in scanning NV center noise-magnetometry systems in order to extract non-Markovian non-Gaussian noise cumulants.
We have further shown how multiple qubits can be used to access spatially non-local non-Gaussian noise cumulants. 
Then, we demonstrated how even in a simple case of a paramagnetic system with a finite density of independent and unpolarized bath spins exhibiting random telegraph noise, signatures of the discrete nature of the bath spins and their non-Gaussian noise statistics could be extracted, and the Gaussian limit could be controllably recovered by varying the qubit-sample distance.
This already points to the great potential that local noise-magnetometry can hold for the study of noise and correlations in quantum materials.
We also considered the role of spatial correlations near a magnetic phase transition, which we argued is particularly promising for observing large non-Gaussian noise signatures. 

Looking forwards, there are a number of fascinating possibilities which can be explored once we are no longer constrained by linear response.
From the perspective of material phenomena, it would be interesting to apply these techniques to study nonlinear vortex motion in a superconductor~\cite{Larkin.1975} or noise at the threshold of depinning~\cite{Larkin.1971}.
Single static vortices have already been imaged using NV centers in Josephson junctions~\cite{Chen.2024ihp} as well as two-dimensional thin-film superconductors~\cite{Thiel.2016,Pelliccione.2016}, and it has been predicted that dynamic vortex noise can unveil aspects of the Berezinskii-Kosterlitz-Thouless transition in sufficiently thin superconducting materials~\cite{Curtis.2024}, as well as different vortex phases in finite field~\cite{Liu.2025}.
Puzzling noise due to vortex motion has also recently been measured~\cite{Jayaram.2025} away from the flux-lattice melting transition.
Specific to the case of the superconducting vortices, magnetic flux is carried in discrete quanta and therefore, the distribution of the flux noise is in some sense expected to follow Poisson statistics~\cite{Koushik.2013}, whereas magnetic noise due to Bogoliubov quasiparticles is not quantized and can vary continuously.
While it is difficult to distinguish between these two scenarios using linear noise spectroscopy~\cite{Curtis.2024}, the technique we propose here is well suited to be able to distinguish between these two sources of magnetic noise.

Along similar lines, it may be possible to use this technique to study shot noise due to magnon transport in magnetic systems at low-temperature, analogously to what was studied in the context of a single-mode cavity photon in Ref.~\cite{Wang.2020}.
This may be useful for determining the statistics of emergent quasiparticles in spin-liquids, which may have elementary excitations that carry fractions of spin-1, such as in the case of a one-dimensional Heisenberg chain wherein the elementary excitations are spinons that carry spin-1/2 and obey fermionic statistics.
Magnon-magnon interactions and hydrodynamics~\cite{Xue.2024} are also expected to show interesting signatures in the non-Gaussian noise cumulants, which effectively measures the magnon-magnon scattering vertex. 

From a nonlinear spectroscopy perspective, there are also interesting connections between the nonlinear noise spectroscopy we propose here and coherent multidimensional spectroscopy techniques such as two-dimensional terahertz spectroscopy~\cite{Liu.202472,Salvador.2024,Salvador.2025,McGinley.2024}, {or multidimensional correlation imprinting~\cite{Zhang.2025}.}
In terms of operators, the fourth-order noise correlation function proposed in this work involves the sequence of nested anticommutators $\sim \Tr \hat{\rho}\{ \hat{\Delta}(t_4),\{\hat{\Delta}(t_3),\{\hat{\Delta}(t_2),\hat{\Delta}(t_1)\}\}\}$, whereas the nonlinear response functions calculated in, e.g. Ref.~\cite{Salvador.2024,Salvador.2025} involve mixed commuators and anticommutators such as $\sim \Tr \hat{\rho}\{ \hat{\Delta}(t_4),[\hat{\Delta}(t_3),[\hat{\Delta}(t_2),\hat{\Delta}(t_1)]]\}$.
As a result, the nonlinear noise spectroscopy and nonlinear response techniques are expected to be related to each other through fluctuation dissipation relations, which relate the nonlinear noise spectrum to the nonlinear response spectrum~\cite{Kamenev.2011}.
Similarly, using quench dynamics~\cite{Wang.2021fej} in conjunction with the protocols we outline here, it may be possible to measure higher-order response functions and the fluctuation dissipation relations experimentally.
It should in principle also be possible to extend the protocol outlined in this work, which utilized echos solely comprised of $\pi$ rotations (which are Clifford gates) to also include $\pi/2$ rotations, which should allow for probing both response and correlation functions.

It is also interesting to explore these ideas from a quantum information theory perspective, especially in the case of multiple spin-qubits, and especially since recent experiments have demonstrated the ability to perform the necessary coincidence measurements on at least two distinct NV centers~\cite{Bernien.2012,Rovny.2022}, as well as the ability to entangle multiple qubits~\cite{Rovny.2025,Zhou.2025}.
In this case, there are interesting questions pertaining to whether multiple qubits can be used to characterize or generate entanglement~\cite{Zou.2022} in the shared magnetic bath and potential violations of Bell inequalities which are tied to the genuine quantum mechanical nature of the wavefunction, or whether entanglement can be used to improve sensitivity in a material setting~\cite{Zhou.2020lzi,Wang.2024,Ji.2024}.
Likewise, given the rise of so-called Noisy Intermediate-Scale Quantum (NISQ) devices, it is interesting to consider whether quantum simulators can be used to effectively simulate quantum noise processes for the purposes of modeling quantum sensing experiments~\cite{Zhou.2020lzi,Bradley.2019,Seetharam.2023,Smart.2022}, or whether sensing protocols such as those outlined here can be used to better characterize device noise and error processes~\cite{Guimaraes.2023,Dag.2024,Berg.2023,Temme.2017,Klimov.2018,Torre.2023}, or even connections to questions of sampling complexity, negative Wigner distributions~\cite{Wang.2020}, thermalization~\cite{Choi.2017,Choi.20170po,Kucsko.2018,Andersen.2024,Rosenberg.2024h49,Kim.2023} or quantum error correction~\cite{Taminiau.2014,Chirame.2025} and how these processes may present in dynamical noise signatures.

\section*{Author Contributions}
This work was developed by J.B.C. and E.D., and carried out by J.B.C. under the guidance and direction of E.D. and A.Y.
The Author's would like to acknowledge fruitful discussions with generative AI ChatGPT. 

\begin{acknowledgements}
The authors would like to acknowledge productive discussions with Yuxin Wang, Ania Bleszynski Jayich, Alex G{\'o}mez Salvador, Nandini Trivedi, Wenchao Xu, Jamir Marino, Sarang Gopalakrishnan, Francisco Machado, Nikola Maksimovic, Nicholas Poniatowski, Ji Zou, and Prineha Narang. 
This work is supported by the Quantum Science Center (QSC), a National Quantum Information Science Research Center of the U.S. Department of Energy (DOE). 
A.~Y. is also partly supported by the Gordon and Betty Moore Foundation through Grant GBMF 9468, and by the U.S. Army Research Office (ARO) MURI project under grant number W911NF-21-2-0147.
J.B.C. and E.D. are supported by the SNSF project 200021 212899, the Swiss State Secretariat for Education, Research and Innovation (SERI) under contract number UeM019-1 and NCCR SPIN, a National Centre of Competence in Research, funded by the Swiss National Science Foundation (grant number 225153).
This work was performed in part at Aspen Center for Physics, which is supported by National Science Foundation grant PHY-2210452.
This research was supported in part by grant NSF PHY-2309135 to the Kavli Institute for Theoretical Physics (KITP). 
\end{acknowledgements}

\appendix

\section{Quantum Noise Sources}
\label{app:quantum}

Here we show that the treatment in terms of classical noise sources essentially remains unchanged if the noise is of quantum origin, by using an appropriate generalization to the Schwinger-Keldysh two-time contour. 
Let us start by considering the qubit-bath Hamiltonian as 
\begin{equation}
    \label{eqn:qubit-bath-Ham}
    \hat{H} = \hat{H}_B + \frac12 \hat{\Delta} \hat{\sigma}_z,
\end{equation}
where $\hat{H}_B$ describes the many-body dynamics of the bath, such that $\hat{\Delta}$ is an operator acting on the bath subsystem, which will acquire fluctuations and time-dependence upon passing to the interaction picture with respect to the bath Hamiltonian $\hat{H}_B$. 
Here we again assume the qubit splitting is on average zero, which can be enacted by passing to a rotating frame.
We will discuss transverse couplings later. 

We consider the evolution of the density matrix of the joint system, which we assume at $t =0$ is of the form 
\begin{equation}
    \hat{\rho}(0) = \hat{\rho}_B(0) \otimes \hat{\rho}_S(0).
\end{equation}
We can write the initial density matrix for the qubit in terms of the $\sigma_z$ eigenstates as 
\begin{equation}
    \hat{\rho}_S(0) = \sum_{s_+,s_-}\rho_{s_+,s_-}(0) |s_+\rangle\langle s_-|.
\end{equation}
Here we have anticipated that the density matrix will evolve with two-contours; the $+$ forward contour for the ket and the $-$ backward contour for the bra. 
This will be done using the Feynman-Vernon influence functional for the two-time contour~\cite{Feynman.1963,Leggett.1987,Caldeira.1983rkp,Kamenev.2011,Sieberer.2016}.

In particular, we are interested in the state of only the qubit at a later time, say $\tau_R$ in the case of a single Ramsey sequence.
We again expand in the eigenbasis of $\sigma_z$ for the two time-contours and trace over the bath degrees of freedom. 
This yields 
\begin{equation}
    \langle s_+'|\hat{\rho}_S(\tau_R)|s_-'\rangle = \\
    \sum_{s_+,s_-}\rho_{s_+,s_-}(0)\mathcal{I}_{s_+,s_-}^{s_+',s_-'}(\tau_R).
\end{equation}
This has been expressed in terms of the influence functional for the bath which in this case reads 
\begin{multline}
    \mathcal{I}_{s_+,s_-}^{s_+',s_-'}(\tau_R) = \delta_{s_+',s_+}\delta_{s_-',s_-} \\
    \Tr_{B}\left[ \mathcal{T}e^{-i\frac{s_+}{2} \int_0^{\tau_R}\tilde{\Delta}(t)dt } \hat{\rho}_B(0) \overline{\mathcal{T}}e^{+i\frac{s_-}{2} \int_0^{\tau_R}\tilde{\Delta}(t)dt }\right] .
\end{multline}
This is expressed now in the interaction picture for the bath Hamiltonian and the time-ordering and anti-time-ordering symbols $\mathcal{T},\mathcal{\bar{T}}$ respectively.
Furthermore, because the spin coupling commutes with the spin Hamiltonian, the value of $s_{\pm}$ cannot change on each contour and so the influence functional is diagonal in on each contour.
We can further express this bath influence functional in terms of the Feynman path-integral for the two time contours over the bath degrees of freedom.
This yields 
\begin{widetext}
\begin{multline}
    \mathcal{I}_{s_+,s_-}^{s_+',s_-'}(\tau_R) = \delta_{s_+',s_+}\delta_{s_-',s_-} \\
    \int\mathcal{D}[\Delta_+(t),\Delta_-(t)]\exp\left(iS_+[\Delta_+(t)] - iS_-[\Delta_-(t)] - i\frac12\int_0^{\tau_R}dt  \left[ s_+\Delta_+(t) - s_-\Delta_-(t)\right]\right) \\
    \langle \Delta_+(0)|\hat{\rho}_B(0)|\Delta_-(0)\rangle \langle \Delta_-(\infty)| \Delta_+(\infty)\rangle.
\end{multline}    
\end{widetext}
Here $S_\pm$ is the intrinsic action governing the bath variable $\Delta_\pm$ which is obtained from the regular path-integral construction for Hamiltonian $\hat{H}_B$.
The last line reflects the initial correlations of the fields which are encoded in the bath density matrix, as well as the trace condition which constrains the fields on the two contours to be matched at the final time-point.
We can extend this final time-point to $t=\infty$ since after time $\tau_R$ the qubit is decoupled and therefore the dynamics can be trivially extended in the bath since the unitaries will simply cancel each other at this point. 
In equilibrium this can be evaluated using the Schwinger-Keldysh formalism~\cite{Feynman.1963,Kamenev.2011}. 
We will then be evaluating the expectation value using this functional method of 
\begin{multline}
    \mathcal{I}_{s_+,s_-}^{s_+',s_-'}(\tau_R) = \delta_{s_+',s_+}\delta_{s_-',s_-} \\
    \bigg\llangle \exp\left(- i\frac12\int_0^{\tau_R}dt  \left[ s_+\Delta_+(t) - s_-\Delta_-(t)\right]\right)\bigg\rrangle.
\end{multline}
Here we use the $\llangle\cdot\rrangle$ to indicate the bath correlation function in absence of the spin. 
We now perform the usual splitting of the field $\Delta_\pm(t)$ into the quantum and classical variables defined by 
\begin{equation}
    \Delta_\pm(t) =  \Delta_{\rm cl}(t) \pm \frac12 \Delta_{\rm q}(t). 
\end{equation}
This then gives 
\begin{multline}
    \mathcal{I}_{s_+,s_-}^{s_+',s_-'}(\tau_R) = \delta_{s_+',s_+}\delta_{s_-',s_-} \\
    \bigg\llangle \exp\left(- i\frac12\int_0^{\tau_R}dt  \left[ (s_+- s_-)\Delta_{\rm cl}(t)  +\frac{s_++s_-}{2} \Delta_{\rm q}(t)\right]\right)\bigg\rrangle.
\end{multline}

Importantly, let us now consider the echo protocol. 
In this case, the initial density matrix, immediately after the first $\pi/2$ rotation, is $\hat{\rho}_S(0) = |+\rangle\langle +| \Leftrightarrow \rho_{s_+,s_-}(0) = \frac12$.
This has contributions from multiple different paths in the influence functional; there are diagonal amplitudes with $s_+ = s_- = \pm 1$ with equal amplitude.
There are also the coherences, which have $s_+ = - s_- = \pm 1$, again with equal amplitude. 
Now, the diagonal entries in the influence functional are 
\begin{equation}
    \mathcal{I}_{s,s}^{s,s}(\tau_R) = \bigg\llangle \exp\left(- i\frac{s}{2}\int_0^{\tau_R}dt \Delta_{\rm q}(t)\right)\bigg\rrangle.
\end{equation}
However, it is a central property of the Keldysh path integral that expectation values of only quantum fields vanish~\cite{Kamenev.2011}; that is $\llangle (\Delta_q )^n \rrangle  =0$ for $n \geq 1$. 
Therefore, this contribution will simply give $\mathcal{I}_{s,s}^{s,s}(\tau_R) = 1$; this is a requirement for the conservation of probability that under pure-dephasing evolution the qubit density matrix diagonals will not evolve.

Now, we turn to the evolution of the cohernces.
These have instead the influence functional 
\begin{equation}
    \mathcal{I}_{s,-s}^{s,-s}(\tau_R) = \bigg\llangle \exp\left(- is\int_0^{\tau_R}dt \Delta_{\rm cl}(t)\right)\bigg\rrangle.
\end{equation}
In contrast, these terms are not constrained and we generically find this given by the cumulants of the quantum noise source, as expressed in terms of the Keldysh path integral.
In fact, this is formally the same result as that implied by the classical noise process, although the specific form of the cumulants of $\Delta_{\rm cl}(t)$ will depend importantly on whether the noise is quantum or classical through, e.g. the occupation function of the noise. 

We can also study the echo sequence and how this contributes {\it via} influence functional formalism. 
To obtain this, we will insert at a time $\tau_E$ the echo pulse on the qubit.
This will have a non-trivial effect on the qubit density matrix; we find the influence functional in the presence of the echo by inserting the echo operator $\hat{\sigma}_x$ in the appropriate spot in the time-ordering
\begin{widetext}
\begin{multline}
    \left[\mathcal{I}^E\right]_{s_+,s_-}^{s_+',s_-'}(\tau_R,\tau_E) = \\
    \Tr_{B}\left[  \langle s_+'|\mathcal{T}e^{-i\frac{s_+'}{2} \int_{\tau_E}^{\tau_R}\tilde{\Delta}(t)dt }\hat{\sigma}_x|s_+ \rangle \mathcal{T}e^{-i\frac{s_+}{2} \int_0^{\tau_E}\tilde{\Delta}(t)dt } \hat{\rho}_B(0) \overline{\mathcal{T}}e^{+i\frac{s_-}{2} \int_0^{\tau_E}\tilde{\Delta}(t)dt } \langle s_-|\hat{\sigma}_x\overline{\mathcal{T}}e^{+i\frac{s_-'}{2} \int_{\tau_E}^{\tau_R}\tilde{\Delta}(t)dt }|s_-'\rangle \right] .
\end{multline}
Via the same construction as before, we see the influence functional with the echo is expressed in terms of the Keldysh functional of the bath and the echo matrix elements as 
\begin{multline}
    \left[\mathcal{I}^E\right]_{s_+,s_-}^{s_+',s_-'}(\tau_R,\tau_E) = \langle s_+'|\hat{\sigma}_x|s_+ \rangle \langle s_-|\hat{\sigma}_x|s_-'\rangle \\
    \bigg\llangle \exp\left(-i\frac12 \int_{0}^{\tau_E} dt \left[s_+\Delta_+(t) - s_-\Delta_-(t)\right] -i\frac12 \int_{\tau_E}^{\tau_R} dt \left[s_+'\Delta_+(t) - s_-'\Delta_-(t)\right] \right)\bigg\rrangle. 
\end{multline}
We likewise express this in terms of the quantum and classical fields as 
\begin{multline}
    \left[\mathcal{I}^E\right]_{s_+,s_-}^{s_+',s_-'}(\tau_R,\tau_E) = \langle s_+'|\hat{\sigma}_x|s_+ \rangle \langle s_-|\hat{\sigma}_x|s_-'\rangle \\
    \bigg\llangle \exp\left(-i\frac12 \int_{0}^{\tau_E} dt \left[(s_+ - s_-)\Delta_{\rm cl}(t) + \frac{s_++s_-}{2}\Delta_{\rm q}(t)\right] -i\frac12 \int_{\tau_E}^{\tau_R} dt \left[(s_+' - s_-')\Delta_{\rm cl}(t) + \frac{s_+'+s_-'}{2}\Delta_{\rm q}(t)\right] \right)\bigg\rrangle. 
\end{multline}
For the echo, we see the matrix elements are purely real and off-diagonal so that 
\begin{multline}
    \left[\mathcal{I}^E\right]_{s_+,s_-}^{s_+',s_-'}(\tau_R,\tau_E) = \delta_{s_+',-s_+}\delta_{s_-',-s_-}\\
    \bigg\llangle \exp\left(-i\frac12 \int_{0}^{\tau_E} dt \left[(s_+ - s_-)\Delta_{\rm cl}(t) + \frac{s_++s_-}{2}\Delta_{\rm q}(t)\right] -i\frac12 \int_{\tau_E}^{\tau_R} dt \left[(s_+' - s_-')\Delta_{\rm cl}(t) + \frac{s_+'+s_-'}{2}\Delta_{\rm q}(t)\right] \right)\bigg\rrangle. 
\end{multline}
\end{widetext}
Nevertheless, this still admits the decomposition into the diagonals and coherences of the density matrix. 
The density matrix diagonals again do not evolve because this involves an exponential of only $\Delta_{\rm q}(t)$, albeit with a more complicated coupling in time. 
On the other hand, the coherences have $s_+=-s_-$ and the same holds for $s_\pm' = -s_\pm$.
We find specifically 
\begin{multline}
    \left[\mathcal{I}^E\right]_{-s,s}^{s,-s}(\tau_R,\tau_E) = \\
    \bigg\llangle \exp\left(-is\left[ \int_{\tau_E}^{\tau_R} dt \Delta_{\rm cl}(t)   -  \int_{0}^{\tau_E} dt \Delta_{\rm cl}(t)\right]\right)\bigg\rrangle. 
\end{multline}
This is exactly the same result we expect from the classical noise derivation, understood in the appropriate Keldysh contour sense. 
This shows that our derivations based on classical noise hold the same for quantum noise sources, provided we express the noise in terms of the classical correlation functions of the noise in the Keldysh formalism. 

All of this followed rather straightforwardly if we considered only longitudinal noise.
However, in reality we should expect in general noise that is both longitudinal and transverse. 
This case is substantially more complicated; here we now only briefly outline what happens when we relax this assumption.

\subsection{Energy Relaxation}
\label{app:t1}
A related, but much more complicated possibility is that the magnetic noise may not solely be longitudinal (i.e. dephasing), but also transverse.
In this case, the state of the spin qubit can change due to the noise and exhibit feedback on the bath, which manifests in the appearance of additional correlation functions, including those which incorporate back reaction effects from the spin qubit on to the bath. 
In principle this can be incorporated into the formalism used here through the Feynman-Vernon influence functional technique, as detailed in Leggett, {\it et al.}~\cite{Leggett.1987,Feynman.1963}, which uses the ``blip and sojourn" formalism to handle the case of both longitudinal and transverse fields acting on the qubit. 
However, if the qubit has a large zero-field splitting as compared to the typical size of the fluctuations, then it is well-known that low-frequency transverse noise effects are strongly suppressed due to energy conservation constraints.
In the case at hand, where spin qubits have typical splittings of order GHz, this is true and therefore we can realistically include transverse relaxation processes through a simple $T_1$ energy relaxation time. 
We model this by describing the dynamics using a Linblad equation for the joint qubit-system density matrix $\hat{\rho}$
\begin{multline}
\label{eqn:lindblad}
    \frac{\partial  \hat{\rho}}{\partial t} = -i [ \hat{H}_B + \frac12\hat{\Delta}\hat{\sigma}_z, \hat{\rho}] \\
    + \Gamma_{\downarrow}\left[ \hat{\sigma}^- \hat{\rho} \hat{\sigma}^+ - \frac12 \{ \hat{\sigma}^- \hat{\sigma}^+,\hat{\rho}\}\right] \\
    + \Gamma_{\uparrow}\left[ \hat{\sigma}^+ \hat{\rho} \hat{\sigma}^- - \frac12 \{ \hat{\sigma}^+ \hat{\sigma}^-,\hat{\rho}\}\right] .
\end{multline}
We have describe the qubit splitting by an operator $\hat{\Delta}$ which acts on the bath degrees of freedom (with their own intrinsic Hamiltonian $\hat{H}_B$), as well as including a constant zero-field splitting of $\Delta_0$.
Here the jump operators describe absorption and emission of quanta with rates $\Gamma_\uparrow,\Gamma_\downarrow$ respectively. 
According to detailed balance, these should satisfy $\Gamma_\uparrow/\Gamma_\downarrow = e^{-\beta \Delta_0}$ with temperature $\beta$.
For a GHz splitting, this is nearly infinite temperature, such that in practice we often find $\Gamma_\uparrow \approx \Gamma_\downarrow$.
Note that these rates are due to transverse noise processes which have dynamics at frequencies comparable to $\Delta_0$, in accordance with Fermi's Golden Rule, but due to the large separation of time scales, this are taken as an independent and effectively Gaussian decoherence channel. 

We diagnose the effect of these terms on our cumulant isolation protocol.
Let us consider the Ramsey protocol.
At time $t =0$ we again initialize the qubit in to the $|+\rangle\langle +|$ state. 
We then compute the Feynman-Vernon influence functional which describes the dynamics under Eq.~\eqref{eqn:lindblad}~\cite{Sieberer.2016,Kamar.2024}.
The action of the bath can be incorporated, as in Sec.~\ref{app:quantum}, by converting the operator $\hat{\Delta}$ in to two fluctuating fields $\Delta_\pm(t)$, one for the evolution of bras and one for kets, which then must be averaged over the bath distribution function as before. 
The contribution to the influence functional due to the coherences is seen by inserting $\rho_S(0) = |s_+\rangle\langle s_-| $ where $s_+ = -s_-$. 
In this case, the action of the jump-operators is diagonal and thus this simply ends up appending an exponential decay rate on to the influence functional so that after time $t_R$ the coherence of the density matrix is given by 
\begin{equation}
    \mathcal{I}_{s,-s}^{s,-s}(\tau_R) = \llangle \exp\left( -is\int_0^{\tau_R}dt \Delta_{\rm cl}(t) - \tau_R/(2T_1) \right) \rrangle. 
\end{equation}
Here we have introduced $T_1^{-1} = (\Gamma_\uparrow + \Gamma_\downarrow).$
We also see that the exponential decay does not depend on which coherence the qubit is in, and therefore the Hahn echo will be effected in the same manner. 
We therefore see that the $T_1$ corrected cumulants $\tilde{\mathcal{C}}$, expressed in terms of the cumulants with $T_1 = \infty$ will be 
\begin{subequations}
    \begin{align}
        & \tilde{\mathcal{C}}_{X_1} = \mathcal{C}_{X_1} - \tau_1/(2T_1) \\
        & \tilde{\mathcal{C}}_{X_2} = \mathcal{C}_{X_2} - \tau_2/(2T_1) \\
        & \tilde{\mathcal{C}}_{X_1+X_2} = \mathcal{C}_{X_1+X_2} - (\tau_1+\tau_2)/(2T_1) \\
        & \tilde{\mathcal{C}}_{X_1-X_2} = \mathcal{C}_{X_1-X_2} - (\tau_1+\tau_2)/(2T_1).
    \end{align}
\end{subequations}
As a result, we see that the contribution to the non-Gaussian statistic $\Gamma_{X_1,X_2}$ will simply cancel between the different pulse sequences since they all add up to the same durations.

Having now demonstrated the broad applicability of the echo spectroscopy for detecting non-Markovian and non-Gaussian dynamics in both classical and quantum baths, we turn our attention to the study of non-Gaussian correlations in space, which we will show can be detected using two qubits. 

\section{Generalizing to Higher Orders}
\label{app:6order}
We here present a formalism which allows us to generalize to higher order cumulant extraction, and in particular a protocol for the sixth-order cumulant. 

Consider a set of $M$ random variables $\{X_{j}\}_{j=1}^{M}$, which will be for us the phases acquired over different pulse sequence time intervals, or possibly different spatially distinct qubits. 
Under the framework we consider, different dynamical decoupling sequences are characterized by different time-domain filter functions which are characterized for the $n$th different pulse sequence by coefficients $\alpha_j^{(n)} \in \{ 0,\pm 1\}$.
$\alpha_j^{(n)}$ will be zero if the $j$th time interval is not present in the $n$th pulse sequence, and is otherwise $+1\ (-1)$ if it is present and has an even (odd) number of $\pi$-pulses after it.
In principle, from this one obtains measurements of the random variable $X^{(n)}= \sum_j \alpha_j^{(n)} X_j$, and in particular the cumulant generating function 
\begin{equation}
    \mathcal{C}^{(n)} = \log \llangle \exp\left( i X^{(n)} \right) \rrangle.
\end{equation}
Now, given a set of $\mathcal{C}^{(n)}$ we need to find a set of coefficients $A_n$ such that the statistic
\begin{equation}
    \Gamma = \sum_n A_n \mathcal{C}^{(n)}
\end{equation}
vanishes to a desired order.
In terms of the multivariate cumulants we find 
\begin{multline}
    \Gamma = \sum_n A_n \sum_N \frac{i^N}{N!} \llangle (X^{(n)} )^N\rrangle_c \\
    =  \sum_{N=1}^{\infty} \frac{i^N}{N!}\sum_{j_1,...,j_N}\sum_n A_n \alpha_{j_1}^{(n)}...\alpha_{j_N}^{(n)} \llangle X_{j_1}...X_{j_N} \rrangle_c. 
\end{multline}

As an example, the main protocol in the paper given is equivalent considering $M = 2$ different pulse durations, with filter function coefficients $\alpha\in \{ (\pm 1,0),(0,\pm 1),(1,\pm 1),(-1,\pm 1)\}$.
We can simplify this by considering a measurement not of the full phase $e^{iX}$ but rather only the real part $\cos X$, which equivalently means we symmetrize between $\alpha^{(n)}_j$ and $-\alpha^{(n)}_j$. 
Then, each individual Ramsey or Hahn sequence corresponds to measuring with coefficients $(1,0)$ and $(0,1)$, while the conjoined sequence is $(1,1)$ and the phase difference is obtained by $(1,-1)$. 
We see that this indeed cancels the first two terms in the series, as we find odd terms vanish when we add $\pm $ pairs, and therefore up to fourth order we find  
\begin{multline}
    \Gamma = -\frac{1}{2!}\sum_{j_1,j_2}\sum_n A_n \alpha_{j_1}^{(n)}\alpha_{j_2}^{(n)}\llangle X_{j_1} X_{j_2} \rrangle_c \\
    + \frac{1}{4!}\sum_{j_1,j_2,j_3,j_4}\sum_n A_n \alpha_{j_1}^{(n)}\alpha_{j_2}^{(n)}\alpha_{j_3}^{(n)}\alpha_{j_4}^{(n)}\llangle X_{j_1} X_{j_2} X_{j_3} X_{j_4} \rrangle_c . 
\end{multline}
We want this to vanish at quadratic order, for all possible cumulant values.
This then imposes a system of linear equations on the $A_n$ coefficients which reads 
\begin{subequations}
    \begin{align}
        & A_{(1,0)} + A_{(1,1)} + A_{(1,-1)} = 0\\
        & A_{(0,1)} + A_{(1,1)} + A_{(1,-1)} = 0\\
        & A_{(1,1)} - A_{(1,-1)}  = 0.
    \end{align}
\end{subequations}
Up to an overall constant, we see our pulse sequence protocol of $A_{(1,0)} = A_{(0,1)} = -2$, $A_{(1,1)} = A_{(1,-1)} = 1$ is the solution.
Then, we find at quartic order the leading contribution of 
\begin{equation}
    \Gamma = \llangle X_1^2 X_{2}^2 \rrangle_c + ...,
\end{equation}
which other than an overall factor of $2$ coming from the symmetrization not being normalized, is the result in the main text. 

We now show how to generalize this to isolate the third, and also sixth cumulants. 
To isolate the third cumulant, we can instead choose $A_{\alpha} = -A_{-\alpha}$, such that all even terms are canceled.
In the case that the mean is non-vanishing, we then must simply cancel the linear term in order to isolate the leading contribution as the cubic cumulant. 
One such set of coefficients is $A_{(1,0)} = 2, A_{(0,1)} = 0, A_{(1,1)} = -1, A_{(1,-1)} = -1$, which isolates then to leading order 
\begin{equation}
    \Gamma = \frac{i}{3} \llangle X_1 X_{2}^2 \rrangle_c + ...,
\end{equation}
which is seen to be imaginary due to being an odd term. 

More interesting is perhaps isolating the sixth cumulant, which is the next leading term in the even series expansion. 
This can be seen to not be possible with only $M=2$ pulses.
If we try with just $4$ pulses, we arrive at the two equations to cancel the quadratic terms of 
\begin{subequations}
    \begin{align}
        & A_{(1,0)} + A_{(1,1)} + A_{(1,-1)} = 0\\
        & A_{(0,1)} + A_{(1,1)} + A_{(1,-1)} = 0\\
        & A_{(1,1)} - A_{(1,-1)}  = 0,
    \end{align}
\end{subequations}
which are the same as before.
Finally, in order to cancel the remaining quartic term, we must also impose 
\begin{equation}
A_{(1,1)} + A_{(1,-1)} = 0. 
\end{equation}
This is a system of four equations for four coefficients and the determinant is seen to be non-vanishing (alternatively this can easily be solved uniquely to yield $A_n = 0$ which is the trivial solution).
We therefore must expand the number of pulses used to $M = 3$.
Not including parity doubles and trivial cases, we then have $(3^3-1)/2 = 13$ different combinations of pulse sequences.

\begin{widetext}
These are 
\begin{multline}
    \alpha \in\\
     \{(1,0,0),(0,1,0),(0,0,1),(1,1,0),(1,0,1),(0,1,1),(0,1,-1),(1,0,-1),(1,-1,0),(1,1,1),(1,1,-1),(1,-1,1),(-1,1,1)\},
\end{multline}
which can be more succinctly labeled as $\{1,2,3,\bar{3},\bar{2},\bar{1},\bar{1}{3}^c,2^c\bar{3},0,3^c,2^c,1^c\}$
Imposing the diagonal quadratic terms vanish in this case imposes 
\begin{subequations}
    \begin{align}
        & 0 = A_1 + A_{\bar{3}}+A_{\bar{2}} + A_{0} + A_{2^c\bar{3}} +A_{\bar{2}3^c} + A_{3^c} + A_{2^c} +A_{1^c}\\
        & 0 = A_2 + A_{\bar{3}}+A_{\bar{1}} + A_{0} + A_{2^c\bar{3}} +A_{\bar{1}3^c} + A_{3^c} + A_{2^c} +A_{1^c} \\
        & 0 = A_3 + A_{\bar{1}}+A_{\bar{2}} + A_{0} + A_{\bar{2}3^c} +A_{\bar{1}3^c} + A_{3^c} + A_{2^c} +A_{1^c} ,
    \end{align}
\end{subequations}    
while vanishing of the quadratic cross terms implies 
\begin{subequations}
    \begin{align}
        & 0 = A_{\bar{3}} + A_{0} - A_{2^c\bar{3}} + A_{3^c} - A_{2^c} -A_{1^c} \\
        & 0 = A_{\bar{2}} + A_{0} - A_{\bar{2}3^c} -A_{3^c} + A_{2^c}  - A_{1^c}\\
        & 0 = A_{\bar{1}} + A_{0} - A_{\bar{1}3^c} - A_{3^c} - A_{2^c} + A_{1^c}.
    \end{align}
\end{subequations}  
This gives us six equations.
Imposing the quartic terms vanish will then yield another set of linear equations.
It can be seen that the coefficients of the $\llangle X_1^4\rrangle_c$ will be proportional to that of $\llangle X_1^2\rrangle_c$ and likewise for all diagonal terms.
Therefore, these are redundant equations.
Similarly, the coefficients of term $\llangle X_1X_2^3\rrangle_c$ will match the coefficient of $\llangle X_1 X_2\rrangle_c$.
These are therefore again redundant.
We find then that the terms which are not redundant will be the biquadratic terms and the totally mixed terms. 
The three biquadratic terms yield 
\begin{subequations}
    \begin{align}
        & 0 = A_{\bar{3}} + A_{0} + A_{2^c\bar{3}} + A_{3^c} + A_{2^c} + A_{1^c} \\
        & 0 = A_{\bar{2}} + A_{0} + A_{\bar{2}3^c} + A_{3^c} + A_{2^c} + A_{1^c}\\
        & 0 = A_{\bar{1}} + A_{0} + A_{\bar{1}3^c} + A_{3^c} + A_{2^c} + A_{1^c}.
    \end{align}
\end{subequations} 
Lastly, there are totally mixed terms $\llangle X_1^2 X_2 X_3\rrangle_c $ and such. 
These will contribute equations 
\begin{subequations}
    \begin{align}
        & 0 = A_0 + A_{3^c} - A_{1^c} - A_{2^c} \\
        & 0 = A_0 + A_{2^c} - A_{1^c} - A_{3^c} \\
        & 0 = A_0 + A_{1^c} - A_{3^c} - A_{2^c}.
    \end{align}
\end{subequations} 
We therefore find we have 12 equations in total for 13 coefficients and this can be solved. 
\end{widetext}
We utilize permutation invariance of $1\to 2 \to 3$ to argue that $A_{1} = A_2 = A_3 = a$, $A_{1^c} = A_{2^c} = A_{3^c} = c$, $A_{\bar{1}} = A_{\bar{2}} = A_{\bar{3}} = b$, and $A_{2^c \bar{3}} =A_{\bar{2} {3}^c}=A_{\bar{1}3^c }= d $.
We then obtain 
\begin{subequations}
    \begin{align}
        & 0 = A_0 -c \\
        & 0 = b + A_0 + d +3c \\
        & 0 = b+ A_0 -d-c \\
        & 0 = a +2b + A_0 + 2d + 3c .
    \end{align}
\end{subequations} 
This is solved by $A_0 = 1$, $A_{1}=A_2 = A_3 = 4 $, $A_{\bar 1}=A_{\bar 2} = A_{\bar 3} = -2 $, $A_{1^c} = A_{2^c} = A_{3^c} = 1$ and $A_{\bar{2}3^c} = A_{\bar{1}3^c} = A_{2^c \bar{3}} = -2$. 
The remaining terms will then be at sixth order.

\section{Angular Averaging for Dipole Tensor}
\label{app:angular}
We want to evaluate the averages 
\begin{equation}
    \overline{V^2} = \sum_j V_j^2,\quad \overline{V^4} = \sum_j V_j^4.
\end{equation}
We will will do this for an ensemble of random spins distributed throughout a two-dimensional plane with density $n_{\rm 2D}$ and random orientation of magnetic moment.
The probe qubit will be located at the origin in the $x-y$ plane and a distance $h$ above the plane; we will consider the quantization axis probe spin to point along the $\mathbf{\hat{e}}_z$ direction. 

We need to average the dipole tensor over random dipole orientations as 
\begin{subequations}
\begin{align}
& \overline{ \left( 3 \mathbf{R}_j\cdot\mathbf{\hat{e}}_z \mathbf{R}_j \cdot\bm{\mu}_j - {R}_j^2 \bm{\mu}_j\cdot\mathbf{\hat{e}}_z \right)^2} \equiv \mu^2 R_j^4 \mathcal{F}_2(\hat{\mathbf{R}}_j)\\
& \overline{ \left(  3 \mathbf{R}_j\cdot\mathbf{\hat{e}}_z \mathbf{R}_j \cdot\bm{\mu}_j - {R}_j^2 \bm{\mu}_j\cdot\mathbf{\hat{e}}_z\right)^4} \equiv \mu^4 R_j^8 \mathcal{F}_4(\hat{\mathbf{R}}_j).
\end{align}
\end{subequations}
Here, $\mathbf{R}_j$ is the location of the bath spin with respect to the probe spin, and has magnitude $R_j = \sqrt{\mathbf{r}_j^2 +h^2}$ where $\mathbf{r}_j$ is the planar coordinate of the bath spin. 
This would then give the averages 
\begin{subequations}
\begin{align}
&\overline{V^2} =  \left(\frac{\mu_0 \mu }{4\pi}\right)^2 \overline{ \frac{1}{R_j^6}\mathcal{F}_2(\vartheta)}\\
& \overline{V^4} = \left(\frac{\mu_0 \mu }{4\pi}\right)^4 \overline{ \frac{1}{R_j^{12}}\mathcal{F}_4(\vartheta)} .
\end{align}
\end{subequations}
Here $\vartheta$ is obtained as $\cos \vartheta = h/\sqrt{r_j^2 + h^2}$. 
The remaining average is taken over the planar position of impurities. 

Let us evaluate this by choosing coordinates where $\mathbf{\hat{R}}_j = (\sin \vartheta,0,\cos\vartheta)$ such that $\mathbf{R}_j$ lies in the $x-z$ plane.
We then obtain
\begin{subequations}
\begin{align}
& \mathcal{F}_2 = \overline{ \left( 3 \cos\vartheta \left[ x \sin \vartheta + z \cos \vartheta\right] - z\right)^2}\\
& \mathcal{F}_4 = \overline{ \left( 3 \cos\vartheta \left[ x \sin \vartheta + z \cos \vartheta\right] - z\right)^4} .
\end{align}
\end{subequations}
Here $x,y,z$ refer to the components of the dipole moment unit vector. 
This simplifies to, in general 
\begin{subequations}
\begin{align}
& \mathcal{F}_2 = \overline{ \left( 3 \cos\vartheta \sin \vartheta x  + z \left[3\cos^2 \vartheta - 1\right]\right)^2}\\
& \mathcal{F}_4 = \overline{\left( 3 \cos\vartheta \sin \vartheta x  + z \left[3\cos^2 \vartheta - 1\right]\right)^4} .
\end{align}
\end{subequations}
The non-vanishing moments in terms of $x,z$ are 
\begin{subequations}
    \begin{align}
        & \overline{z^2} = \overline{x^2} = \frac{1}{3}\\
        & \overline{z^4} = \overline{x^4}= \frac15 \\
        & \overline{z^2 x^2} = \frac{1}{15}.
        \end{align}
\end{subequations}
This yields 
\begin{widetext}
\begin{subequations}
\begin{align}
& \mathcal{F}_2 = \frac13 \left(3 \cos\vartheta \sin \vartheta\right)^2 +\frac13 \left(3\cos^2 \vartheta - 1\right)^2\\
&     \mathcal{F}_4 =\frac15 \left(3 \cos\vartheta \sin \vartheta \right)^4  + \frac15\left(3\cos^2\vartheta -1 \right)^4 + \frac{6}{15}  \left( 3 \cos\vartheta \sin \vartheta \right)^2 \left( 3\cos^2\vartheta -1 \right)^2.
\end{align}
\end{subequations}
\end{widetext}
These further simplify to 
\begin{subequations}
\begin{align}
& \mathcal{F}_2(\vartheta_j) = \frac13 \left[3 \cos\vartheta_j^2  +1\right]\\
& \mathcal{F}_4(\vartheta_j) =\frac15 \left[3 \cos\vartheta_j^2  +1\right]^2.
\end{align}
\end{subequations}
Here, $\cos\vartheta_j = \mathbf{\hat{n}}\cdot\hat{\mathbf{R}}_j$ is the (cosine of the) angle between the quantization axis and the location of the magnetic moment at $\mathbf{R}_j$. 
These only depend on the polar angle of the bath spin $\vartheta_j$.

We will then also have to deal with the average over two-level system location as 
\begin{subequations}
\begin{align}
& \overline{V^2}= \sum_j \left(\frac{ \mu_0 \mu}{4\pi} \right)^2 \frac{1}{R_j^6} \mathcal{F}_2(\vartheta_j) \\
& \overline{V^4}= \sum_j \left(\frac{ \mu_0 \mu}{4\pi} \right)^4 \frac{1}{R_j^{12}} \mathcal{F}_4(\vartheta_j) .
\end{align}
\end{subequations}

Let us now replace the sum over bath spins by an integral over their location in the two-dimensional plane via
\begin{equation}
    \sum_j f(\mathbf{R}_j) = n_{\rm 2D} \int d^2 R f(\mathbf{R}).
\end{equation}
This gives 
\begin{subequations}
\begin{align}
& \overline{V^2}= n_{\rm 2D}\int d^2r  \left(\frac{ \mu_0 \mu}{4\pi} \right)^2 \frac{1}{R^6} \mathcal{F}_2(\vartheta) \\
& \overline{V^4}=  n_{\rm 2D} \int d^2 r \left(\frac{ \mu_0 \mu}{4\pi} \right)^4 \frac{1}{R^{12}} \mathcal{F}_4(\vartheta) .
\end{align}
\end{subequations}

We first evaluate the second moment.
We find 
\begin{equation}
    \overline{V^2} = \pi n_{\rm 2D}\left(\frac{\mu_0\mu}{4\pi}\right)^2 \frac{1}{2h^4} .
\end{equation}
Next, we compute the fourth moment. 
After averaging over the plane we get 
\begin{equation}
    \overline{V^4} = \pi n_{\rm 2D}\left(\frac{\mu_0\mu}{4\pi}\right)^4 \frac{87}{175 h^{10}}.
\end{equation}
We have numerically checked that for different quantization axis angles, this does not vary dramatically. 

\section{Nonperturbative Cumulant Generating Functional for Two-Level System Noise}
\label{app:TLS_cumulant}
We need to evaluate the cumulant generating functional 
\begin{equation}
    \mathcal{C}_\lambda[f] = \log \llangle \exp\left( i\lambda \int dt f(t) \sigma(t) \right) \rrangle,
\end{equation}
where $f(t)$ encodes the pulse sequence.
We will here focus only on Ramsey sequences of duration $\tau_R$ or Hahn echo of duration $\tau_E$ on each half-echo. 
To evaluate this, we will introduce the conditional kernels 
\begin{equation}
    M_\sigma(t) = \llangle \exp\left(i\lambda \int_0^t ds \sigma(s) \right) \rrangle\bigg|_{\sigma(0) = \sigma}. 
\end{equation}
We have the initial condition 
\begin{equation}
    M_\sigma(0) = 1. 
\end{equation}
Now, consider evolving this for a short time $dt$.
We need
\begin{equation}
    M_\sigma(t+dt) = \llangle \exp\left(i\lambda \int_0^{t+dt} ds \sigma(s) \right) \rrangle\bigg|_{\sigma(0) = \sigma} .
\end{equation}
\begin{widetext}
This can be expanded using the additivity of the integral as 
\begin{equation}
    M_\sigma(t+dt) = \llangle \exp\left(i\lambda \int_0^{dt} ds \sigma(s) \right) \exp\left(i\lambda \int_{dt}^{t+dt} ds \sigma(s) \right) \rrangle\bigg|_{\sigma(0) = \sigma}. 
\end{equation}
The initial short-time evolution can be expanded in terms of the probability for either no flips, with probability $1-\gamma dt$ where $\gamma$ is the switching rate, or one flip with probability $\gamma dt$, with higher-order processes being irrelevant at short times. 
This yields 
\begin{multline}
    M_\sigma(t+dt) = \\
    \left[ 1- \gamma dt \right] e^{i\lambda \sigma(0)} \llangle  \exp\left(i\lambda \int_{dt}^{t+dt} ds \sigma(s) \right) \rrangle\bigg|_{\sigma(0) = \sigma} + \gamma dt \gamma dt e^{-i\lambda \sigma(0)}\llangle  \exp\left(i\lambda \int_{dt}^{t+dt} ds \sigma(s) \right) \delta[\sigma(0) = \sigma]\rrangle\bigg|_{\sigma(0) = \sigma}.
\end{multline}
The remaining kernels can be re-expressed using the Markovian property to shift $t \to t - dt$, noting that in the second term since a flip has occurred the spin at time $dt$ is in the state $\sigma(dt) = -\sigma$, as 
\begin{equation}
    M_\sigma(t+dt) = \left[ 1- \gamma dt \right] e^{i\lambda \sigma(0)dt } \llangle  \exp\left(i\lambda \int_{0}^{t} ds \sigma(s) \right) \rrangle\bigg|_{\sigma(0) = \sigma} + \gamma dt e^{-i\lambda \sigma(0)dt }\llangle  \exp\left(i\lambda \int_{0}^{t} ds \sigma(s) \right) \rrangle\bigg|_{\sigma(0) = -\sigma}.
\end{equation}
But these remaining terms are simply $M_\sigma(t)$ and $M_{-\sigma}(t)$ respectively.
Then, subtracting to get the change $dM_\sigma$ and dividing by $dt$ as $dt \to 0$ we obtain the differential equation
\begin{equation}
    \frac{dM_\sigma(t)}{dt} = \left[ i\lambda \sigma - \gamma \right] M_\sigma(t) + \gamma M_{-\sigma}(t) .
\end{equation}
We then obtain the solution 
\begin{equation}
    M_\sigma(t) = e^{-\gamma t} \left[ \cosh \Omega t + \frac{\gamma + i\sigma\lambda}{\Omega}\sinh \Omega t\right],
\end{equation}
with 
\begin{equation}
   \Omega = \sqrt{\gamma^2 - \lambda^2}. 
\end{equation}

Let us consider a Ramsey sequence of duration $\tau_R$. 
This will be obtained by setting $t=\tau_R$ and averaging over the initial conditions with equal probability, such that 
\begin{multline}
    \mathcal{C}_\lambda[{\rm Ramsey}, \tau_R] = \log e^{-\gamma \tau_R} \left[ \cosh \Omega \tau_R + \frac{\gamma }{\Omega}\sinh \Omega \tau_R\right] \\
    = -\gamma \tau_R+\log \left[ \cosh \Omega \tau_R + \frac{\gamma }{\Omega}\sinh \Omega \tau_R\right].
\end{multline}

We also need the result of the Hahn echo.
This can be obtained in a similar way, by computing the result also conditioned on final state. 
We introduce 
\begin{equation}
    Q_{\sigma',\sigma}(t) = \llangle e^{ i\lambda \int_0^t ds \sigma(s) }\delta[\sigma(t) = \sigma'] \rrangle\bigg|_{\sigma(0) = \sigma}.
\end{equation}
Again performing a short time expansion on the intial evolution of time $dt$ we get 
\begin{multline}
    Q_{\sigma',\sigma}(t+dt) = \llangle e^{ i\lambda \int_0^{t+dt} ds \sigma(s) }\delta[\sigma(t+dt) = \sigma'] \rrangle\bigg|_{\sigma(0) = \sigma} \\
    = e^{i\lambda \sigma dt} (1-\gamma dt) \llangle e^{ i\lambda \int_{dt}^{t+dt} ds \sigma(s) }\delta[\sigma(t+dt) = \sigma'] \rrangle\bigg|_{\sigma(dt) = \sigma} + \gamma dt e^{-i\lambda \sigma dt} \llangle e^{ i\lambda \int_{dt}^{t+dt} ds \sigma(s) }\delta[\sigma(t+dt) = \sigma'] \rrangle\bigg|_{\sigma(dt) = -\sigma}.
\end{multline}
We can shift the time $t \to t-dt$, bearing in mind the flipped initial condition on the second term.
This gives 
\begin{equation}
    Q_{\sigma',\sigma}(t+dt)  = e^{i\lambda \sigma dt} (1-\gamma dt) \llangle e^{ i\lambda \int_{0}^{t} ds \sigma(s) }\delta[\sigma(t) = \sigma'] \rrangle\bigg|_{\sigma(0) = \sigma} + \gamma dt e^{-i\lambda \sigma dt} \llangle e^{ i\lambda \int_{0}^{} ds \sigma(s) }\delta[\sigma(t) = \sigma'] \rrangle\bigg|_{\sigma(0) = -\sigma}.
\end{equation}
We can then again obtain the differential equation 
\begin{equation}
    \frac{d}{dt}Q_{\sigma',\sigma}(t)  = (i\lambda \sigma-\gamma ) Q_{\sigma',\sigma}(t)  + \gamma dt Q_{\sigma',-\sigma}(t).
\end{equation}
This is the same, except we have for each initial condition also two final conditions. 
These are obtained by enforcing the initial condition that 
\begin{equation}
    Q_{\sigma',\sigma}(0)  = \delta_{\sigma'=\sigma}.
\end{equation}
This solution in this case is 
\begin{equation}
    Q_{\sigma',\sigma}(t)  = e^{-\gamma t} \begin{pmatrix}
        \cosh \Omega t + i\frac{\lambda}{\Omega} \sinh\Omega t & \frac{\gamma}{\Omega} \sinh\Omega  \\
        \frac{\gamma}{\Omega} \sinh\Omega t & \cosh \Omega t - i\frac{\lambda}{\Omega} \sinh\Omega t \\
    \end{pmatrix}.
\end{equation}

For the Hahn echo, we have initial condition that $\sigma$ is averaged over with equal probability, but at the midpoint of the echo we impose a spin-flip.
This means that the overall amplitude is 
\begin{equation}
    \mathcal{C}_\lambda[{\rm Hahn}, \tau_E] = \log \sum_{\sigma,\sigma',\sigma''}\frac12 Q_{\sigma'',-\sigma'}(\tau_E)Q_{\sigma',\sigma}(\tau_E).
\end{equation}
We find 
\begin{equation}
    \mathcal{C}_\lambda[{\rm Hahn}, \tau_E] = \log e^{-2\gamma \tau_E }\left[ \left(\cosh \Omega \tau_E + \frac{\gamma}{\Omega }\sinh \Omega \tau_E \right)^2 + \left(\frac{\lambda}{\Omega }\sinh \Omega \tau_E \right)^2\right].
\end{equation}

Now, we combine the results for our non-Gaussian echo, which involves 
\begin{multline}
    \Gamma_{1,2}(\lambda) = \mathcal{C}_\lambda[{\rm Hahn}, \tau_R] + \mathcal{C}_\lambda[{\rm Ramsey}, 2\tau_R] - 4 \mathcal{C}_\lambda[{\rm Ramsey}, \tau_R]\\
    =\log \left[ \left(\cosh \Omega \tau_R + \frac{\gamma}{\Omega }\sinh \Omega \tau_R\right)^2 + \left(\frac{\lambda}{\Omega }\sinh \Omega \tau_E \right)^2\right] + \log \left[ \cosh 2\Omega \tau_R + \frac{\gamma }{\Omega}\sinh 2\Omega \tau_R\right]\\
    -4 \log \left[ \cosh \Omega \tau_R + \frac{\gamma }{\Omega}\sinh \Omega \tau_R\right].
\end{multline}
After some algebra we find 
\begin{equation}
\Gamma_{1,2}(\lambda) = \log \left[1 - \left( \frac{\lambda \tanh\Omega \tau_R}{\Omega +\gamma \tanh \Omega \tau_R } \right)^4 \right].
\end{equation}
We note the expansion to $O(\lambda^4)$ is given by 
\begin{equation}
\Gamma^{(4)}(\lambda) \sim  -(\lambda/\gamma)^4 \left( \frac{ \tanh\gamma \tau_R}{1+ \tanh \gamma \tau_R } \right)^4 .
\end{equation}

In the main text, we need to integrate over the coupling strength $\lambda$, in accordance with the expression 
\begin{equation}
\Gamma_{1,2} = 2\pi n_{\rm 2D}\int dr r\log\left[ 1 - \left( \frac{\lambda(r) \tanh \sqrt{\gamma^2 - \lambda^2(r)}\tau_R}{\sqrt{\gamma^2 - \lambda^2(r)} + \gamma \tanh \sqrt{\gamma^2 - \lambda^2(r)}\tau_R}\right)^4\right],
\end{equation}
where we have 
\begin{equation}
    \lambda(r) = \frac{\mu \mu_0 g\mu_B }{4\pi }\frac{3h^2/(h^2 +r^2) -1}{(r^2 + h^2)^{3/2}}.
\end{equation}
\end{widetext}

It is useful in evaluating this to define $(r/h)^2 + 1 = v$ and introduce unitless time $\theta = \tau_R \gamma$ and coupling parameter 
\begin{equation}
    \alpha = \frac{\mu \mu_0 g\mu_B }{2\pi\gamma h^3 }.
\end{equation}
We then obtain the integral 
\begin{multline}
\Gamma_{1,2} = \pi n_{\rm 2D}h^2\int_1^\infty dv \\
\log\left[ 1 - \left( \frac{\alpha f(v) \tanh[ \sqrt{1 - \alpha^2 f^2(v)}\theta]}{\sqrt{1 - \alpha^2 f^2(v)} +  \tanh[ \sqrt{1 - \alpha^2 f^2(v)}\theta ]}\right)^4\right],
\end{multline}
where 
\begin{equation}
    f(v) = \frac{3/v-1}{2v^{3/2}}.
\end{equation}
Evidently, when $\alpha = 1$ this encounters a branch cut in the integrand, signaling the potential emergence of non-perturbative phenomena. 
In principle, we find that perturbation theory is justified in terms of an expansion in $\alpha$ for 
\begin{equation}
    \alpha \ll 1 \Rightarrow  h \gg \left( \frac{\mu \mu_0 g\mu_B}{2\pi \gamma}\right)^{\frac13}. 
\end{equation}

\section{Ising Model Ginzburg-Landau}
\label{app:GL}
Here we present the details of the derivation of the Ginzburg-Landau coefficients from the two-dimensional Ising model. 
We start from the two-dimensional Ising model on a square lattice with partition function 
\begin{equation}
    Z = \sum_{\{S\}} \exp\left( \beta J \sum_{<j,k>}S^z_j S^z_k  + \beta \mu_B S \sum_j S^z_j h_j \right),
\end{equation}
where the Ising spins satisfy $(S^z_j)^2 = 1 $ and the external field $h_j$ couples via the spin moment $\mu_B S$.
We will employ a Hubbard-Stratonovich transformation to express this in terms of an effective energy for the order parameter. 
First, we introduce a Gaussian integral over a real field $m_{\bf q}$ in momentum space such that up to constants 
\begin{widetext}
\begin{equation}
    Z = \sum_{\{S\}}\int \mathcal{D}[m] \exp\left( \beta J \sum_{<j,k>}S^z_j S^z_k  - \frac{1}{2\beta J} \sum_{\bf q} C(\mathbf{q}) m_{\bf q}m_{-\bf q}+ \beta \mu_B S \sum_j S^z_j h_j  \right).
\end{equation}    
\end{widetext}

We will still need to choose an appropriate $C(\mathbf{q})$. 
Now, we can shift in our integral the field $m_{\bf q}$ by 
\begin{equation}
    m_{\bf q} \to m_{\bf q} + \beta J \frac{1}{C(\mathbf{q})} \sum_j e^{-i\mathbf{q}\cdot\mathbf{R}_j }S^z_j.
\end{equation}
This will yield 
\begin{widetext}
\begin{multline}
    Z = \sum_{\{S\}}\int \mathcal{D}[m] \\
    \exp\left( \beta J \sum_{<j,k>}S^z_j S^z_k  - \frac{1}{2\beta J} \sum_{\bf q} C(\mathbf{q}) m_{\bf q}m_{-\bf q} - \sum_{j}  m_j S^z_j  - \frac12 \beta J \sum_{jk\bf q} C^{-1}(\mathbf{q})e^{i\mathbf{q}\cdot(\mathbf{R}_j-\mathbf{R}_k)}S^z_jS^z_k+ \beta \mu_B S \sum_j S^z_j h_j \right).
\end{multline}   
\end{widetext}
We now see that if we choose $C(\mathbf{q})$ such that 
\begin{equation}
    \beta J \sum_{<j,k>}S^z_j S^z_k  - \frac12 \beta J \sum_{jk\bf q} C^{-1}(\mathbf{q})e^{i\mathbf{q}\cdot(\mathbf{R}_j-\mathbf{R}_k)}S^z_jS^z_k  = 0,
\end{equation}  
we will be able to decouple the spin-spin interaction and integrate out the spins in favor of the order parameter field. 
Up to a constant (due to the $\mathbb{Z}_2$ nature of the spins), we see this is satisfied by 
\begin{equation}
    C^{-1}(\mathbf{q}) = \cos q_x + \cos q_y + 2 . 
\end{equation}
The constant ensures that this is never zero, and can be added without penalty as $(S^z_j)^2 = $const. can be discarded from the free energy.
Now, we can integrate out the spins via 
\begin{multline}
    Z = \sum_{\{S\}}\exp\left( \beta \mu_B S \sum_j S^z_j h_j- \sum_{j}  m_j S^z_j  \right)  \\
    = \prod_j 2 \cosh (m_j-\beta \mu_B S h_j).
\end{multline}   
This is a local energy functional, and therefore can be evaluated. 
We then find the {\bf exact} effective free-energy for the order parameter of 
\begin{widetext}
\begin{equation}
    F_{\rm eff}  = \frac{T^2}{2J} \sum_{\bf q} \frac{1}{2 + \cos q_x + \cos q_y} m_{\bf q}m_{-\bf q} - T \sum_j \log\cosh( m_j - \beta \mu_B S h_j)  .
\end{equation} 
We now expand in the order parameter up to quartic order in $m$, and up to quadratic order in spatial derivatives, and linear order in external field to obtain 
\begin{equation}
    F_{\rm eff}  = \frac12 \sum_{\bf q} \left[ \frac{T^2}{4J} + \frac{T^2\mathbf{q}^2}{32J}  \right]m_{\bf q}m_{-\bf q} + \sum_j -\frac{T}{2}m_j^2   + \frac{T}{12} m_j^4 + \mu_B S m_j h_j.
\end{equation} 
Replacing the lattice constant $a$ this corresponds to a continuum free energy of 
\begin{equation}
    F_{\rm eff}  = \int d^2 r  \left[ \frac12\frac{T_C}{8} (\nabla m)^2  + \frac12\frac{(T - T_C)}{a^2} m^2 + \frac{\mu_B S}{a^2} m h   +\frac14  \frac{T_C}{3a^2} m^4 \right].
\end{equation} 
\end{widetext}
This is in terms of the mean-field critical temperature 
\begin{equation}
    T_C = 4J.
\end{equation} 
The derivative with respect to external field identifies the magnetization density as $\mu_B S /a^2 \times m(\mathbf{r},t)$.
We find the Ginzburg-Landau coefficients of 
\begin{subequations}
    \begin{align}
        & K =\frac{T_C}{8} \\
        & r =\frac{T- T_C}{a^2}\\
        & u = \frac{T_C}{3a^2} . 
    \end{align}
\end{subequations}
Then, to leading order in $T-T_C$ we find the prefactor of the magnetic noise scales as 
\begin{multline}
12 u \left(\frac{ 2\Gamma T}{K}\right)^3 \left( \frac{g S\mu_B^2\mu_0}{2Ka^2} \right)^4  \tau_c \xi_c^{4} \\
= 4\times 8  \times 4^4  T_C^3 \frac{T_C}{a^2}\left( \frac{gS\mu_B^2\mu_0}{T_C a^2}\right)^4  \tau_c^4 \xi_c^{-2}\\
= 2^{13} \frac{\tau_c^4}{(\xi_c/a)^2} \left(\frac{gS\mu_B^2\mu_0}{a^3}\right)^4  
\end{multline}

\section{Model A Integration}
\label{app:modelA}

Here we present the details on the evaluation of the integrals for the noise for the model A relaxational dynamics.
The noise cumulant to be calculated can be written as 
\begin{equation}
    \Gamma^{(4)} = -12 u \left(\frac{ 2\Gamma T}{K}\right)^3 \left( \frac{g S\mu_B^2\mu_0}{2Ka^2} \right)^4  \tau_c \xi_c^{4}F(z/\xi_c,\tau_R/\tau_c),
\end{equation}
where the unitless function is expressed as an integral 
\begin{multline}
F(x,s) \equiv \frac{1}{(2\pi)^9} \frac{s}{x^{10}} \int d^2 \tilde{q}_1 d^2 \tilde{q}_2 d^2 \tilde{q}_3\int d\tilde{\omega}_1 d\tilde{\omega}_2d \tilde{\omega}_3 \\
\tilde{q}_1  e^{-\tilde{q}_1}\tilde{q}_2  e^{-\tilde{q}_2}\tilde{q}_3  e^{-\tilde{q}_3} \tilde{Q} e^{- \tilde{Q}}\\
\textrm{sinc}(\tilde{\omega}_1/2)\textrm{sinc}(\tilde{\omega}_2/2)\textrm{sinc}(\tilde{\omega}_3/2)\textrm{sinc}(\tilde{\Omega}/2) \\
\frac13 \sum_{j<k=1}^{3} \cos(\tilde{\omega}_j + \tilde{\omega}_k) \\
 \frac{1 + \tilde{Q}^2/x^2}{ ( \tilde{\Omega}/s)^2 + ( 1 + \tilde{Q}^2/x^2 )^2}\prod_{j=1}^{3} \frac{1}{ ( \tilde{\omega}_j/s)^2 + ( 1 + \tilde{q}_j^2/x^2 )^2}.
\end{multline}

This integral is naively nine dimensional and oscillatory, making it very poorly behaved for numerical integration.
In particular, we are interested in the behavior in the regime of $z \lesssim \xi_c $ to study the effects on nonlocal correlations.
In this work we will be content in evaluating this within the regime of $\tau_R \ll \tau_c$, although studying the effects of nonlocal correlations in time and space both would be interesting for future works. 
In this regime we can first evaluate the frequency integrals by writing them as 
\begin{multline}
\frac{1}{(2\pi)^3} s \int d\tilde{\omega}_1 d\tilde{\omega}_2d \tilde{\omega}_3 \\
\textrm{sinc}(\tilde{\omega}_1/2)\textrm{sinc}(\tilde{\omega}_2/2)\textrm{sinc}(\tilde{\omega}_3/2)\textrm{sinc}(\tilde{\Omega}/2) \\
\frac13 \sum_{j<k=1}^{3} \cos(\tilde{\omega}_j + \tilde{\omega}_k) \\
 \frac{1 + \tilde{Q}^2/x^2}{ ( \tilde{\Omega}/s)^2 + ( 1 + \tilde{Q}^2/x^2 )^2}\prod_{j=1}^{3} \frac{1}{ ( \tilde{\omega}_j/s)^2 + ( 1 + \tilde{q}_j^2/x^2 )^2} \\
 = \frac{s^4}{(2\pi)^4} \int d{\omega}_1 d{\omega}_2d{\omega}_3 d\omega_4  \int dt e^{-i\sum_j \omega_j t}\\
\textrm{sinc}(s{\omega}_1/2)\textrm{sinc}(s{\omega}_2/2)\textrm{sinc}(s{\omega}_3/2)\textrm{sinc}(s{\omega}_4/2) \\
\frac16 \sum_{j<k=1}^{4} \cos(s{\omega}_j + s{\omega}_k) \\
 \frac{1 + \tilde{Q}^2/x^2}{ ( {\omega_4})^2 + ( 1 + \tilde{Q}^2/x^2 )^2}\prod_{j=1}^{3} \frac{1}{ ( {\omega}_j)^2 + ( 1 + \tilde{q}_j^2/x^2 )^2} .
\end{multline}
If we take the limit of $s \to 0$ we find the leading contribution grows as $s^4$ at short times, with behavior 
\begin{multline}
\frac{1}{(2\pi)^3} s \int d\tilde{\omega}_1 d\tilde{\omega}_2d \tilde{\omega}_3 \\
\textrm{sinc}(\tilde{\omega}_1/2)\textrm{sinc}(\tilde{\omega}_2/2)\textrm{sinc}(\tilde{\omega}_3/2)\textrm{sinc}(\tilde{\Omega}/2) \\
\frac13 \sum_{j<k=1}^{3} \cos(\tilde{\omega}_j + \tilde{\omega}_k) \\
 \frac{1 + \tilde{Q}^2/x^2}{ ( \tilde{\Omega}/s)^2 + ( 1 + \tilde{Q}^2/x^2 )^2}\prod_{j=1}^{3} \frac{1}{ ( \tilde{\omega}_j/s)^2 + ( 1 + \tilde{q}_j^2/x^2 )^2} \\
 = \frac{s^4}{(2\pi)^4} \int d{\omega}_1 d{\omega}_2d{\omega}_3 d\omega_4  \int dt e^{-i\sum_j \omega_j t} \\
\frac{1 + \tilde{Q}^2/x^2}{ ( {\omega_4})^2 + ( 1 + \tilde{Q}^2/x^2 )^2}\prod_{j=1}^{3} \frac{1}{ ( {\omega}_j)^2 + ( 1 + \tilde{q}_j^2/x^2 )^2} .
\end{multline}
We can now evaluate the frequency integrals as 
\[
\int d\omega \frac{1}{\omega^2 + \gamma^2} e^{-i\omega t} = \frac{\pi}{\gamma}e^{-\gamma|t|},
\]
so that 
\begin{multline}
\frac{1}{(2\pi)^3} s \int d\tilde{\omega}_1 d\tilde{\omega}_2d \tilde{\omega}_3 \\
\textrm{sinc}(\tilde{\omega}_1/2)\textrm{sinc}(\tilde{\omega}_2/2)\textrm{sinc}(\tilde{\omega}_3/2)\textrm{sinc}(\tilde{\Omega}/2) \\
\frac13 \sum_{j<k=1}^{3} \cos(\tilde{\omega}_j + \tilde{\omega}_k) \\
 \frac{1 + \tilde{Q}^2/x^2}{ ( \tilde{\Omega}/s)^2 + ( 1 + \tilde{Q}^2/x^2 )^2}\prod_{j=1}^{3} \frac{1}{ ( \tilde{\omega}_j/s)^2 + ( 1 + \tilde{q}_j^2/x^2 )^2} \\
 = \frac{s^4}{16} \int dt   e^{- \sum_j (1 + q_j^2/x^2)|t| } \prod_{j=1}^{3} \frac{1}{ 1 + \tilde{q}_j^2/x^2 } \\
 = \frac{s^4}{8} \frac{1}{\sum_{j=1}^{4} (1 + q_j^2/x^2) } \prod_{j=1}^{3} \frac{1}{ 1 + \tilde{q}_j^2/x^2 } .
\end{multline}
This then yields the momentum integrals 
\begin{multline}
F(x,s) \sim \frac{1}{(2\pi)^6} \frac{s^4}{x^{10}} \int d^2 \tilde{q}_1 d^2 \tilde{q}_2 d^2 \tilde{q}_3\\
\tilde{q}_1  e^{-\tilde{q}_1}\tilde{q}_2  e^{-\tilde{q}_2}\tilde{q}_3  e^{-\tilde{q}_3} \tilde{Q} e^{- \tilde{Q}}\\
\frac{1}{8} \frac{1}{\sum_{j=1}^{4} (1 + q_j^2/x^2) } \prod_{j=1}^{3} \frac{1}{ 1 + \tilde{q}_j^2/x^2 }
\end{multline}
Here we have labeled $q_4 = Q$.

This is still a high-dimensional integral; we can however relatively easily compute this using importance sampling as follows. 
First, we sample $N$ momentum points, which are 3 2-dimensional vectors each of which is sampled according to a distribution with 
\begin{equation}
    dP(\mathbf{q}) \propto dq d\theta q^2 e^{-q}. 
\end{equation}
We then find 
\begin{multline}
F(x,s) \sim \frac{1}{(2\pi)^6} \frac{s^4}{x^{10}} A \frac{1}{N}\sum_{j=1}^{N}  \tilde{Q} e^{- \tilde{Q}}\\
\frac{1}{8} \frac{1}{\sum_{j=1}^{4} (1 + q_j^2/x^2) } \prod_{j=1}^{3} \frac{1}{ 1 + \tilde{q}_j^2/x^2 }
\end{multline}
This then fixes the normalization constant $A$ such that 
\begin{equation}
    A = (4\pi)^3. 
\end{equation}
We then find the overall noise, in terms of the Ginzburg-Landau estimates and the sampling points as 
\begin{multline}
    \Gamma^{(4)} = - \frac{2^{10} }{\pi^3} \frac{1}{(\xi_c/a)^2} \left(\frac{\tau_c gS \mu_B^2 \mu_0}{a^3}\right)^4 \frac{s^4}{x^{10}} \\
    \frac{1}{N} \sum_{j=1}^{N}  \tilde{Q} e^{- \tilde{Q}}\\
    \frac{1}{\sum_{j=1}^{4} (1 + q_j^2/x^2) } \prod_{j=1}^{3} \frac{1}{ 1 + \tilde{q}_j^2/x^2 }.
\end{multline}
Let us introduce the ``dipole coupling" time 
\begin{equation}
    \tau_{\rm dp} =  a^3/(g S \mu_B^2 \mu_0).
\end{equation}
For $g = S = 1$ and a lattice constant of $5\AA$ we get times of order $\tau_{\rm dp} \sim 100 $ps.

\bibliographystyle{quantum}
\bibliography{refs}

\begin{thebibliography}{100}

\bibitem{Rovny.2024}
Jared Rovny, Sarang Gopalakrishnan, Ania C~Bleszynski Jayich, Patrick
  Maletinsky, Eugene Demler, and Nathalie P~de Leon.
\newblock ``{New opportunities in condensed matter physics for nanoscale
  quantum sensors}''~(2024).
\newblock  \href{http://arxiv.org/abs/2403.13710}{arXiv:2403.13710}.

\bibitem{Curtis.2024}
Jonathan~B. Curtis, Nikola Maksimovic, Nicholas~R. Poniatowski, Amir Yacoby,
  Bertrand Halperin, Prineha Narang, and Eugene Demler.
\newblock ``{Probing the Berezinskii-Kosterlitz-Thouless vortex unbinding
  transition in two-dimensional superconductors using local noise
  magnetometry}''.
\newblock \href{https://dx.doi.org/10.1103/physrevb.110.144518}{Physical Review
  B {\bf 110}, 144518}~(2024).
\newblock  \href{http://arxiv.org/abs/2404.06147}{arXiv:2404.06147}.

\bibitem{Dolgirev.2022}
Pavel~E. Dolgirev, Shubhayu Chatterjee, Ilya Esterlis, Alexander~A. Zibrov,
  Mikhail~D. Lukin, Norman~Y. Yao, and Eugene Demler.
\newblock ``{Characterizing two-dimensional superconductivity via nanoscale
  noise magnetometry with single-spin qubits}''.
\newblock \href{https://dx.doi.org/10.1103/physrevb.105.024507}{Physical Review
  B {\bf 105}, 024507}~(2022).
\newblock  \href{http://arxiv.org/abs/2106.05283}{arXiv:2106.05283}.

\bibitem{Chatterjee.2022}
Shubhayu Chatterjee, Pavel~E. Dolgirev, Ilya Esterlis, Alexander~A. Zibrov,
  Mikhail~D. Lukin, Norman~Y. Yao, and Eugene Demler.
\newblock ``{Single-spin qubit magnetic spectroscopy of two-dimensional
  superconductivity}''.
\newblock \href{https://dx.doi.org/10.1103/physrevresearch.4.l012001}{Physical
  Review Research {\bf 4}, L012001}~(2022).
\newblock  \href{http://arxiv.org/abs/2106.03859}{arXiv:2106.03859}.

\bibitem{Kelly.2024}
Shane~P Kelly and Yaroslav Tserkovnyak.
\newblock ``{Superconductivity-enhanced magnetic field noise}''~(2024).
\newblock  \href{http://arxiv.org/abs/2412.05465}{arXiv:2412.05465}.

\bibitem{Konig.2020}
E~J K{\"o}nig, Piers Coleman, and A~M Tsvelik.
\newblock ``{Spin magnetometry as a probe of stripe superconductivity in
  twisted bilayer graphene}''.
\newblock \href{https://dx.doi.org/10.1103/PhysRevB.102.104514}{Phys. Rev. B
  {\bf 102}, 104514}~(2020).
\newblock  \href{http://arxiv.org/abs/2006.10684}{arXiv:2006.10684}.

\bibitem{Liu.2025}
Zhongyuan Liu, Ruotian Gong, Jaewon Kim, Oriana~K Diessel, Qiaozhi Xu, Zack
  Rehfuss, Xinyi Du, Guanghui He, Abhishek Singh, Yun~Suk Eo, Erik~A Henriksen,
  G~D Gu, Norman~Y Yao, Francisco Machado, Sheng Ran, Shubhayu Chatterjee, and
  Chong Zu.
\newblock ``{Quantum noise spectroscopy of superconducting critical dynamics
  and vortex fluctuations in a high-temperature cuprate}''~(2025).
\newblock  \href{http://arxiv.org/abs/2502.04439}{arXiv:2502.04439}.

\bibitem{Khoo.2022}
Jun~Yong Khoo, Falko Pientka, Patrick~A. Lee, and Inti~Sodemann Villadiego.
\newblock ``{Probing the quantum noise of the spinon Fermi surface with NV
  centers}''.
\newblock \href{https://dx.doi.org/10.1103/physrevb.106.115108}{Physical Review
  B {\bf 106}, 115108}~(2022).
\newblock  \href{http://arxiv.org/abs/2205.06822}{arXiv:2205.06822}.

\bibitem{De.2024}
Suman~Jyoti De, Tami Pereg-Barnea, and Kartiek Agarwal.
\newblock ``{Nanoscale defects as probes of time reversal symmetry
  breaking}''~(2024).
\newblock  \href{http://arxiv.org/abs/2406.14648}{arXiv:2406.14648}.

\bibitem{Dolgirev.2023}
Pavel~E Dolgirev, Ilya Esterlis, Alexander~A Zibrov, Mikhail~D Lukin, Thierry
  Giamarchi, and Eugene Demler.
\newblock ``{Local noise spectroscopy of Wigner crystals in two-dimensional
  materials}''.
\newblock \href{https://dx.doi.org/10.1103/PhysRevLett.132.246504}{Phys. Rev.
  Lett. {\bf 132}, 246504}~(2024).

\bibitem{Zhang.2022o8f}
Shu Zhang and Yaroslav Tserkovnyak.
\newblock ``{Flavors of magnetic noise in quantum materials}''.
\newblock \href{https://dx.doi.org/10.1103/physrevb.106.l081122}{Physical
  Review B {\bf 106}, L081122}~(2022).
\newblock  \href{http://arxiv.org/abs/2108.07305}{arXiv:2108.07305}.

\bibitem{Quan.2005}
H.~T. Quan, Z.~Song, X.~F. Liu, P.~Zanardi, and C.~P. Sun.
\newblock ``{Decay of Loschmidt Echo Enhanced by Quantum Criticality}''.
\newblock \href{https://dx.doi.org/10.1103/physrevlett.96.140604}{Physical
  Review Letters {\bf 96}, 140604}~(2006).
\newblock
  \href{http://arxiv.org/abs/quant-ph/0509007}{arXiv:quant-ph/0509007}.

\bibitem{Wei.2012}
Bo-Bo Wei and Ren-Bao Liu.
\newblock ``{Lee-Yang Zeros and Critical Times in Decoherence of a Probe Spin
  Coupled to a Bath}''.
\newblock \href{https://dx.doi.org/10.1103/physrevlett.109.185701}{Physical
  Review Letters {\bf 109}, 185701}~(2012).
\newblock  \href{http://arxiv.org/abs/1206.2077}{arXiv:1206.2077}.

\bibitem{Chen.2013}
Shao-Wen Chen, Zhan-Feng Jiang, and Ren-Bao Liu.
\newblock ``{Quantum criticality at high temperature revealed by spin echo}''.
\newblock \href{https://dx.doi.org/10.1088/1367-2630/15/4/043032}{New Journal
  of Physics {\bf 15}, 043032}~(2013).
\newblock  \href{http://arxiv.org/abs/1202.4958}{arXiv:1202.4958}.

\bibitem{Machado.2022}
Francisco Machado, Eugene~A Demler, Norman~Y Yao, and Shubhayu Chatterjee.
\newblock ``{Quantum noise spectroscopy of dynamical critical phenomena}''.
\newblock \href{https://dx.doi.org/10.1103/PhysRevLett.131.070801}{Phys. Rev.
  Lett. {\bf 131}, 070801}~(2023).

\bibitem{Agarwal.2017}
Kartiek Agarwal, Richard Schmidt, Bertrand Halperin, Vadim Oganesyan, Gergely
  Zar{\'a}nd, Mikhail~D. Lukin, and Eugene Demler.
\newblock ``{Magnetic noise spectroscopy as a probe of local electronic
  correlations in two-dimensional systems}''.
\newblock \href{https://dx.doi.org/10.1103/physrevb.95.155107}{Physical Review
  B {\bf 95}, 155107}~(2017).
\newblock  \href{http://arxiv.org/abs/1608.03278}{arXiv:1608.03278}.

\bibitem{Rodriguez-Nieva.2018}
Joaquin~F. Rodriguez-Nieva, Kartiek Agarwal, Thierry Giamarchi, Bertrand~I.
  Halperin, Mikhail~D. Lukin, and Eugene Demler.
\newblock ``{Probing one-dimensional systems via noise magnetometry with single
  spin qubits}''.
\newblock \href{https://dx.doi.org/10.1103/physrevb.98.195433}{Physical Review
  B {\bf 98}, 195433}~(2018).
\newblock  \href{http://arxiv.org/abs/1803.01521}{arXiv:1803.01521}.

\bibitem{Zhang.2024}
Yifan Zhang, Rhine Samajdar, and Sarang Gopalakrishnan.
\newblock ``{Nanoscale sensing of spatial correlations in nonequilibrium
  current noise}''~(2024).
\newblock  \href{http://arxiv.org/abs/2404.15398}{arXiv:2404.15398}.

\bibitem{Kolkowitz.2015}
S.~Kolkowitz, A.~Safira, A.~A. High, R.~C. Devlin, S.~Choi, Q.~P.
  Unterreithmeier, D.~Patterson, A.~S. Zibrov, V.~E. Manucharyan, H.~Park, and
  M.~D. Lukin.
\newblock ``{Probing Johnson noise and ballistic transport in normal metals
  with a single-spin qubit}''.
\newblock \href{https://dx.doi.org/10.1126/science.aaa4298}{Science {\bf 347},
  1129--1132}~(2015).

\bibitem{McLaughlin.2022}
Nathan~J. McLaughlin, Chaowei Hu, Mengqi Huang, Shu Zhang, Hanyi Lu, Gerald~Q.
  Yan, Hailong Wang, Yaroslav Tserkovnyak, Ni~Ni, and Chunhui~Rita Du.
\newblock ``{Quantum Imaging of Magnetic Phase Transitions and Spin
  Fluctuations in Intrinsic Magnetic Topological Nanoflakes}''.
\newblock \href{https://dx.doi.org/10.1021/acs.nanolett.2c01390}{Nano Letters
  {\bf 22}, 5810--5817}~(2022).
\newblock  \href{http://arxiv.org/abs/2112.09863}{arXiv:2112.09863}.

\bibitem{Xue.2024}
Ruolan Xue, Nikola Maksimovic, Pavel~E Dolgirev, Li-Qiao Xia, Aaron M{\"u}ller,
  Ryota Kitagawa, Francisco Machado, Dahlia~R Klein, David MacNeill, Kenji
  Watanabe, Takashi Taniguchi, Pablo Jarillo-Herrero, Mikhail~D Lukin, Eugene
  Demler, and Amir Yacoby.
\newblock ``{Magnon hydrodynamics in an atomically-thin ferromagnet}''.
\newblock \href{https://dx.doi.org/10.1126/science.adp2397}{Science {\bf 392},
  873}~(2026).

\bibitem{Ziffer.2024}
Mark~E Ziffer, Francisco Machado, Benedikt Ursprung, Artur Lozovoi, Aya~Batoul
  Tazi, Zhiyang Yuan, Michael~E Ziebel, Tom Delord, Nanyu Zeng, Evan Telford,
  Daniel~G Chica, Dane~W deQuilettes, Xiaoyang Zhu, James~C Hone, Kenneth~L
  Shepard, Xavier Roy, Nathalie P~de Leon, Emily~J Davis, Shubhayu Chatterjee,
  Carlos~A Meriles, Jonathan~S Owen, P~James Schuck, and Abhay~N Pasupathy.
\newblock ``{Quantum Noise Spectroscopy of Critical Slowing Down in an
  Atomically Thin Magnet}''~(2024).
\newblock  \href{http://arxiv.org/abs/2407.05614}{arXiv:2407.05614}.

\bibitem{Andersen.2019}
Trond~I. Andersen, Bo~L. Dwyer, Javier~D. Sanchez-Yamagishi, Joaquin~F.
  Rodriguez-Nieva, Kartiek Agarwal, Kenji Watanabe, Takashi Taniguchi,
  Eugene~A. Demler, Philip Kim, Hongkun Park, and Mikhail~D. Lukin.
\newblock ``{Electron-phonon instability in graphene revealed by global and
  local noise probes}''.
\newblock \href{https://dx.doi.org/10.1126/science.aaw2104}{Science {\bf 364},
  154--157}~(2019).

\bibitem{Rovny.2022}
Jared Rovny, Zhiyang Yuan, Mattias Fitzpatrick, Ahmed~I. Abdalla, Laura
  Futamura, Carter Fox, Matthew~Carl Cambria, Shimon Kolkowitz, and Nathalie
  P.~de Leon.
\newblock ``{Nanoscale covariance magnetometry with diamond quantum sensors}''.
\newblock \href{https://dx.doi.org/10.1126/science.ade9858}{Science {\bf 378},
  1301--1305}~(2022).
\newblock  \href{http://arxiv.org/abs/2209.08703}{arXiv:2209.08703}.

\bibitem{Huxter.2024}
William~S Huxter, Federico Dalmagioni, and Christian~L Degen.
\newblock ``{Multiplexed scanning microscopy with dual-qubit spin sensors}''.
\newblock \href{https://dx.doi.org/10.1103/y14n-44h3}{Phys. Rev. Lett. {\bf
  135}, 153801}~(2025).

\bibitem{Ji.2024}
Wentao Ji, Zhaoxin Liu, Yuhang Guo, Zhihao Hu, Jingyang Zhou, Siheng Dai,
  Yu~Chen, Pei Yu, Mengqi Wang, Kangwei Xia, Fazhan Shi, Ya~Wang, and Jiangfeng
  Du.
\newblock ``{Correlated sensing with a solid-state quantum multisensor system
  for atomic-scale structural analysis}''.
\newblock \href{https://dx.doi.org/10.1038/s41566-023-01352-4}{Nature Photonics
  {\bf 18}, 230--235}~(2024).

\bibitem{Li.2025}
Xin Li, Jamir Marino, Darrick~E Chang, and Benedetta Flebus.
\newblock ``{Solid-state platform for cooperative quantum dynamics driven by
  correlated emission}''.
\newblock \href{https://dx.doi.org/10.1103/physrevb.111.064424}{Phys. Rev. B
  {\bf 111}, 064424}~(2025).
\newblock  \href{http://arxiv.org/abs/2309.08991}{arXiv:2309.08991}.

\bibitem{Choi.2017}
Soonwon Choi, Joonhee Choi, Renate Landig, Georg Kucsko, Hengyun Zhou, Junichi
  Isoya, Fedor Jelezko, Shinobu Onoda, Hitoshi Sumiya, Vedika Khemani, Curt von
  Keyserlingk, Norman~Y. Yao, Eugene Demler, and Mikhail~D. Lukin.
\newblock ``{Observation of discrete time-crystalline order in a disordered
  dipolar many-body system}''.
\newblock \href{https://dx.doi.org/10.1038/nature21426}{Nature {\bf 543},
  221---225}~(2017).
\newblock  \href{http://arxiv.org/abs/1610.08057}{arXiv:1610.08057}.

\bibitem{Davis.2023}
E.~J. Davis, B.~Ye, F.~Machado, S.~A. Meynell, W.~Wu, T.~Mittiga, W.~Schenken,
  M.~Joos, B.~Kobrin, Y.~Lyu, Z.~Wang, D.~Bluvstein, S.~Choi, C.~Zu,
  A.~C.~Bleszynski Jayich, and N.~Y. Yao.
\newblock ``{Probing many-body dynamics in a two-dimensional dipolar spin
  ensemble}''.
\newblock \href{https://dx.doi.org/10.1038/s41567-023-01944-5}{Nature Physics
  {\bf 19}, 836--844}~(2023).
\newblock  \href{http://arxiv.org/abs/2103.12742}{arXiv:2103.12742}.

\bibitem{Hughes.2024}
Lillian~B Hughes, Simon~A Meynell, Weijie Wu, Shreyas Parthasarathy, Lingjie
  Chen, Zhiran Zhang, Zilin Wang, Emily~J Davis, Kunal Mukherjee, Norman~Y Yao,
  and Ania C~Bleszynski Jayich.
\newblock ``{Strongly interacting, two-dimensional, dipolar spin ensemble in
  (111)-oriented diamond}''.
\newblock \href{https://dx.doi.org/10.1103/PhysRevX.15.021035}{Phys. Rev. X
  {\bf 15}, 021035}~(2025).
\newblock  \href{http://arxiv.org/abs/2404.10075}{arXiv:2404.10075}.

\bibitem{Zhou.2020lzi}
Hengyun Zhou, Joonhee Choi, Soonwon Choi, Renate Landig, Alexander~M. Douglas,
  Junichi Isoya, Fedor Jelezko, Shinobu Onoda, Hitoshi Sumiya, Paola
  Cappellaro, Helena~S. Knowles, Hongkun Park, and Mikhail~D. Lukin.
\newblock ``{Quantum Metrology with Strongly Interacting Spin Systems}''.
\newblock \href{https://dx.doi.org/10.1103/physrevx.10.031003}{Physical Review
  X {\bf 10}, 031003}~(2020).
\newblock  \href{http://arxiv.org/abs/1907.10066}{arXiv:1907.10066}.

\bibitem{Bramwell.2009}
Steven~T. Bramwell.
\newblock ``{The distribution of spatially averaged critical properties}''.
\newblock \href{https://dx.doi.org/10.1038/nphys1268}{Nature Physics {\bf 5},
  444--447}~(2009).

\bibitem{Brown.1954}
R.~Hanbury Brown and R.Q. Twiss.
\newblock ``{A new type of interferometer for use in radio astronomy}''.
\newblock \href{https://dx.doi.org/10.1080/14786440708520475}{Philosophical
  Magazine {\bf 45}, 663--682}~(1954).

\bibitem{Kimble.1977}
H.~J. Kimble, M.~Dagenais, and L.~Mandel.
\newblock ``{Photon Antibunching in Resonance Fluorescence}''.
\newblock \href{https://dx.doi.org/10.1103/physrevlett.39.691}{Physical Review
  Letters {\bf 39}, 691--695}~(1977).

\bibitem{Folling.2005}
Simon F{\"o}lling, Fabrice Gerbier, Artur Widera, Olaf Mandel, Tatjana Gericke,
  and Immanuel Bloch.
\newblock ``{Spatial quantum noise interferometry in expanding ultracold atom
  clouds}''.
\newblock \href{https://dx.doi.org/10.1038/nature03500}{Nature {\bf 434},
  481--484}~(2005).
\newblock
  \href{http://arxiv.org/abs/cond-mat/0503587}{arXiv:cond-mat/0503587}.

\bibitem{Rom.2006}
T.~Rom, Th. Best, D.~van Oosten, U.~Schneider, S.~F{\"o}lling, B.~Paredes, and
  I.~Bloch.
\newblock ``{Free fermion antibunching in a degenerate atomic Fermi gas
  released from an optical lattice}''.
\newblock \href{https://dx.doi.org/10.1038/nature05319}{Nature {\bf 444},
  733--736}~(2006).
\newblock
  \href{http://arxiv.org/abs/cond-mat/0611561}{arXiv:cond-mat/0611561}.

\bibitem{Altman.2004}
Ehud Altman, Eugene Demler, and Mikhail~D. Lukin.
\newblock ``{Probing many-body states of ultracold atoms via noise
  correlations}''.
\newblock \href{https://dx.doi.org/10.1103/physreva.70.013603}{Physical Review
  A {\bf 70}, 013603}~(2004).
\newblock
  \href{http://arxiv.org/abs/cond-mat/0306226}{arXiv:cond-mat/0306226}.

\bibitem{Greiner.2005}
M.~Greiner, C.~A. Regal, J.~T. Stewart, and D.~S. Jin.
\newblock ``{Probing Pair-Correlated Fermionic Atoms through Correlations in
  Atom Shot Noise}''.
\newblock \href{https://dx.doi.org/10.1103/physrevlett.94.110401}{Physical
  Review Letters {\bf 94}, 110401}~(2005).
\newblock
  \href{http://arxiv.org/abs/cond-mat/0502411}{arXiv:cond-mat/0502411}.

\bibitem{Jeltes.2007}
T.~Jeltes, J.~M. McNamara, W.~Hogervorst, W.~Vassen, V.~Krachmalnicoff,
  M.~Schellekens, A.~Perrin, H.~Chang, D.~Boiron, A.~Aspect, and C.~I.
  Westbrook.
\newblock ``{Comparison of the Hanbury Brown-Twiss effect for bosons and
  fermions}''.
\newblock \href{https://dx.doi.org/10.1038/nature05513}{Nature {\bf 445},
  402--405}~(2007).
\newblock
  \href{http://arxiv.org/abs/cond-mat/0612278}{arXiv:cond-mat/0612278}.

\bibitem{Hofferberth.2008}
S.~Hofferberth, I.~Lesanovsky, T.~Schumm, A.~Imambekov, V.~Gritsev, E.~Demler,
  and J.~Schmiedmayer.
\newblock ``{Probing quantum and thermal noise in an interacting many-body
  system}''.
\newblock \href{https://dx.doi.org/10.1038/nphys941}{Nature Physics {\bf 4},
  489--495}~(2008).
\newblock  \href{http://arxiv.org/abs/0710.1575}{arXiv:0710.1575}.

\bibitem{Schweigler.2021}
Thomas Schweigler, Marek Gluza, Mohammadamin Tajik, Spyros Sotiriadis, Federica
  Cataldini, Si-Cong Ji, Frederik~S. M{\o}ller, Jo{\~a}o Sabino, Bernhard
  Rauer, Jens Eisert, and J{\"o}rg Schmiedmayer.
\newblock ``{Decay and recurrence of non-Gaussian correlations in a quantum
  many-body system}''.
\newblock \href{https://dx.doi.org/10.1038/s41567-020-01139-2}{Nature Physics
  {\bf 17}, 559--563}~(2021).
\newblock  \href{http://arxiv.org/abs/2003.01808}{arXiv:2003.01808}.

\bibitem{Levitov.1996}
Leonid~S. Levitov, Hyunwoo Lee, and Gordey~B. Lesovik.
\newblock ``{Electron counting statistics and coherent states of electric
  current}''.
\newblock \href{https://dx.doi.org/10.1063/1.531672}{Journal of Mathematical
  Physics {\bf 37}, 4845--4866}~(1996).
\newblock
  \href{http://arxiv.org/abs/cond-mat/9607137}{arXiv:cond-mat/9607137}.

\bibitem{Ubbelohde.2012}
Niels Ubbelohde, Christian Fricke, Christian Flindt, Frank Hohls, and Rolf~J.
  Haug.
\newblock ``{Measurement of finite-frequency current statistics in a
  single-electron transistor}''.
\newblock \href{https://dx.doi.org/10.1038/ncomms1620}{Nature Communications
  {\bf 3}, 612}~(2012).
\newblock  \href{http://arxiv.org/abs/1201.2163}{arXiv:1201.2163}.

\bibitem{Bednorz.2010}
Adam Bednorz and Wolfgang Belzig.
\newblock ``{Quasiprobabilistic Interpretation of Weak Measurements in
  Mesoscopic Junctions}''.
\newblock \href{https://dx.doi.org/10.1103/physrevlett.105.106803}{Physical
  Review Letters {\bf 105}, 106803}~(2010).
\newblock  \href{http://arxiv.org/abs/1002.4984}{arXiv:1002.4984}.

\bibitem{Kuhne.2015}
J.K. K{\"u}hne, I.V. Protopopov, Y.~Oreg, and A.D. Mirlin.
\newblock ``{Measuring the Luttinger liquid parameter with shot noise}''.
\newblock \href{https://dx.doi.org/10.1016/j.physe.2015.08.010}{Physica E:
  Low-dimensional Systems and Nanostructures {\bf 74}, 651--658}~(2015).

\bibitem{Joubaud.2008}
S.~Joubaud, A.~Petrosyan, S.~Ciliberto, and N.~B. Garnier.
\newblock ``{Experimental Evidence of Non-Gaussian Fluctuations near a Critical
  Point}''.
\newblock \href{https://dx.doi.org/10.1103/physrevlett.100.180601}{Physical
  Review Letters {\bf 100}, 180601}~(2008).

\bibitem{Nambiar.2024}
Gautam Nambiar, Andrey Grankin, and Mohammad Hafezi.
\newblock ``{Diagnosing electronic phases of matter using photonic correlation
  functions}''.
\newblock \href{https://dx.doi.org/10.1103/67zs-hqf3}{Phys. Rev. X {\bf 15},
  041020}~(2025).
\newblock  \href{http://arxiv.org/abs/2410.24215}{arXiv:2410.24215}.

\bibitem{Cheung.2024}
Brian Chung~Hang Cheung and Ren-Bao Liu.
\newblock ``{Quantum Nonlinear Spectroscopy via Correlations of Weak
  Faraday-Rotation Measurements}''.
\newblock \href{https://dx.doi.org/10.1002/qute.202300286}{Advanced Quantum
  Technologies {\bf 8}, 2300286}~(2025).

\bibitem{Randi.2017}
Francesco Randi, Martina Esposito, Francesca Giusti, Oleg Misochko, Fulvio
  Parmigiani, Daniele Fausti, and Martin Eckstein.
\newblock ``{Probing the Fluctuations of Optical Properties in Time-Resolved
  Spectroscopy}''.
\newblock \href{https://dx.doi.org/10.1103/physrevlett.119.187403}{Physical
  Review Letters {\bf 119}, 187403}~(2017).
\newblock  \href{http://arxiv.org/abs/1705.08523}{arXiv:1705.08523}.

\bibitem{Kass.2024}
Benjamin Kass, Spenser Talkington, Ajit Srivastava, and Martin Claassen.
\newblock ``{Many-Body Photon Blockade and Quantum Light Generation from Cavity
  Quantum Materials}''~(2024).
\newblock  \href{http://arxiv.org/abs/2411.08964}{arXiv:2411.08964}.

\bibitem{Feng.2023}
Shi Feng, Adhip Agarwala, Subhro Bhattacharjee, and Nandini Trivedi.
\newblock ``{Anyon dynamics in field-driven phases of the anisotropic Kitaev
  model}''.
\newblock \href{https://dx.doi.org/10.1103/physrevb.108.035149}{Physical Review
  B {\bf 108}, 035149}~(2023).
\newblock  \href{http://arxiv.org/abs/2206.12990}{arXiv:2206.12990}.

\bibitem{McGinley.2024}
Max McGinley, Michele Fava, and S.~A. Parameswaran.
\newblock ``{Anomalous thermal relaxation and pump-probe spectroscopy of
  two-dimensional topologically ordered systems}''.
\newblock \href{https://dx.doi.org/10.1103/physrevb.109.075108}{Physical Review
  B {\bf 109}, 075108}~(2024).
\newblock  \href{http://arxiv.org/abs/2402.06484}{arXiv:2402.06484}.

\bibitem{Potts.2024}
Mark Potts, Roderich Moessner, and Owen Benton.
\newblock ``{Signatures of Spinon Dynamics and Phase Structure of
  Dipolar-Octupolar Quantum Spin Ices in Two-Dimensional Coherent
  Spectroscopy}''.
\newblock \href{https://dx.doi.org/10.1103/physrevlett.133.226701}{Physical
  Review Letters {\bf 133}, 226701}~(2024).

\bibitem{Galperin.2006}
Y.~M. Galperin, B.~L. Altshuler, J.~Bergli, and D.~V. Shantsev.
\newblock ``{Non-Gaussian Low-Frequency Noise as a Source of Qubit
  Decoherence}''.
\newblock \href{https://dx.doi.org/10.1103/physrevlett.96.097009}{Physical
  Review Letters {\bf 96}, 097009}~(2006).
\newblock
  \href{http://arxiv.org/abs/cond-mat/0312490}{arXiv:cond-mat/0312490}.

\bibitem{Paladino.2002}
E.~Paladino, L.~Faoro, G.~Falci, and Rosario Fazio.
\newblock ``{Decoherence and 1/f Noise in Josephson Qubits}''.
\newblock \href{https://dx.doi.org/10.1103/physrevlett.88.228304}{Physical
  Review Letters {\bf 88}, 228304}~(2002).
\newblock
  \href{http://arxiv.org/abs/cond-mat/0201449}{arXiv:cond-mat/0201449}.

\bibitem{Cai.2020}
Xiangji Cai.
\newblock ``{Quantum dephasing induced by non-Markovian random telegraph
  noise}''.
\newblock \href{https://dx.doi.org/10.1038/s41598-019-57081-8}{Scientific
  Reports {\bf 10}, 88}~(2020).
\newblock  \href{http://arxiv.org/abs/1910.14077}{arXiv:1910.14077}.

\bibitem{Cywinski.2008}
{\L}ukasz Cywi{\'n}ski, Roman~M. Lutchyn, Cody~P. Nave, and S.~Das Sarma.
\newblock ``{How to enhance dephasing time in superconducting qubits}''.
\newblock \href{https://dx.doi.org/10.1103/physrevb.77.174509}{Physical Review
  B {\bf 77}, 174509}~(2008).
\newblock  \href{http://arxiv.org/abs/0712.2225}{arXiv:0712.2225}.

\bibitem{Bergli.2007}
Joakim Bergli and Lara Faoro.
\newblock ``{Exact solution for the dynamical decoupling of a qubit with
  telegraph noise}''.
\newblock \href{https://dx.doi.org/10.1103/physrevb.75.054515}{Physical Review
  B {\bf 75}, 054515}~(2007).
\newblock
  \href{http://arxiv.org/abs/cond-mat/0609073}{arXiv:cond-mat/0609073}.

\bibitem{Klimov.2018}
P.~V. Klimov, J.~Kelly, Z.~Chen, M.~Neeley, A.~Megrant, B.~Burkett, R.~Barends,
  K.~Arya, B.~Chiaro, Yu~Chen, A.~Dunsworth, A.~Fowler, B.~Foxen, C.~Gidney,
  M.~Giustina, R.~Graff, T.~Huang, E.~Jeffrey, Erik Lucero, J.~Y. Mutus,
  O.~Naaman, C.~Neill, C.~Quintana, P.~Roushan, Daniel Sank, A.~Vainsencher,
  J.~Wenner, T.~C. White, S.~Boixo, R.~Babbush, V.~N. Smelyanskiy, H.~Neven,
  and John~M. Martinis.
\newblock ``{Fluctuations of Energy-Relaxation Times in Superconducting
  Qubits}''.
\newblock \href{https://dx.doi.org/10.1103/physrevlett.121.090502}{Physical
  Review Letters {\bf 121}, 090502}~(2018).
\newblock  \href{http://arxiv.org/abs/1809.01043}{arXiv:1809.01043}.

\bibitem{Wang.2019}
Ping Wang, Chong Chen, Xinhua Peng, J{\"o}rg Wrachtrup, and Ren-Bao Liu.
\newblock ``{Characterization of Arbitrary-Order Correlations in Quantum Baths
  by Weak Measurement}''.
\newblock \href{https://dx.doi.org/10.1103/physrevlett.123.050603}{Physical
  Review Letters {\bf 123}, 050603}~(2019).
\newblock  \href{http://arxiv.org/abs/1902.03606}{arXiv:1902.03606}.

\bibitem{Norris.2016}
Leigh~M. Norris, Gerardo~A. Paz-Silva, and Lorenza Viola.
\newblock ``{Qubit Noise Spectroscopy for Non-Gaussian Dephasing
  Environments}''.
\newblock \href{https://dx.doi.org/10.1103/physrevlett.116.150503}{Physical
  Review Letters {\bf 116}, 150503}~(2016).
\newblock  \href{http://arxiv.org/abs/1512.01575}{arXiv:1512.01575}.

\bibitem{Dong.2023nel}
Wenzheng Dong, Gerardo~A. Paz-Silva, and Lorenza Viola.
\newblock ``{Resource-efficient digital characterization and control of
  classical non-Gaussian noise}''.
\newblock \href{https://dx.doi.org/10.1063/5.0153530}{Applied Physics Letters
  {\bf 122}, 244001}~(2023).
\newblock  \href{http://arxiv.org/abs/2304.03735}{arXiv:2304.03735}.

\bibitem{Szankowski.2016}
Piotr Sza{\'n}kowski, Marek Trippenbach, and {\L}ukasz Cywi{\'n}ski.
\newblock ``{Spectroscopy of cross correlations of environmental noises with
  two qubits}''.
\newblock \href{https://dx.doi.org/10.1103/physreva.94.012109}{Physical Review
  A {\bf 94}, 012109}~(2016).
\newblock  \href{http://arxiv.org/abs/1507.03897}{arXiv:1507.03897}.

\bibitem{Kuffer.2024}
Martin Kuffer, Anal{\'i}a Zwick, and Gonzalo~A {\'A}lvarez.
\newblock ``{Sensing Out-of-Equilibrium and Quantum Non-Gaussian environments
  via induced Time-Reversal Symmetry Breaking on the quantum-probe dynamics}''.
\newblock \href{https://dx.doi.org/10.1103/PRXQuantum.6.020320}{Phys. Rev. X
  Quantum {\bf 6}, 020320}~(2025).

\bibitem{Cywinski.2014}
{\L}ukasz Cywi{'n}ski.
\newblock ``{Dynamical-decoupling noise spectroscopy at an optimal working
  point of a qubit}''.
\newblock \href{https://dx.doi.org/10.1103/physreva.90.042307}{Physical Review
  A {\bf 90}, 042307}~(2014).
\newblock  \href{http://arxiv.org/abs/1308.3102}{arXiv:1308.3102}.

\bibitem{Sung.2019}
Youngkyu Sung, F{\'e}lix Beaudoin, Leigh~M. Norris, Fei Yan, David~K. Kim,
  Jack~Y. Qiu, Uwe~von L{\"u}pke, Jonilyn~L. Yoder, Terry~P. Orlando, Simon
  Gustavsson, Lorenza Viola, and William~D. Oliver.
\newblock ``{Non-Gaussian noise spectroscopy with a superconducting qubit
  sensor}''.
\newblock \href{https://dx.doi.org/10.1038/s41467-019-11699-4}{Nature
  Communications {\bf 10}, 3715}~(2019).
\newblock  \href{http://arxiv.org/abs/1903.01043}{arXiv:1903.01043}.

\bibitem{Meinel.2022}
Jonas Meinel, Vadim Vorobyov, Ping Wang, Boris Yavkin, Mathias Pfender, Hitoshi
  Sumiya, Shinobu Onoda, Junichi Isoya, Ren-Bao Liu, and J.~Wrachtrup.
\newblock ``{Quantum nonlinear spectroscopy of single nuclear spins}''.
\newblock \href{https://dx.doi.org/10.1038/s41467-022-32610-8}{Nature
  Communications {\bf 13}, 5318}~(2022).
\newblock  \href{http://arxiv.org/abs/2109.11170}{arXiv:2109.11170}.

\bibitem{Laraoui.2011}
Abdelghani Laraoui, Jonathan~S. Hodges, Colm~A. Ryan, and Carlos~A. Meriles.
\newblock ``{Diamond nitrogen-vacancy center as a probe of random fluctuations
  in a nuclear spin ensemble}''.
\newblock \href{https://dx.doi.org/10.1103/physrevb.84.104301}{Physical Review
  B {\bf 84}, 104301}~(2011).
\newblock  \href{http://arxiv.org/abs/1104.2546}{arXiv:1104.2546}.

\bibitem{Xia.2025}
Fan Xia, Shuowei Li, Xin-Yu Pan, Ping Wang, and Gang-Qin Liu.
\newblock ``{Gaussian-to-Non-Gaussian Transition of a Quantum Spin Bath
  Revealed by Fourth-Order Correlation}''.
\newblock \href{https://dx.doi.org/10.1103/4fdk-dblq}{Physical Review Letters
  {\bf 134}, 253601}~(2025).

\bibitem{Zhang.2025}
Tao Zhang, Wentao Ji, Yuhang Guo, Mengqi Wang, Bo~Chong, Xing Rong, Fazhan Shi,
  Ping Wang, and Ya~Wang.
\newblock ``{Nanoscale Detection of Many-Body Entanglement via Multidimensional
  Correlation Imprinting}''.
\newblock \href{https://dx.doi.org/10.1103/zfk1-7t4d}{Physical Review Letters
  {\bf 135}, 240402}~(2025).

\bibitem{Kotler.2013}
Shlomi Kotler, Nitzan Akerman, Yinnon Glickman, and Roee Ozeri.
\newblock ``{Nonlinear Single-Spin Spectrum Analyzer}''.
\newblock \href{https://dx.doi.org/10.1103/physrevlett.110.110503}{Physical
  Review Letters {\bf 110}, 110503}~(2013).
\newblock  \href{http://arxiv.org/abs/1208.4017}{arXiv:1208.4017}.

\bibitem{Boter.2020}
Jelmer~M. Boter, Xiao Xue, Tobias Kr{\"a}henmann, Thomas~F. Watson, Vickram~N.
  Premakumar, Daniel~R. Ward, Donald~E. Savage, Max~G. Lagally, Mark Friesen,
  Susan~N. Coppersmith, Mark~A. Eriksson, Robert Joynt, and Lieven M.~K.
  Vandersypen.
\newblock ``{Spatial noise correlations in a Si/SiGe two-qubit device from Bell
  state coherences}''.
\newblock \href{https://dx.doi.org/10.1103/physrevb.101.235133}{Physical Review
  B {\bf 101}, 235133}~(2020).
\newblock  \href{http://arxiv.org/abs/1906.02731}{arXiv:1906.02731}.

\bibitem{Lupke.2020}
Uwe~von L{\"u}pke, F{\'e}lix Beaudoin, Leigh~M Norris, Youngkyu Sung, Roni
  Winik, Jack~Y Qiu, Morten Kjaergaard, David Kim, Jonilyn Yoder, Simon
  Gustavsson, Lorenza Viola, and William~D Oliver.
\newblock ``{Two-Qubit Spectroscopy of Spatiotemporally Correlated Quantum
  Noise in Superconducting Qubits}''.
\newblock \href{https://dx.doi.org/10.1103/prxquantum.1.010305}{PRX Quantum{\bf
  1}}~(2020).
\newblock  \href{http://arxiv.org/abs/1912.04982}{arXiv:1912.04982}.

\bibitem{Rovny.2025}
Jared Rovny, Shimon Kolkowitz, and Nathalie P~de Leon.
\newblock ``{Multi-qubit nanoscale sensing with entanglement as a resource}''.
\newblock \href{https://dx.doi.org/10.1038/s41586-025-09760-y}{Nature {\bf
  647}, 876}~(2025).
\newblock  \href{http://arxiv.org/abs/2504.12533}{arXiv:2504.12533}.

\bibitem{Hosseinabadi.2025}
H.~Hosseinabadi et~al.
\newblock ``{Unpublished preprint}''~(2025).

\bibitem{Le.2025}
Xuan~Hoang Le, Pavel~E Dolgirev, Piotr Put, Eric~L Peterson, Arjun Pillai,
  Alexander~A Zibrov, Eugene Demler, Hongkun Park, and Mikhail~D Lukin.
\newblock ``{Wideband covariance magnetometry below the diffraction limit}''.
\newblock \href{https://dx.doi.org/10.1103/7xlj-52xt}{Phys. Rev. Lett. {\bf
  135}, 170803}~(2025).
\newblock  \href{http://arxiv.org/abs/2505.00260}{arXiv:2505.00260}.

\bibitem{Zhou.2025}
Xu~Zhou, Mengqi Wang, Xiangyu Ye, Haoyu Sun, Yuhang Guo, Han Shuo, Zihua Chai,
  Wentao Ji, Kangwei Xia, Fazhan Shi, Ya~Wang, and Jiangfeng Du.
\newblock ``{Entanglement-Enhanced Nanoscale Single-Spin Sensing}''~(2025).
\newblock  \href{http://arxiv.org/abs/2504.21715}{arXiv:2504.21715}.

\bibitem{Bernien.2012}
Hannes Bernien, Lilian Childress, Lucio Robledo, Matthew Markham, Daniel
  Twitchen, and Ronald Hanson.
\newblock ``{Two-Photon Quantum Interference from Separate Nitrogen Vacancy
  Centers in Diamond}''.
\newblock \href{https://dx.doi.org/10.1103/physrevlett.108.043604}{Physical
  Review Letters {\bf 108}, 043604}~(2012).
\newblock  \href{http://arxiv.org/abs/1110.3329}{arXiv:1110.3329}.

\bibitem{Gasbarri.2018}
G.~Gasbarri and L.~Ferialdi.
\newblock ``{Stochastic unravelings of non-Markovian completely positive and
  trace-preserving maps}''.
\newblock \href{https://dx.doi.org/10.1103/physreva.98.042111}{Physical Review
  A {\bf 98}, 042111}~(2018).
\newblock  \href{http://arxiv.org/abs/1804.03742}{arXiv:1804.03742}.

\bibitem{Mukamel.1995}
Shaul Mukamel.
\newblock ``{Principles of Nonlinear Optical Spectroscopy}''.
\newblock Oxford University Press. ~(1995).

\bibitem{Hamm.2011}
Peter Hamm and Martin Zanni.
\newblock ``{Concepts and Methods of 2D Infrared Spectroscopy}''.
\newblock
  \href{https://dx.doi.org/https://doi.org/10.1017/CBO9780511675935}{Cambridge
  University Press}. New York~(2011).

\bibitem{Casola.2018}
Francesco Casola, Toeno van~der Sar, and Amir Yacoby.
\newblock ``{Probing condensed matter physics with magnetometry based on
  nitrogen-vacancy centres in diamond}''.
\newblock \href{https://dx.doi.org/10.1038/natrevmats.2017.88}{Nature Reviews
  Materials {\bf 3}, 17088}~(2018).
\newblock  \href{http://arxiv.org/abs/1804.08742}{arXiv:1804.08742}.

\bibitem{Wang.2021fej}
Yu-Xin Wang and Aashish~A. Clerk.
\newblock ``{Intrinsic and induced quantum quenches for enhancing qubit-based
  quantum noise spectroscopy}''.
\newblock \href{https://dx.doi.org/10.1038/s41467-021-26868-7}{Nature
  Communications {\bf 12}, 6528}~(2021).
\newblock  \href{http://arxiv.org/abs/2104.02047}{arXiv:2104.02047}.

\bibitem{Zheng.2022}
Tian-Xing Zheng, Anran Li, Jude Rosen, Sisi Zhou, Martin Koppenh{\"o}fer, Ziqi
  Ma, Frederic~T. Chong, Aashish~A. Clerk, Liang Jiang, and Peter~C. Maurer.
\newblock ``{Preparation of metrological states in dipolar-interacting spin
  systems}''.
\newblock \href{https://dx.doi.org/10.1038/s41534-022-00667-4}{npj Quantum
  Information {\bf 8}, 150}~(2022).
\newblock  \href{http://arxiv.org/abs/2203.03084}{arXiv:2203.03084}.

\bibitem{Fukami.2021}
Masaya Fukami, Denis~R. Candido, David~D. Awschalom, and Michael~E. Flatt{\'e}.
\newblock ``{Opportunities for Long-Range Magnon-Mediated Entanglement of Spin
  Qubits via On- and Off-Resonant Coupling}''.
\newblock \href{https://dx.doi.org/10.1103/prxquantum.2.040314}{PRX Quantum
  {\bf 2}, 040314}~(2021).
\newblock  \href{http://arxiv.org/abs/2101.09220}{arXiv:2101.09220}.

\bibitem{Humphreys.2018}
Peter~C. Humphreys, Norbert Kalb, Jaco P.~J. Morits, Raymond~N. Schouten,
  Raymond F.~L. Vermeulen, Daniel~J. Twitchen, Matthew Markham, and Ronald
  Hanson.
\newblock ``{Deterministic delivery of remote entanglement on a quantum
  network}''.
\newblock \href{https://dx.doi.org/10.1038/s41586-018-0200-5}{Nature {\bf 558},
  268--273}~(2018).
\newblock  \href{http://arxiv.org/abs/1712.07567}{arXiv:1712.07567}.

\bibitem{Pompili.2021}
M.~Pompili, S.~L.~N. Hermans, S.~Baier, H.~K.~C. Beukers, P.~C. Humphreys,
  R.~N. Schouten, R.~F.~L. Vermeulen, M.~J. Tiggelman, L.~dos~Santos Martins,
  B.~Dirkse, S.~Wehner, and R.~Hanson.
\newblock ``{Realization of a multinode quantum network of remote solid-state
  qubits}''.
\newblock \href{https://dx.doi.org/10.1126/science.abg1919}{Science {\bf 372},
  259--264}~(2021).
\newblock  \href{http://arxiv.org/abs/2102.04471}{arXiv:2102.04471}.

\bibitem{Knaut.2024}
C.~M. Knaut, A.~Suleymanzade, Y.-C. Wei, D.~R. Assumpcao, P.-J. Stas, Y.~Q.
  Huan, B.~Machielse, E.~N. Knall, M.~Sutula, G.~Baranes, N.~Sinclair,
  C.~De-Eknamkul, D.~S. Levonian, M.~K. Bhaskar, H.~Park, M.~Lon{\v c}ar, and
  M.~D. Lukin.
\newblock ``{Entanglement of nanophotonic quantum memory nodes in a telecom
  network}''.
\newblock \href{https://dx.doi.org/10.1038/s41586-024-07252-z}{Nature {\bf
  629}, 573--578}~(2024).
\newblock  \href{http://arxiv.org/abs/2310.01316}{arXiv:2310.01316}.

\bibitem{Delteil.2016}
Aymeric Delteil, Zhe Sun, Wei-bo Gao, Emre Togan, Stefan Faelt, and Ata{\c c}
  Imamo{\v g}lu.
\newblock ``{Generation of heralded entanglement between distant hole spins}''.
\newblock \href{https://dx.doi.org/10.1038/nphys3605}{Nature Physics {\bf 12},
  218--223}~(2016).
\newblock  \href{http://arxiv.org/abs/1507.00465}{arXiv:1507.00465}.

\bibitem{Huang.2020h0e}
Hsin-Yuan Huang, Richard Kueng, and John Preskill.
\newblock ``{Predicting many properties of a quantum system from very few
  measurements}''.
\newblock \href{https://dx.doi.org/10.1038/s41567-020-0932-7}{Nature Physics
  {\bf 16}, 1050--1057}~(2020).
\newblock  \href{http://arxiv.org/abs/2002.08953}{arXiv:2002.08953}.

\bibitem{Wang.2024}
Yu-Xin Wang, Jacob Bringewatt, Alireza Seif, Anthony~J Brady, Changhun Oh, and
  Alexey~V Gorshkov.
\newblock ``{Exponential entanglement advantage in sensing correlated
  noise}''~(2024).
\newblock  \href{http://arxiv.org/abs/2410.05878}{arXiv:2410.05878}.

\bibitem{Lange.2012}
Gijs~de Lange, Toeno van~der Sar, Machiel Blok, Zhi-Hui Wang, Viatcheslav
  Dobrovitski, and Ronald Hanson.
\newblock ``{Controlling the quantum dynamics of a mesoscopic spin bath in
  diamond}''.
\newblock \href{https://dx.doi.org/10.1038/srep00382}{Scientific Reports {\bf
  2}, 382}~(2012).
\newblock  \href{http://arxiv.org/abs/1104.4648}{arXiv:1104.4648}.

\bibitem{Hohenberg.1977}
P.~C. Hohenberg and B.~I. Halperin.
\newblock ``{Theory of dynamic critical phenomena}''.
\newblock \href{https://dx.doi.org/10.1103/revmodphys.49.435}{Rev. Mod. Phys.
  {\bf 49}, 435---479}~(1977).

\bibitem{Thiel.2019}
L.~Thiel, Z.~Wang, M.~A. Tschudin, D.~Rohner, I.~Guti{\'e}rrez-Lezama,
  N.~Ubrig, M.~Gibertini, E.~Giannini, A.~F. Morpurgo, and P.~Maletinsky.
\newblock ``{Probing magnetism in 2D materials at the nanoscale with
  single-spin microscopy}''.
\newblock \href{https://dx.doi.org/10.1126/science.aav6926}{Science {\bf 364},
  973--976}~(2019).
\newblock  \href{http://arxiv.org/abs/1902.01406}{arXiv:1902.01406}.

\bibitem{Ghiasi.2023}
Talieh~S. Ghiasi, Michael Borst, Samer Kurdi, Brecht~G. Simon, Iacopo Bertelli,
  Carla Boix-Constant, Samuel Ma{\~n}as-Valero, Herre S. J. van~der Zant, and
  Toeno van~der Sar.
\newblock ``{Nitrogen-vacancy magnetometry of CrSBr by diamond membrane
  transfer}''.
\newblock \href{https://dx.doi.org/10.1038/s41699-023-00423-y}{npj 2D Materials
  and Applications {\bf 7}, 62}~(2023).
\newblock  \href{http://arxiv.org/abs/2307.01129}{arXiv:2307.01129}.

\bibitem{Hohenberg.1967}
P~C Hohenberg.
\newblock ``{Existence of Long-Range Order in One and Two Dimensions}''.
\newblock \href{https://dx.doi.org/10.1103/physrev.158.383}{Phys. Rev. {\bf
  158}, 383---386}~(1967).

\bibitem{Mermin.1966}
N~D Mermin and H~Wagner.
\newblock ``{Absence of Ferromagnetism or Antiferromagnetism in One- or
  Two-Dimensional Isotropic Heisenberg Models}''.
\newblock \href{https://dx.doi.org/10.1103/physrevlett.17.1133}{Phys. Rev.
  Lett. {\bf 17}, 1133---1136}~(1966).

\bibitem{Kamenev.2011}
Alex Kamenev.
\newblock ``{Field Theory of Non-Equilibrium Systems}''.
\newblock \href{https://dx.doi.org/10.1017/cbo9781139003667}{Cambridge
  University Press}. Cambridge~(2011).

\bibitem{Larkin.1975}
A.I. Larkin and Yu.N. Ovchinikov.
\newblock ``{Nonlinear conductivity of superconductors in the mixed state}''.
\newblock J.E.T.P. {\bf 41}, 960~(1975).
\newblock  url:~\url{https://www.jetp.ras.ru/cgi-bin/dn/e_041_05_0960.pdf}.

\bibitem{Larkin.1971}
A.I. Larkin and Yu.N. Ovchinikov.
\newblock ``{INFLUENCE OF INHOMOGENEITIES ON SUPERCONDUCTOR PROPERTIES}''.
\newblock J.E.T.P. {\bf 34}, 651~(1972).
\newblock  url:~\url{https://jetp.ras.ru/cgi-bin/dn/e_034_03_0651.pdf}.

\bibitem{Chen.2024ihp}
Shaowen Chen, Seunghyun Park, Uri Vool, Nikola Maksimovic, David~A. Broadway,
  Mykhailo Flaks, Tony~X. Zhou, Patrick Maletinsky, Ady Stern, Bertrand~I.
  Halperin, and Amir Yacoby.
\newblock ``{Current induced hidden states in Josephson junctions}''.
\newblock \href{https://dx.doi.org/10.1038/s41467-024-52271-z}{Nature
  Communications {\bf 15}, 8059}~(2024).

\bibitem{Thiel.2016}
L.~Thiel, D.~Rohner, M.~Ganzhorn, P.~Appel, E.~Neu, B.~M{\"u}ller, R.~Kleiner,
  D.~Koelle, and P.~Maletinsky.
\newblock ``{Quantitative nanoscale vortex imaging using a cryogenic quantum
  magnetometer}''.
\newblock \href{https://dx.doi.org/10.1038/nnano.2016.63}{Nature Nanotechnology
  {\bf 11}, 677--681}~(2016).
\newblock  \href{http://arxiv.org/abs/1511.02873}{arXiv:1511.02873}.

\bibitem{Pelliccione.2016}
Matthew Pelliccione, Alec Jenkins, Preeti Ovartchaiyapong, Christopher Reetz,
  Eve Emmanouilidou, Ni~Ni, and Ania C.~Bleszynski Jayich.
\newblock ``{Scanned probe imaging of nanoscale magnetism at cryogenic
  temperatures with a single-spin quantum sensor}''.
\newblock \href{https://dx.doi.org/10.1038/nnano.2016.68}{Nature Nanotechnology
  {\bf 11}, 700--705}~(2016).

\bibitem{Jayaram.2025}
Sreehari Jayaram, Malik Lenger, Dong Zhao, Lucas Pupim, Takashi Taniguchi,
  Kenji Watanabe, Ruoming Peng, Marc Scheffler, Rainer St{\"o}hr, Mathias~S
  Scheurer, Jurgen Smet, and J{\"o}rg Wrachtrup.
\newblock ``{Probing Vortex Dynamics in 2D Superconductors with Scanning
  Quantum Microscope}''.
\newblock \href{https://dx.doi.org/10.1103/1fzm-pb1d}{Phys. Rev. Lett. {\bf
  135}, 126001}~(2025).
\newblock  \href{http://arxiv.org/abs/2505.03003}{arXiv:2505.03003}.

\bibitem{Koushik.2013}
R.~Koushik, Siddhartha Kumar, Kazi~Rafsanjani Amin, Mintu Mondal, John
  Jesudasan, Aveek Bid, Pratap Raychaudhuri, and Arindam Ghosh.
\newblock ``{Correlated Conductance Fluctuations Close to the
  Berezinskii-Kosterlitz-Thouless Transition in Ultrathin NbN Films}''.
\newblock \href{https://dx.doi.org/10.1103/physrevlett.111.197001}{Physical
  Review Letters {\bf 111}, 197001}~(2013).
\newblock  \href{http://arxiv.org/abs/1308.4234}{arXiv:1308.4234}.

\bibitem{Wang.2020}
Yu-Xin Wang and A.~A. Clerk.
\newblock ``{Spectral characterization of non-Gaussian quantum noise: Keldysh
  approach and application to photon shot noise}''.
\newblock \href{https://dx.doi.org/10.1103/physrevresearch.2.033196}{Physical
  Review Research {\bf 2}, 033196}~(2020).
\newblock  \href{http://arxiv.org/abs/2003.03926}{arXiv:2003.03926}.

\bibitem{Liu.202472}
A.~Liu, D.~Pavi{\'c}evi{\'c}, M.~H. Michael, A.~G. Salvador, P.~E. Dolgirev,
  M.~Fechner, A.~S. Disa, P.~M. Lozano, Q.~Li, G.~D. Gu, E.~Demler, and
  A.~Cavalleri.
\newblock ``{Probing inhomogeneous cuprate superconductivity by terahertz
  Josephson echo spectroscopy}''.
\newblock \href{https://dx.doi.org/10.1038/s41567-024-02643-5}{Nature Physics
  {\bf 20}, 1751--1756}~(2024).
\newblock  \href{http://arxiv.org/abs/2308.14849}{arXiv:2308.14849}.

\bibitem{Salvador.2024}
Alex~G{\'o}mez Salvador, Pavel~E. Dolgirev, Marios~H. Michael, Albert Liu,
  Danica Pavicevic, Michael Fechner, Andrea Cavalleri, and Eugene Demler.
\newblock ``{Principles of two-dimensional terahertz spectroscopy of collective
  excitations: The case of Josephson plasmons in layered superconductors}''.
\newblock \href{https://dx.doi.org/10.1103/physrevb.110.094514}{Physical Review
  B {\bf 110}, 094514}~(2024).
\newblock  \href{http://arxiv.org/abs/2401.05503}{arXiv:2401.05503}.

\bibitem{Salvador.2025}
Alex~G{\'o}mez Salvador, Ivan Morera, Marios~H Michael, Pavel~E Dolgirev,
  Danica Pavicevic, Albert Liu, Andrea Cavalleri, and Eugene Demler.
\newblock ``{Two-dimensional spectroscopy of bosonic collective excitations in
  disordered many-body systems}''~(2025).
\newblock  \href{http://arxiv.org/abs/2501.16856}{arXiv:2501.16856}.

\bibitem{Zou.2022}
Ji~Zou, Shu Zhang, and Yaroslav Tserkovnyak.
\newblock ``{Bell-state generation for spin qubits via dissipative coupling}''.
\newblock \href{https://dx.doi.org/10.1103/physrevb.106.l180406}{Physical
  Review B {\bf 106}, L180406}~(2022).
\newblock  \href{http://arxiv.org/abs/2108.07365}{arXiv:2108.07365}.

\bibitem{Bradley.2019}
C.~E. Bradley, J.~Randall, M.~H. Abobeih, R.~C. Berrevoets, M.~J. Degen, M.~A.
  Bakker, M.~Markham, D.~J. Twitchen, and T.~H. Taminiau.
\newblock ``{A Ten-Qubit Solid-State Spin Register with Quantum Memory up to
  One Minute}''.
\newblock \href{https://dx.doi.org/10.1103/physrevx.9.031045}{Physical Review X
  {\bf 9}, 031045}~(2019).
\newblock  \href{http://arxiv.org/abs/1905.02094}{arXiv:1905.02094}.

\bibitem{Seetharam.2023}
Kushal Seetharam, Debopriyo Biswas, Crystal Noel, Andrew Risinger, Daiwei Zhu,
  Or~Katz, Sambuddha Chattopadhyay, Marko Cetina, Christopher Monroe, Eugene
  Demler, and Dries Sels.
\newblock ``{Digital quantum simulation of NMR experiments}''.
\newblock \href{https://dx.doi.org/10.1126/sciadv.adh2594}{Science Advances
  {\bf 9}, eadh2594}~(2023).
\newblock  \href{http://arxiv.org/abs/2109.13298}{arXiv:2109.13298}.

\bibitem{Smart.2022}
Scott~E. Smart, Zixuan Hu, Sabre Kais, and David~A. Mazziotti.
\newblock ``{Relaxation of stationary states on a quantum computer yields a
  unique spectroscopic fingerprint of the computer's noise}''.
\newblock \href{https://dx.doi.org/10.1038/s42005-022-00803-8}{Communications
  Physics {\bf 5}, 28}~(2022).
\newblock  \href{http://arxiv.org/abs/2104.14552}{arXiv:2104.14552}.

\bibitem{Guimaraes.2023}
Jos{\'e}~D. Guimar{\~a}es, James Lim, Mikhail~I. Vasilevskiy, Susana~F. Huelga,
  and Martin~B. Plenio.
\newblock ``{Noise-Assisted Digital Quantum Simulation of Open Systems Using
  Partial Probabilistic Error Cancellation}''.
\newblock \href{https://dx.doi.org/10.1103/prxquantum.4.040329}{PRX Quantum
  {\bf 4}, 040329}~(2023).
\newblock  \href{http://arxiv.org/abs/2302.14592}{arXiv:2302.14592}.

\bibitem{Dag.2024}
Ceren~B Dag, Hanzhen Ma, P~Myles Eugenio, Fang Fang, and Susanne~F Yelin.
\newblock ``{Emergent disorder and sub-ballistic dynamics in quantum
  simulations of the Ising model using Rydberg atom arrays}''~(2024).
\newblock  \href{http://arxiv.org/abs/2411.13643}{arXiv:2411.13643}.

\bibitem{Berg.2023}
Ewout van~den Berg, Zlatko~K. Minev, Abhinav Kandala, and Kristan Temme.
\newblock ``{Probabilistic error cancellation with sparse Pauli-Lindblad models
  on noisy quantum processors}''.
\newblock \href{https://dx.doi.org/10.1038/s41567-023-02042-2}{Nature Physics
  {\bf 19}, 1116--1121}~(2023).
\newblock  \href{http://arxiv.org/abs/2201.09866}{arXiv:2201.09866}.

\bibitem{Temme.2017}
Kristan Temme, Sergey Bravyi, and Jay~M. Gambetta.
\newblock ``{Error Mitigation for Short-Depth Quantum Circuits}''.
\newblock \href{https://dx.doi.org/10.1103/physrevlett.119.180509}{Physical
  Review Letters {\bf 119}, 180509}~(2017).
\newblock  \href{http://arxiv.org/abs/1612.02058}{arXiv:1612.02058}.

\bibitem{Torre.2023}
Emanuele G~Dalla Torre and Mor~M Roses.
\newblock ``{Dissipative mean-field theory of IBM utility experiment}''~(2023).
\newblock  \href{http://arxiv.org/abs/2308.01339}{arXiv:2308.01339}.

\bibitem{Choi.20170po}
Joonhee Choi, Soonwon Choi, Georg Kucsko, Peter~C. Maurer, Brendan~J. Shields,
  Hitoshi Sumiya, Shinobu Onoda, Junichi Isoya, Eugene Demler, Fedor Jelezko,
  Norman~Y. Yao, and Mikhail~D. Lukin.
\newblock ``{Depolarization Dynamics in a Strongly Interacting Solid-State Spin
  Ensemble}''.
\newblock \href{https://dx.doi.org/10.1103/physrevlett.118.093601}{Physical
  Review Letters {\bf 118}, 093601}~(2017).
\newblock  \href{http://arxiv.org/abs/1608.05471}{arXiv:1608.05471}.

\bibitem{Kucsko.2018}
G.~Kucsko, S.~Choi, J.~Choi, P.~C. Maurer, H.~Zhou, R.~Landig, H.~Sumiya,
  S.~Onoda, J.~Isoya, F.~Jelezko, E.~Demler, N.~Y. Yao, and M.~D. Lukin.
\newblock ``{Critical Thermalization of a Disordered Dipolar Spin System in
  Diamond}''.
\newblock \href{https://dx.doi.org/10.1103/physrevlett.121.023601}{Physical
  Review Letters {\bf 121}, 023601}~(2018).
\newblock  \href{http://arxiv.org/abs/1609.08216}{arXiv:1609.08216}.

\bibitem{Andersen.2024}
Trond~I Andersen, Nikita Astrakhantsev, Amir~H Karamlou, Julia Berndtsson,
  Johannes Motruk, Aaron Szasz, Jonathan~A Gross, Alexander Schuckert, Tom
  Westerhout, Yaxing Zhang, Ebrahim Forati, Dario Rossi, Bryce Kobrin,
  Agustin~Di Paolo, Andrey~R Klots, Ilya Drozdov, Vladislav~D Kurilovich, Andre
  Petukhov, Lev~B Ioffe, Andreas Elben, Aniket Rath, Vittorio Vitale, Benoit
  Vermersch, Rajeev Acharya, Laleh~Aghababaie Beni, Kyle Anderson, Markus
  Ansmann, Frank Arute, Kunal Arya, Abraham Asfaw, Juan Atalaya, Brian Ballard,
  Joseph~C Bardin, Andreas Bengtsson, Alexander Bilmes, Gina Bortoli, Alexandre
  Bourassa, Jenna Bovaird, Leon Brill, Michael Broughton, David~A Browne, Brett
  Buchea, Bob~B Buckley, David~A Buell, Tim Burger, Brian Burkett, Nicholas
  Bushnell, Anthony Cabrera, Juan Campero, Hung-Shen Chang, Zijun Chen, Ben
  Chiaro, Jahan Claes, Agnetta~Y Cleland, Josh Cogan, Roberto Collins, Paul
  Conner, William Courtney, Alexander~L Crook, Sayan Das, Dripto~M Debroy,
  Laura~De Lorenzo, Alexander Del~Toro Barba, Sean Demura, Paul Donohoe, Andrew
  Dunsworth, Clint Earle, Alec Eickbusch, Aviv~Moshe Elbag, Mahmoud Elzouka,
  Catherine Erickson, Lara Faoro, Reza Fatemi, Vinicius~S Ferreira,
  Leslie~Flores Burgos, Austin~G Fowler, Brooks Foxen, Suhas Ganjam, Robert
  Gasca, William Giang, Craig Gidney, Dar Gilboa, Marissa Giustina, Raja
  Gosula, Alejandro~Grajales Dau, Dietrich Graumann, Alex Greene, Steve
  Habegger, Michael~C Hamilton, Monica Hansen, Matthew~P Harrigan, Sean~D
  Harrington, Stephen Heslin, Paula Heu, Gordon Hill, Markus~R Hoffmann,
  Hsin-Yuan Huang, Trent Huang, Ashley Huff, William~J Huggins, Sergei~V
  Isakov, Evan Jeffrey, Zhang Jiang, Cody Jones, Stephen Jordan, Chaitali
  Joshi, Pavol Juhas, Dvir Kafri, Hui Kang, Kostyantyn Kechedzhi, Trupti
  Khaire, Tanuj Khattar, Mostafa Khezri, M{\'a}ria Kieferov{\'a}, Seon Kim,
  Alexei Kitaev, Paul~V Klimov, Alexander~N Korotkov, Fedor Kostritsa,
  John~Mark Kreikebaum, David Landhuis, Brandon~W Langley, Pavel Laptev,
  Kim-Ming Lau, Lo{\"i}ck~Le Guevel, Justin Ledford, Joonho Lee, Kenny Lee,
  Yuri~D Lensky, Brian~J Lester, Wing~Yan Li, Alexander~T Lill, Wayne Liu,
  William~P Livingston, Aditya Locharla, Daniel Lundahl, Aaron Lunt, Sid
  Madhuk, Ashley Maloney, Salvatore Mandr{\`a}, Leigh~S Martin, Orion Martin,
  Steven Martin, Cameron Maxfield, Jarrod~R McClean, Matt McEwen, Seneca Meeks,
  Kevin~C Miao, Amanda Mieszala, Sebastian Molina, Shirin Montazeri, Alexis
  Morvan, Ramis Movassagh, Charles Neill, Ani Nersisyan, Michael Newman,
  Anthony Nguyen, Murray Nguyen, Chia-Hung Ni, Murphy~Yuezhen Niu, William~D
  Oliver, Kristoffer Ottosson, Alex Pizzuto, Rebecca Potter, Orion Pritchard,
  Leonid~P Pryadko, Chris Quintana, Matthew~J Reagor, David~M Rhodes, Gabrielle
  Roberts, Charles Rocque, Eliott Rosenberg, Nicholas~C Rubin, Negar Saei,
  Kannan Sankaragomathi, Kevin~J Satzinger, Henry~F Schurkus, Christopher
  Schuster, Michael~J Shearn, Aaron Shorter, Noah Shutty, Vladimir Shvarts,
  Volodymyr Sivak, Jindra Skruzny, Spencer Small, W~Clarke Smith, Sofia
  Springer, George Sterling, Jordan Suchard, Marco Szalay, Alex Sztein, Douglas
  Thor, Alfredo Torres, M~Mert Torunbalci, Abeer Vaishnav, Sergey Vdovichev,
  Benjamin Villalonga, Catherine~Vollgraff Heidweiller, Steven Waltman,
  Shannon~X Wang, Theodore White, Kristi Wong, Bryan~W Woo, Cheng Xing, Z~Jamie
  Yao, Ping Yeh, Bicheng Ying, Juhwan Yoo, Noureldin Yosri, Grayson Young, Adam
  Zalcman, Ningfeng Zhu, Nicholas Zobrist, Hartmut Neven, Ryan Babbush, Sergio
  Boixo, Jeremy Hilton, Erik Lucero, Anthony Megrant, Julian Kelly, Yu~Chen,
  Vadim Smelyanskiy, Guifre Vidal, Pedram Roushan, Andreas~M Lauchli, Dmitry~A
  Abanin, and Xiao Mi.
\newblock ``{Thermalization and Criticality on an Analog-Digital Quantum
  Simulator}''.
\newblock \href{https://dx.doi.org/10.1038/s41586-024-08460-3}{Nature {\bf
  638}, 79}~(2025).
\newblock  \href{http://arxiv.org/abs/2405.17385}{arXiv:2405.17385}.

\bibitem{Rosenberg.2024h49}
E.~Rosenberg, T.~I. Andersen, R.~Samajdar, A.~Petukhov, J.~C. Hoke, D.~Abanin,
  A.~Bengtsson, I.~K. Drozdov, C.~Erickson, P.~V. Klimov, X.~Mi, A.~Morvan,
  M.~Neeley, C.~Neill, R.~Acharya, R.~Allen, K.~Anderson, M.~Ansmann, F.~Arute,
  K.~Arya, A.~Asfaw, J.~Atalaya, J.~C. Bardin, A.~Bilmes, G.~Bortoli,
  A.~Bourassa, J.~Bovaird, L.~Brill, M.~Broughton, B.~B. Buckley, D.~A. Buell,
  T.~Burger, B.~Burkett, N.~Bushnell, J.~Campero, H.-S. Chang, Z.~Chen,
  B.~Chiaro, D.~Chik, J.~Cogan, R.~Collins, P.~Conner, W.~Courtney, A.~L.
  Crook, B.~Curtin, D.~M. Debroy, A.~Del~Toro Barba, S.~Demura, A.~Di Paolo,
  A.~Dunsworth, C.~Earle, L.~Faoro, E.~Farhi, R.~Fatemi, V.~S. Ferreira,
  L.~Flores Burgos, E.~Forati, A.~G. Fowler, B.~Foxen, G.~Garcia,
  {\'E}.~Genois, W.~Giang, C.~Gidney, D.~Gilboa, M.~Giustina, R.~Gosula,
  A.~Grajales Dau, J.~A. Gross, S.~Habegger, M.~C. Hamilton, M.~Hansen, M.~P.
  Harrigan, S.~D. Harrington, P.~Heu, G.~Hill, M.~R. Hoffmann, S.~Hong,
  T.~Huang, A.~Huff, W.~J. Huggins, L.~B. Ioffe, S.~V. Isakov, J.~Iveland,
  E.~Jeffrey, Z.~Jiang, C.~Jones, P.~Juhas, D.~Kafri, T.~Khattar, M.~Khezri,
  M.~Kieferov{\'a}, S.~Kim, A.~Kitaev, A.~R. Klots, A.~N. Korotkov,
  F.~Kostritsa, J.~M. Kreikebaum, D.~Landhuis, P.~Laptev, K.-M. Lau, L.~Laws,
  J.~Lee, K.~W. Lee, Y.~D. Lensky, B.~J. Lester, A.~T. Lill, W.~Liu,
  A.~Locharla, S.~Mandr{\`a}, O.~Martin, S.~Martin, J.~R. McClean, M.~McEwen,
  S.~Meeks, K.~C. Miao, A.~Mieszala, S.~Montazeri, R.~Movassagh,
  W.~Mruczkiewicz, A.~Nersisyan, M.~Newman, J.~H. Ng, A.~Nguyen, M.~Nguyen,
  M.~Y. Niu, T.~E. O'Brien, S.~Omonije, A.~Opremcak, R.~Potter, L.~P. Pryadko,
  C.~Quintana, D.~M. Rhodes, C.~Rocque, N.~C. Rubin, N.~Saei, D.~Sank,
  K.~Sankaragomathi, K.~J. Satzinger, H.~F. Schurkus, C.~Schuster, M.~J.
  Shearn, A.~Shorter, N.~Shutty, V.~Shvarts, V.~Sivak, J.~Skruzny, W.~Clarke
  Smith, R.~D. Somma, G.~Sterling, D.~Strain, M.~Szalay, D.~Thor, A.~Torres,
  G.~Vidal, B.~Villalonga, C.~Vollgraff Heidweiller, T.~White, B.~W.~K. Woo,
  C.~Xing, Z.~Jamie Yao, P.~Yeh, J.~Yoo, G.~Young, A.~Zalcman, Y.~Zhang,
  N.~Zhu, N.~Zobrist, H.~Neven, R.~Babbush, D.~Bacon, S.~Boixo, J.~Hilton,
  E.~Lucero, A.~Megrant, J.~Kelly, Y.~Chen, V.~Smelyanskiy, V.~Khemani,
  S.~Gopalakrishnan, T.~Prosen, and P.~Roushan.
\newblock ``{Dynamics of magnetization at infinite temperature in a Heisenberg
  spin chain}''.
\newblock \href{https://dx.doi.org/10.1126/science.adi7877}{Science {\bf 384},
  48--53}~(2024).
\newblock  \href{http://arxiv.org/abs/2306.09333}{arXiv:2306.09333}.

\bibitem{Kim.2023}
Youngseok Kim, Andrew Eddins, Sajant Anand, Ken~Xuan Wei, Ewout van~den Berg,
  Sami Rosenblatt, Hasan Nayfeh, Yantao Wu, Michael Zaletel, Kristan Temme, and
  Abhinav Kandala.
\newblock ``{Evidence for the utility of quantum computing before fault
  tolerance}''.
\newblock \href{https://dx.doi.org/10.1038/s41586-023-06096-3}{Nature {\bf
  618}, 500--505}~(2023).

\bibitem{Taminiau.2014}
T.~H. Taminiau, J.~Cramer, T.~van~der Sar, V.~V. Dobrovitski, and R.~Hanson.
\newblock ``{Universal control and error correction in multi-qubit spin
  registers in diamond}''.
\newblock \href{https://dx.doi.org/10.1038/nnano.2014.2}{Nature Nanotechnology
  {\bf 9}, 171--176}~(2014).
\newblock  \href{http://arxiv.org/abs/1309.5452}{arXiv:1309.5452}.

\bibitem{Chirame.2025}
Sanket Chirame, Fiona~J. Burnell, Sarang Gopalakrishnan, and Abhinav Prem.
\newblock ``{Stable Symmetry-Protected Topological Phases in Systems with
  Heralded Noise}''.
\newblock \href{https://dx.doi.org/10.1103/physrevlett.134.010403}{Physical
  Review Letters {\bf 134}, 010403}~(2025).
\newblock  \href{http://arxiv.org/abs/2404.16962}{arXiv:2404.16962}.

\bibitem{Feynman.1963}
R.P Feynman and F.L Vernon.
\newblock ``{The theory of a general quantum system interacting with a linear
  dissipative system}''.
\newblock \href{https://dx.doi.org/10.1016/0003-4916(63)90068-x}{Annals of
  Physics {\bf 24}, 118--173}~(1963).

\bibitem{Leggett.1987}
A.~J. Leggett, S.~Chakravarty, A.~T. Dorsey, Matthew P.~A. Fisher, Anupam Garg,
  and W.~Zwerger.
\newblock ``{Dynamics of the dissipative two-state system}''.
\newblock \href{https://dx.doi.org/10.1103/revmodphys.59.1}{Reviews of Modern
  Physics {\bf 59}, 1--85}~(1987).

\bibitem{Caldeira.1983rkp}
A.O Caldeira and A.J Leggett.
\newblock ``{Quantum tunnelling in a dissipative system}''.
\newblock \href{https://dx.doi.org/10.1016/0003-4916(83)90202-6}{Annals of
  Physics {\bf 149}, 374--456}~(1983).

\bibitem{Sieberer.2016}
L~M Sieberer, M~Buchhold, and S~Diehl.
\newblock ``{Keldysh field theory for driven open quantum systems}''.
\newblock \href{https://dx.doi.org/10.1088/0034-4885/79/9/096001}{Reports on
  Progress in Physics {\bf 79}, 096001}~(2016).
\newblock  \href{http://arxiv.org/abs/1512.00637}{arXiv:1512.00637}.

\bibitem{Kamar.2024}
Naushad~Ahmad Kamar, Daniel~A. Paz, and Mohammad~F. Maghrebi.
\newblock ``{Spin-boson model under dephasing: Markovian versus non-Markovian
  dynamics}''.
\newblock \href{https://dx.doi.org/10.1103/physrevb.110.075126}{Physical Review
  B {\bf 110}, 075126}~(2024).
\newblock  \href{http://arxiv.org/abs/2305.00110}{arXiv:2305.00110}.

\end{thebibliography}

\end{document}